\documentclass[12pt]{article}

\def\be{\begin{equation}}
\def\ee{\end{equation}}
\def\bea{\begin{eqnarray}}
\def\eea{\end{eqnarray}}
\def\vp{\varphi}
\def\K{K{\"a}hler}
\newcommand{\rf}[1]{(\ref{#1})}

\usepackage{epsfig}

\usepackage{amssymb,latexsym}
\usepackage{graphicx}
\usepackage{amssymb, amsmath, amsopn, amsthm}
\usepackage[section]{placeins}

\makeatletter \@addtoreset{equation}{section}
\makeatother

\begin{document}
\thispagestyle{empty}
\hskip 1 cm

\vspace{25pt}
\begin{center}
    { \LARGE{\bf Inflationary Cosmology after Planck 2013}\footnote{
Based on the lectures at the Les Houches School ``Post-Planck Cosmology,'' 2013}}
    \vspace{33pt}

  {\large  {\bf  Andrei Linde}}

    \vspace{10pt}

    \vspace{10pt} {Department of Physics,
    Stanford University, Stanford, CA 94305}

    \vspace{20pt}
 \end{center}

\begin{abstract}

I give a general review of inflationary cosmology and of its present status, in view of the 2013 data release by the Planck satellite.
A specific emphasis is given to the new broad class of theories, the cosmological attractors, which have nearly model-independent predictions converging at the sweet spot of the Planck data in the  $(n_{s},r)$ plane. I also discuss the problem of initial conditions for the theories favored by the Planck data.

\end{abstract}

\

\tableofcontents

\newpage

\parskip 6pt

\section{Introduction}
We were waiting for the recent Planck data release for many years. There were persistent rumors that this data are going to confirm the existence of relatively large non-Gaussian perturbations, which would rule out the simplest single-field inflationary models, but would open the door towards complicated multi-field models of inflation, which would provide lots of work for many generations of cosmologists. For several years, many scientists were preparing themselves to this grave eventuality. But instead of that, Planck 2013 found that the non-Gaussian perturbations of the most dangerous type are practically absent \cite{Ade:2013uln}, and therefore one could return to investigation of the simplest and nicest inflationary models. Some of us, me included, were ecstatic to hear the news, but for many others it was a disappointment. It could seem that now we have nothing new to say about inflation since everything related to its simplest versions was already known, classified, forgotten, and rediscovered over and over again during the last 30 years. However, as I will argue in these lectures, the new results obtained by Planck pushed us in an unexpected direction, which stimulated development of new concepts and new broad classes of inflationary models with most interesting properties. In addition to that, the split between the inflationary cosmology and its alternatives became even more apparent after the Planck 2013. Whereas the Planck team emphasized that their results confirm basic principles of inflation and rule out many alternative theories \cite{Ade:2013uln}, some of the proponents of these alternatives arrived at an absolutely opposite conclusion. Thus life after Planck 2013 became quite exciting. Hopefully these lectures may help to understand the essence of the recent developments and debates. But in order to do it, we must start from the very beginning. For the first chapters, I will borrow some materials from my lectures written several years ago \cite{Linde:2007fr}, and then we will fast forward to the very recent developments.
Finally, I should note that some issues closely related to the subjects covered in my lectures have been discussed by other speakers at the Les Houches school. In particular, the approach to cosmology based on supergravity and superconformal theory was described by Renata Kallosh \cite{RenataLecture}, and string theory approach to  inflation was presented by Eva Silverstein \cite{Silverstein:2013wua}.

\section {Brief history of inflation}

Several ingredients of inflationary cosmology were discovered in the beginning of the 70's. The first realization was that the energy density of a scalar field plays the role of the vacuum energy/cosmological constant \cite{Linde:1974at}, which was changing during the cosmological phase transitions \cite{Kirzhnits:1972ut}. In certain cases these changes  occur discontinuously, due to first order phase transitions from a supercooled vacuum state (false vacuum) \cite{Kirzhnits:1976ts}.

In 1978, Gennady Chibisov and I tried to use these facts to construct a cosmological model involving exponential expansion of the universe in a supercooled vacuum as a source of the entropy of the universe, but we immediately realized that the universe becomes very inhomogeneous after the bubble wall collisions. I mentioned our work in my review article \cite{Linde:1978px}, but did not pursue this idea any further.

The first semi-realistic model of inflationary type was proposed by Alexei
Starobinsky in 1980 \cite{Starobinsky:1980te}. It was based on investigation of
conformal anomalies in quantum gravity.  His model was rather complicated, and its goals were in a certain sense opposite to the goals of inflationary cosmology. Instead of attempting to solve the homogeneity and isotropy problems, Starobinsky considered the model of  the universe which was homogeneous and isotropic from the very beginning, and emphasized that his scenario was ``the extreme opposite of Misner's initial chaos''   \cite{Starobinsky:1980te}.  

On the other hand, Starobinsky's model did not suffer from the graceful exit problem, and it was the first model predicting  gravitational waves with a flat spectrum \cite{Starobinsky:1980te}. The first mechanism of production of adiabatic perturbations of the metric with a nearly flat spectrum, which are responsible for galaxy production, and which were found by the observations of the CMB anisotropy,  was proposed by  Mukhanov  and Chibisov \cite{Mukh} in the context of this model. They also pointed out that the spectrum should not be exactly flat, the fact that was confirmed by Planck 2013 at the 5$\sigma$ level.

A much simpler inflationary model with a very clear physical
motivation was proposed by Alan Guth in 1981 \cite{Guth}.  His model,
which is now called ``old inflation,'' was based on the theory of
supercooling during the cosmological phase transitions
\cite{Kirzhnits:1976ts}. Even though this scenario did not work in its original form,  it
played a profound role in the development of inflationary
cosmology since it contained a very clear explanation how
inflation may solve the major cosmological problems.

According to this scenario,  inflation is as   exponential
expansion of the universe in a supercooled false vacuum state.
False vacuum is a metastable state without any fields or particles
but with large energy density. Imagine a universe filled with such
``heavy nothing.'' When the universe expands, empty space remains
empty, so its energy density does not change. The universe with a
constant energy density expands exponentially, thus we have
inflation in the false vacuum. This expansion makes the universe very big and very flat. Then the false vacuum decays, the
bubbles of the new phase collide, and our universe becomes hot.

Unfortunately, this simple and intuitive picture of inflation in the false vacuum state is somewhat misleading. If the probability of the bubble formation is large, bubbles of the new phase are formed near each other, inflation is too short to solve any problems, and the bubble wall collisions make the universe extremely inhomogeneous.  If they are formed far away from each other, which is the case if the probability of their formation is small and inflation is long, each of these bubbles represents a separate open universe with a vanishingly small $\Omega$. Both options are unacceptable, which has lead to the conclusion that this scenario does not work and cannot be improved (graceful exit problem) \cite{Guth,Hawking:1982ga,Guth:1982pn}.

A solution of this problem was found in 1981 with the invention of the new inflationary theory \cite{New}. (Three months later, the same solution was also proposed in \cite{New2}, with a reference to my earlier work \cite{New}.) In this theory, inflation may
begin either in the false vacuum,  or in an unstable state at the
top of the effective potential. Then the inflaton field $\phi$
slowly rolls down to the minimum of its effective potential.  The
motion of the field away from the false vacuum is of crucial
importance: density perturbations produced during the slow-roll
inflation are inversely proportional to $\dot \phi$
\cite{Mukh,Hawk,Mukh2}. Thus the key difference between the new
inflationary scenario and the old one is that the useful part of
inflation in the new scenario, which is responsible for the
homogeneity of our universe, does {\it not} occur in the false
vacuum state, where $\dot\phi =0$.

Soon after the invention of the new inflationary scenario it became very popular. Unfortunately, this scenario was plagued by its own problems. In most versions of this scenario the inflaton field must have an extremely small coupling
constant, so it could not be in thermal equilibrium with other
matter fields, as required in  \cite{Guth,New}. The theory of cosmological phase transitions, which
was the basis for old and new inflation, did not work in this case. Moreover, thermal equilibrium requires many particles
interacting with each other. This means that new inflation could
explain why our universe was so large only if it was very large
and contained many particles from the very beginning \cite{Linde:2005ht}.

Old and new inflation represented a substantial but incomplete
modification of the big bang theory. It was still assumed that the
universe was in a state of thermal equilibrium from the very
beginning, that it was relatively homogeneous and large enough to
survive until the beginning of inflation, and that the stage of
inflation was just an intermediate stage of the evolution of the
universe. In the beginning of the 80's these assumptions seemed
most natural and practically unavoidable. On the basis of all
available observations (CMB, abundance of light elements)
everybody believed that the universe was created in a hot big
bang. That is why it was so difficult to overcome a certain
psychological barrier and abandon all of these assumptions. This
was done in 1983 with the invention of the chaotic inflation scenario
\cite{Linde:1983gd}. This scenario resolved all problems of old and new
inflation. According to this scenario, inflation may begin even if there was no thermal
equilibrium in the early universe, and  it may occur even in
the theories with simplest potentials such as $V(\phi) \sim
\phi^2$. But it is not limited to the theories with polynomial potentials: chaotic inflation occurs in {\it any} theory where the potential has a sufficiently flat region, which allows the existence of the slow-roll regime \cite{Linde:1983gd}.

This change of perspective happened 30 years ago, and since that time cosmologists almost never returned to the original idea of inflation due to high temperature phase transitions in grand unified theories. This idea was abandoned by all experts actively working in this area of physics, but unfortunately it is still present in virtually all textbooks on astrophysics, which incorrectly describe inflation as an exponential expansion in a supercooled false vacuum state during the cosmological phase transitions in grand unified theories. Apparently the front lines of the development of this branch of science moved forward too fast, so let us remember what exactly happened with the development of chaotic inflation.

\section{Chaotic Inflation: the simplest models}

Consider  the simplest model of a scalar field $\phi$ with a mass
$m$ and with the potential energy density $V(\phi)  = {m^2\over 2}
\phi^2$. Since this function has a minimum at $\phi = 0$,  one may
expect that the scalar field $\phi$ should oscillate near this
minimum. This is indeed the case if the universe does not expand,
in which case equation of motion for the scalar field  coincides
with equation for harmonic oscillator, $\ddot\phi = -m^2\phi$.

\begin{figure}[h!t!]
\centering{
\includegraphics[height=8.3cm]{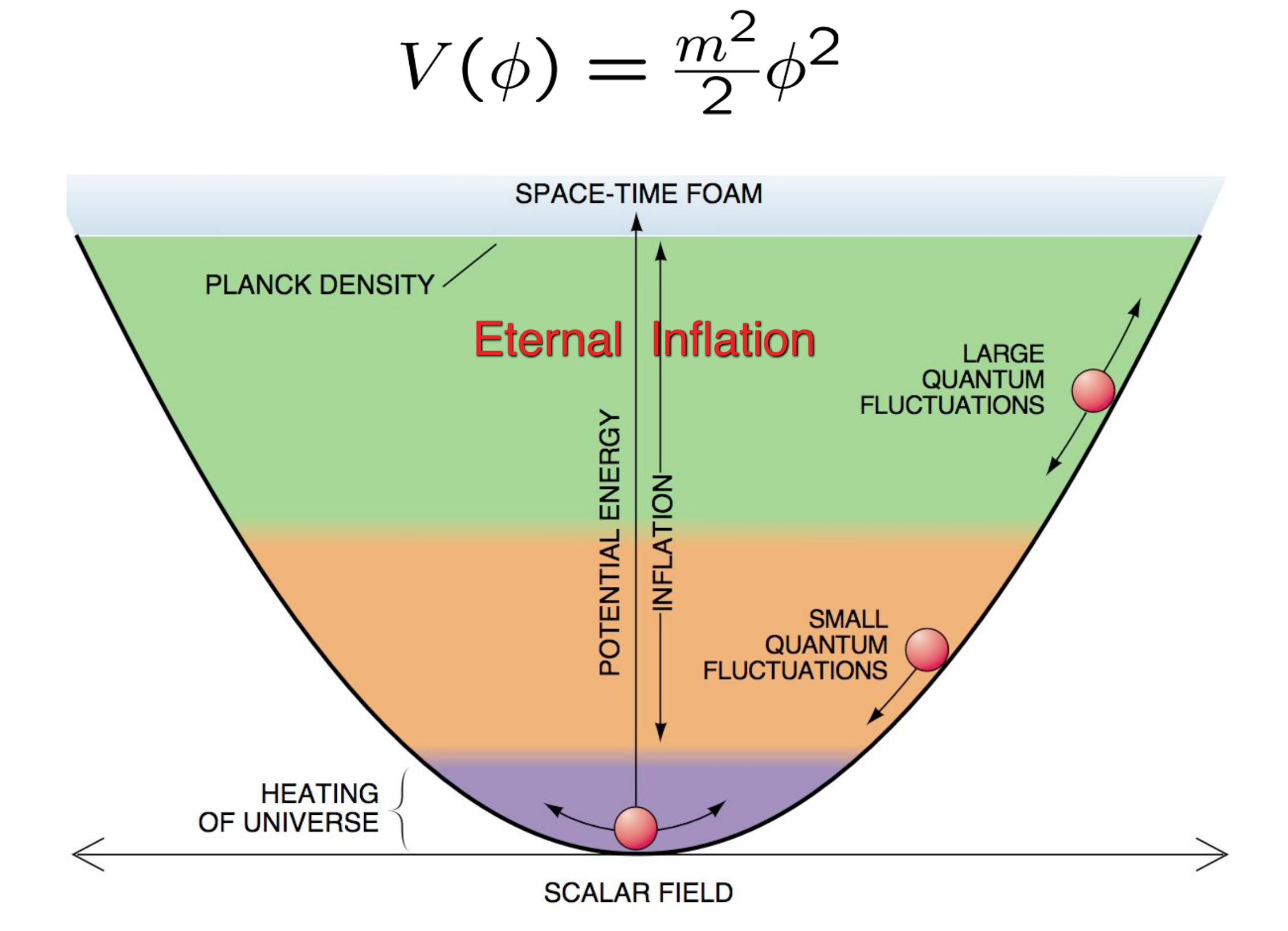}}
\caption{\footnotesize Motion of the scalar field in the theory with $V(\phi) =
{m^2\over 2} \phi^2$. Several different regimes are possible,  depending
on the value of the field $\phi$. If the potential energy density of the
field is greater than the Planck density $M_p^4 = 1$, $\phi \gtrsim m^{-1}$,
quantum fluctuations of space-time are so strong that one cannot describe
it in usual terms. At a somewhat smaller energy density  (for $m  \lesssim V(\phi) \lesssim 1$,\, $m^{-1/2} \lesssim \phi \lesssim  m^{-1}$)
 quantum fluctuations of space-time are small, but quantum fluctuations of
the scalar field $\phi$ may be large. Jumps of the scalar field due to these
quantum fluctuations lead to a process of eternal self-reproduction of
inflationary universe which we are going to discuss later. At even
smaller values of $V(\phi)$ (for $m^2  \lesssim V(\phi) \lesssim m$,\, $1 \lesssim \phi \lesssim  m^{-1/2}$)
fluctuations of the field $\phi$ are small; it slowly moves down as a
ball in a viscous liquid. Inflation occurs for $1 \lesssim \phi \lesssim  m^{-1}$. Finally, near the minimum of $V(\phi)$ (for $\phi \lesssim 1$) the scalar
field rapidly oscillates, creates elementary particles, and the
universe becomes hot.} \label{fig:Fig1}
\end{figure}

However, because of the expansion of the universe with Hubble
constant $H = \dot a/a $, an additional  term $3H\dot\phi$ appears
in the harmonic oscillator equation:
\begin{equation}\label{1x}
 \ddot\phi + 3H\dot\phi = -m^2\phi \ .
\end{equation}
The term $3H\dot\phi$ can be interpreted as a friction term. The
Einstein equation for a homogeneous universe containing scalar
field $\phi$ looks as follows:
\begin{equation}\label{2x}
H^2 +{k\over a^2} ={1\over 6}\, \left(\dot \phi ^2+m^2 \phi^2)
\right) \ .
\end{equation}
Here $k = -1, 0, 1$ for an open, flat or closed universe
respectively. We work in units $M_p^{-2} = 8\pi G = 1$.

If   the scalar field $\phi$  initially was large,   the Hubble
parameter $H$ was large too, according to the second equation.
This means that the friction term $3H\dot\phi$ was very large, and
therefore    the scalar field was moving   very slowly, as a ball
in a viscous liquid. Therefore at this stage the energy density of
the scalar field, unlike the  density of ordinary matter, remained
almost constant, and expansion of the universe continued with a
much greater speed than in the old cosmological theory. Due to the
rapid growth of the scale of the universe and a slow motion of the
field $\phi$, soon after the beginning of this regime one has
$\ddot\phi \ll 3H\dot\phi$, $H^2 \gg {k\over a^2}$, $ \dot \phi
^2\ll m^2\phi^2$, so  the system of equations can be simplified:
\begin{equation}\label{E04}
H= {\dot a \over a}   ={ m\phi\over \sqrt 6}\ , ~~~~~~  \dot\phi =
-m\  \sqrt{2\over 3}     .
\end{equation}
The first equation shows that if the field $\phi$ changes slowly,
the size of the universe in this regime grows approximately as
$e^{Ht}$, where $H = {m\phi\over\sqrt 6}$. This is the stage of
inflation, which ends when the field $\phi$ becomes much smaller
than $M_p=1$. Solution of these equations shows that after a long
stage of inflation  the universe initially filled with the field
$\phi   \gg 1$  grows  exponentially \cite{Linde:2005ht}, 
\begin{equation}\label{E04aa}
 a= a_0
\ e^{\phi^2/4} \   .
\end{equation}

Thus, inflation does not require initial state of thermal equilibrium, 
supercooling and tunneling from the false  vacuum. It appears in the theories that can be as simple as a theory of a harmonic oscillator \cite{Linde:1983gd}. Only when it was
realized, it became clear  that inflation is not just a trick
necessary to fix  problems of the old big bang theory, but a
generic cosmological regime.


The first models of chaotic inflation were based on the theories
with polynomial potentials, such as $\phi^{n}$ or $\pm {m^2\over 2}
\phi^2 +{\lambda\over 4} \phi^4$. But, as emphasized in  \cite{Linde:1983gd}, the main idea of this scenario is quite generic. One may consider {\it any}
potential $V(\phi)$, polynomial or not, with or without
spontaneous symmetry breaking, and study all possible initial
conditions without assuming that the universe was in a state of
thermal equilibrium, and that the field $\phi$ was in the minimum
of its effective potential from the very beginning.

This scenario strongly deviated from the standard lore of the hot
big bang theory and was psychologically difficult to accept.
Therefore during the first few years after invention of chaotic
inflation many authors claimed that the idea of chaotic initial
conditions is unnatural, and made attempts to realize the new
inflation scenario based on the theory of high-temperature phase
transitions, despite numerous problems associated with it. Some
authors believed that the theory must satisfy so-called `thermal
constraints' which were necessary to ensure that the minimum of
the effective potential at large $T$ should be at $\phi=0$
\cite{OvrStein}, even though the scalar field in the models  they
considered was not in a state of thermal equilibrium with other
particles. 

The issue of thermal initial conditions played the central role in the  long debate  about new inflation versus chaotic inflation in the 80's. This debate continued for many years, and a significant part of my book written in 1990 was dedicated to it  \cite{Linde:2005ht}. By now  the debate is over: no realistic versions of new inflation based on the theory of thermal phase transitions and supercooling have been proposed so far. Gradually it became clear that the idea of chaotic initial conditions is most general, and it is much easier to construct a consistent cosmological theory without making unnecessary assumptions about thermal equilibrium and high temperature phase transitions in the early universe. 

As a result, the corresponding terminology changed.  Chaotic inflation, as defined in  \cite{Linde:1983gd}, describes inflation in {\it all} models with sufficiently flat potentials, including the potentials with a flat maximum, originally used in new inflation \cite{New}.  Now the versions of  inflationary scenario with such potentials for simplicity are often called `new inflation', even though inflation begins there not as in the original new inflation scenario, but as in the chaotic inflation
scenario. To avoid this terminological misunderstanding, some authors call the version of chaotic inflation scenario, where inflation occurs near the top of the scalar potential, a `hilltop inflation'.

\section{Initial conditions in the simplest models of chaotic inflation}\label{ini}

To check whether this regime is indeed generic, let us consider a closed universe of a smallest initial  size $l \sim 1$
(in Planck units), which emerges  from the
space-time foam, or from singularity, or from `nothing'  in a state
with  the Planck density $\rho \sim 1$. Only starting from this moment, i.e. at $\rho
\lesssim 1$, can we describe this domain as  a {\it classical} universe.  Thus,
at this initial moment the sum of the kinetic energy density, gradient energy
density, and the potential energy density  is of the order unity:\, ${1\over
2} \dot\phi^2 + {1\over 2} (\partial_i\phi)^2 +V(\phi) \sim 1$.

There are no {\it a priori} constraints on
the initial value of the scalar field in this domain, except for the
constraint ${1\over 2} \dot\phi^2 + {1\over 2} (\partial_i\phi)^2 +V(\phi) \sim
1$.  Let us consider for a moment a theory with $V(\phi) = const$. This theory
is invariant under the {\it shift symmetry}  $\phi\to \phi + c$. Therefore, in such a
theory {\it all} initial values of the homogeneous component of the scalar field
$\phi$ are equally probable.  

The only constraint on the  amplitude of the field appears if the
effective potential is not constant, but grows and becomes greater
than the Planck density at $\phi > \phi_p$, where  $V(\phi_p) = 1$. This
constraint implies that $\phi \lesssim \phi_p$,
but there is no reason to expect that initially
$\phi$ must be much smaller than $\phi_p$. This suggests that the  typical initial value
 of the field $\phi$ in such a  theory is  $\phi 
\sim  \phi_p$.

Thus, we expect that typical initial conditions correspond to
${1\over 2}
\dot\phi^2 \sim {1\over 2} (\partial_i\phi)^2\sim V(\phi) = O(1)$.
If ${1\over 2} \dot\phi^2 + {1\over 2} (\partial_i\phi)^2
 \lesssim V(\phi)$
in the domain under consideration, then inflation begins,
and then within
the Planck time the terms  ${1\over 2} \dot\phi^2$ and ${1\over 2}
(\partial_i\phi)^2$ become much smaller than $V(\phi)$, which ensures
continuation of inflation.  It seems therefore that chaotic inflation
occurs under rather natural initial conditions, if it can begin at $V(\phi)
\sim 1$ \cite{Linde:1985ub,Linde:2005ht}. 

One can get a different perspective on this issue by studying the probability of quantum creation of the universe from ``nothing.'' The basic idea is that quantum fluctuations can create a small universe from nothing if it can be done quickly, in agreement with the quantum uncertainty relation $\Delta E\cdot \Delta t \lesssim 1$. The total energy of scalar field in a closed inflationary universe is proportional to its minimal volume $H^{-3} \sim V^{-3/2}$ multiplied by the energy density $V(\phi)$:~ $E \sim V^{-1/2}$.  Therefore such a universe can appear quantum mechanically within the time $ \Delta t \gtrsim 1$ if  $V(\phi)$ is not too much smaller than the Planck density $O(1)$. 

This qualitative conclusion agrees with the result of the investigation in the context of quantum cosmology. Indeed, according to \cite{Linde:1983mx,Vilenkin:1984wp}, the probability of quantum creation of a closed universe is proportional to
\be
P \sim \exp \left(-{24\pi^{2}\over V}\right)\ ,
\ee
which means that the universe can be created if $V$ is not too much smaller than the Planck density. The Euclidean approach to the quantum creation of the universe is based on the analytical continuation of the Euclidean de Sitter solution to the real time. This continuation is possible if $\dot \phi = 0$ at the moment of quantum creation of the universe. Thus in the simplest chaotic inflation model with $V(\phi)  = {m^2\over 2}
\phi^2$ the universe  is created in a state with $V(\phi) \sim 1$, $\phi \sim m^{-1} \gg 1$ and $\dot \phi = 0$, which is a perfect initial condition for inflation in this model \cite{Linde:1983mx,Linde:2005ht}.

One should note that there are many other attempts to evaluate the probability of initial conditions for inflation. For example, if one interprets the square of the  Hartle-Hawking wave function \cite{Hartle:1983ai} as a probability of initial condition, one obtains a paradoxical answer $P \sim \exp \left({24\pi^{2}\over V}\right)$, which could seem to imply that it is easier to create the universe with  $V\to 0$ and with an infinitely large total energy $E \sim V^{-1/2} \to \infty$. There were many attempts to improve this anti-intuitive answer, but from my perspective these attempts were misplaced: The Hartle-Hawking wave function was derived in  \cite{Hartle:1983ai} as a wave function for the {\it ground state of the universe}, and therefore it describes the most probable {\it final} state of the universe, instead of the probability of initial conditions; see a discussion of this issue in \cite{Open,Linde:2005ht,Linde:2006nw}.

Another attempt to study this problem was made few years ago by Gibbons and Turok  \cite{Gibbons:2006pa}. They studied classical solutions describing a combined evolution of a scalar field and the scale factor of a homogeneous universe, ignoring gradients of the scalar field. An advantage of inflationary regime is that kinetic energy of a scalar field in an expanding universe rapidly decreases, and the general solution of equations of motion for the scalar field rapidly approaches the inflationary slow-roll attractor solutions, which practically does not depend on initial velocity of the inflaton field. However, the authors of  \cite{Gibbons:2006pa} imposed ``initial conditions'' not at the beginning of inflation but at its end. Since one can always reverse the direction of time in the solutions, one can always relate the conditions at the end of inflation to the conditions at its beginning. If one assumes that certain conditions at the end of inflation are equally probable, then one may conclude that the probability of initial conditions suitable for inflation is very small \cite{Gibbons:2006pa}.  

From our perspective \cite{Linde:2007fr,Kofman:2002cj}, we have here the same paradox which is encountered in the discussion of the  growth of entropy. If one take any thermodynamical system, its entropy will always grow. However, if we make a movie of this process, and play it back starting from the end of the process, then the final conditions for the original system become the initial conditions for the time-reversed system, and we will see the entropy decreasing. That is why replacing initial conditions by final conditions can be very misleading. If one replaces initial conditions for inflation by final conditions after inflation and then runs the same process back in time, the inflationary attractor trajectory will look as a repulser. This is the main reason of the negative conclusion of Ref.  \cite{Gibbons:2006pa}. One can easily show that if one uses the same methods and the same probability measure as \cite{Gibbons:2006pa}, but imposes the initial conditions prior to the beginning of inflation (as one should do), rather than at its end (as it was done in  \cite{Gibbons:2006pa}), one finds that inflation is most probable, in agreement with the arguments given in the first part of this section. 

But the main problem with  \cite{Gibbons:2006pa} is not in the replacing initial conditions by the final ones. The methods developed there are valid only for the classical evolution of the universe. But we are talking about initial conditions for the classical evolution. By definition, initial conditions for the classical evolution are determined by processes {\it prior} to the stage of classical evolution, at the quantum epoch near the singularity. Thus the initial conditions have been formed at the quantum cosmology stage where the methods of  \cite{Gibbons:2006pa} are inapplicable. If one uses quantum cosmology to determine the most probable initial conditions, then, independently of the choice of the  tunneling wave function or the Hartle-Hawking wave function, one finds that initially $\dot \phi = 0$. This contradicts the claims of  \cite{Gibbons:2006pa} that initially $\dot\phi^{2} \gg V(\phi)$ \cite{Linde:2007fr}.

 Let us summarize our conclusions so far:  Initial conditions for inflation are quite natural if we consider inflationary models where inflation can occur for $V \sim 1$.
This condition is satisfied in the simplest versions of chaotic inflation, such as the models with $V\sim \phi^{n}$. But what if inflation can occur only at $V \ll 1$?  We will discuss this issue in section \ref{torus} in application to models of a single scalar field, and we will continue this discussion latter in the context of models with many scalars.

\section{Solving the cosmological problems}

As we will see shortly, the realistic value of the mass $m$ is about $6\times 10^{-6}$, in Planck units. Therefore, according to Eq. (\ref{E04aa}), the total amount of inflation achieved starting from $V(\phi) \sim 1$ is of the order $10^{10^{10}}$. The total duration of
inflation in this model is about $10^{-30}$ seconds. When inflation
ends, the scalar field $\phi$ begins to   oscillate near the
minimum of $V(\phi)$. As any rapidly oscillating classical field,
it looses its energy by creating pairs of elementary particles.
These particles interact with each other and come to a state of
thermal equilibrium with some temperature $T_{r}$ \cite{oldtheory}-\cite{thermalization}.
From this time on, the universe can be described by the usual big
bang theory.

The main difference between inflationary theory and the old
cosmology becomes clear when one calculates the size of a typical
inflationary domain at the end of inflation. Investigation of this
question    shows that even if  the initial size of   inflationary
universe  was as small as the Planck size $l_P \sim 10^{-33}$ cm,
after $10^{-30}$ seconds of inflation   the universe acquires a
huge size of   $l \sim 10^{10^{10}}$ cm! This number is
model-dependent, but in all realistic models the  size of the
universe after inflation appears to be many orders of magnitude
greater than the size of the part of the universe which we can see
now, $l \sim 10^{28}$ cm. This immediately solves most of the
problems of the old cosmological theory \cite{Linde:1983gd,Linde:2005ht}.

Our universe is almost exactly homogeneous on  large scale because
all inhomogeneities were exponentially stretched during inflation.
The density of  primordial monopoles  and other undesirable
``defects'' becomes exponentially diluted by inflation.   The
universe   becomes enormously large. Even if it was a closed
universe of a size
 $\sim 10^{-33}$ cm, after inflation the distance between its ``South'' and
``North'' poles becomes many orders of magnitude greater than
$10^{28}$ cm. We see only a tiny part of the huge cosmic balloon.
That is why nobody  has ever seen how parallel lines cross. That
is why the universe looks so flat.

If our universe initially consisted of many domains with
chaotically distributed scalar field  $\phi$ (or if one considers
different universes with different values of the field), then
domains in which the scalar field was too small never inflated.
The main contribution to the total volume of the universe will be
given by those domains which originally contained large scalar
field $\phi$. Inflation of such domains creates huge homogeneous
islands out of initial chaos. (That is why I called this scenario
``chaotic inflation.'') Each  homogeneous domain in this scenario
after inflation becomes much greater than the size of the observable part of the
universe.

It is instructive to compare this scenario with the standard hot big bang model. In that model, the universe was born at the cosmological singularity, but it becomes possible to describe it in terms of classical space-time only when time becomes greater than the Planck time $t_{p} \sim M_{p}^{-1} \sim 1$. At that time,  the temperature of matter was given by the Planck temperature $T_{p} \sim 1$, and the density of the universe was given by the Planck density $\rho_{p} \sim T_{p}^{4} \sim  1$. The size of the causally connected part of the universe at the Planck time was $ct_{p} \sim 1$. Each such part contained a single particle with the Planck temperature. The evolution of the universe was supposed to be nearly adiabatic, so the total number of particles in the universe was nearly conserved. We do not know how many particles are in the whole universe, but their number must be greater than the total number of particles in the observable part of the universe, which is approximately $10^{{90}}$. The adiabaticity condition implies that at the Planck time the universe also contained about $10^{90}$ particles, and therefore it consisted of $10^{90}$ causally disconnected parts. The probability that all of these independent parts emerged from singularity at the same time with the same energy density and pressure is smaller than $\exp\bigl({-10^{90}}\bigr)$, and the probability that our universe from the very beginning was uniform with an accuracy better than $10^{-4}$ is even much smaller.

This should be contrasted with the simplest versions of the chaotic inflation scenario. As discussed in the previous section, the main condition required there was the existence of a single  domain of size $O(1)$ where the kinetic and gradient energy of the scalar field is few times smaller than its potential energy $V(\phi) \sim 1$. For sufficiently flat potentials, it leads to inflation, and then the whole universe appears as a result of expansion of a single Planck size domain. As we argued in the previous section, the probability of this process is not exponentially suppressed.

\section{Creation of matter after inflation: reheating and preheating}

The theory of reheating of the universe after inflation is the most
important application of the quantum theory of particle creation, since
almost all matter constituting the universe  was created during this
process.

At the stage of inflation all energy is concentrated in a classical
slowly moving inflaton field $\phi$. Soon after the end of inflation this
field begins to oscillate near the minimum of its effective potential.
Eventually it produces many elementary particles, they interact  with
each other and come to a state of thermal equilibrium with some
temperature $T_r$.

Early discussions of reheating of the universe after inflation  \cite{oldtheory} were based on the idea that the homogeneous inflaton field can be represented as a collection of the particles of the field $\phi$. Each of these particles decayed independently. This process can be studied by the usual perturbative approach to particle decay. Typically, it takes thousands of oscillations of the inflaton field until it decays into usual elementary particles by this mechanism.  More recently, however, it was discovered that coherent field effects such as parametric resonance can lead to the decay of the homogeneous field much faster than would have been predicted by perturbative methods, within few dozen oscillations \cite{KLS}. These coherent effects produce high energy, non-thermal fluctuations that could have
significance for understanding developments in the early universe, such as baryogenesis.  This early stage of rapid nonperturbative decay  was called `preheating.' 
In \cite{tach} it was found that another effect known as tachyonic preheating can lead to even faster decay than parametric
resonance. This effect occurs whenever the homogeneous field rolls down a tachyonic ($V''<0$) region of its
potential. When this happens, a tachyonic instability leads to exponentially rapid growth of all long wavelength modes with 
$k^2<|V''|$. This growth can often drain all of the energy from the homogeneous field within a single oscillation.

We are now in a position to classify the dominant mechanisms by which the homogeneous inflaton field decays in different classes of
inflationary models. Even though all of these models, strictly speaking,  belong to the general class of chaotic inflation (none of them is based on the theory of thermal initial conditions), one can  break them into three classes: small field, or new inflation models \cite{New}, large field, or chaotic inflation models of the type of the model $m^2\phi^2/2$ \cite{Linde:1983gd}, and multi-field, or hybrid models \cite{Hybrid,F,D}. This classification is imprecise, but still rather helpful.

In the simplest versions of chaotic inflation, the stage of preheating is generally dominated by
parametric resonance, although there are parameter ranges where this
can not occur \cite{KLS}.    In \cite{tach} it was shown that tachyonic preheating
dominates the preheating phase in hybrid models of inflation. New inflation in this respect occupies an intermediate position between chaotic inflation and hybrid inflation:  If spontaneous symmetry breaking in this scenario is very large, reheating occurs due to parametric resonance and perturbative decay. However, for the models with spontaneous symmetry breaking at or below the GUT scale, $\phi \ll 10^{{-2}} M_p$, preheating occurs due to a combination of tachyonic preheating and parametric resonance. The resulting effect is very strong, so that the homogeneous mode of the inflaton field typically decays within few oscillations \cite{Desroche:2005yt}. 

A detailed investigation of preheating usually requires lattice simulations, which can be achieved following \cite{latticeold,latticeeasy}.  Note that preheating is not the last stage of reheating; it is followed by a period of turbulence \cite{thermalization}, by a much slower perturbative decay described by the methods developed in \cite{oldtheory}, and by eventual thermalization.

\section{Quantum fluctuations and density perturbations
\label{Perturb}}

The average amplitude of inflationary perturbations generated during a
typical time interval $H^{-1}$ is given by
\cite{Vilenkin:wt,Linde:uu}
\begin{equation}\label{E23}
|\delta\phi(x)| \approx \frac{H}{2\pi}\ .
\end{equation}

These fluctuations lead to density perturbations that later
produce galaxies. The theory of this effect  is very complicated
\cite{Mukh,Hawk}, and it was fully understood only in the second
part of the 80's \cite{Mukh2}. The main idea can be described as
follows:

Fluctuations of the field $\phi$ lead to a local delay of the time
of the end of inflation,  $\delta t = {\delta\phi\over \dot\phi}
\sim {H\over 2\pi \dot \phi}$. Once the usual post-inflationary
stage begins, the density of the universe starts to decrease as
$\rho = 3 H^2$, where $H \sim t^{-1}$. Therefore a local delay of
expansion leads to a local density increase $\delta_H$ such that
$\delta_H \sim \delta\rho/\rho \sim  {\delta t/t}$. Combining
these estimates together yields the famous result
\cite{Mukh,Hawk,Mukh2}
\begin{equation}\label{E24}
\delta_H \sim \frac{\delta\rho}{\rho} \sim {H^2\over 2\pi\dot\phi}  = {V^{3/2}\over 2\sqrt 3 \pi V_{\phi}}
\ .
\end{equation}
where $V_{\phi} = {dV\over d\phi}$. The field $\phi$ during inflation
changes very slowly, so the quantity ${H^2\over 2\pi\dot\phi}$
remains almost constant over exponentially large range of
wavelengths. This means that the spectrum of perturbations of
metric is almost exactly flat.

When the fluctuations of the scalar field $\phi$ are first produced (frozen), their wavelength is given by $H(\phi)^{-1}$. At the end of inflation in the simplest model ${m^{2}\over 2}\phi^{2}$, the wavelength grows by the factor of $e^{\phi^2/4}$, see Eq. (\ref{E04aa}). In other words,  the logarithm of the wavelength $l$ of the perturbations of the metric is proportional to the value of $\phi^2$ at the moment when these perturbations were produced. As a result, according to  (\ref{E24}), the amplitude of perturbations of the metric in this model depends on the wavelength  logarithmically: $\delta_H \sim   {m\, \ln l} $. A similar logarithmic dependence (with different powers of the logarithm) appears in other versions of chaotic inflation with $V \sim \phi^{n}$.

At the first glance, this logarithmic deviation from scale invariance, which was first discovered in \cite{Mukh}, could seem inconsequential. However, in a certain sense it is similar to the famous logarithmic dependence of the coupling constants in QCD, where it leads to asymptotic freedom at high energies, instead of simple scaling invariance \cite{Gross:1973id,Politzer:1973fx}. In QCD, the slow growth of the coupling constants at small momenta/large distances is responsible for nonperturbative effects resulting in quark confinement. In  inflationary theory, the  slow growth of the amplitude of perturbations of the metric at large distances is equally important. It leads to the existence of the regime of eternal inflation and to the fractal structure of the universe on  super-large scales, see Section \ref{eternalinfl}.

Observations provide us with  information about a rather limited
range of $l$. Therefore it is often possible to parametrize the scale dependence
of density perturbations by a simple power law, $\delta_H \sim
l^{(1-n_{s})/2} \sim k^{(n_{s}-2)/2}$, where $k$ is the corresponding momentum. An exactly flat spectrum, called Harrison-Zeldovich spectrum, would correspond to $n_{s} = 1$.

To make it easier to compare inflationary predictions with the observational results obtained by Planck, we will list here the definitions and relations used in the  Planck data release of 2013  \cite{Ade:2013uln}. 
The square of the amplitude of scalar perturbations $\delta_H$ produced at the time when the inflaton field was equal to some value $\phi$ is given by
\be
 A_{s} = {V^{3}(\phi)\over 12\pi^{2} V^{2}_{\phi}(\phi)} \ .
\ee
For tensors, one has
\be
 A_{t} = {2V(\phi)\over 3\pi^{2}} \ .
\ee

By relating $\phi$ and $k$, one can write, approximately,
\begin{eqnarray}
 \label{eq:P_R} A_{s}(k)&=& A_{s}(k_*)
  \left(\frac{k}{k_*}\right)^{n_s-1}, \\ \label{eq:P_h} A_{t}(k)&=& A_{t}(k_*)
 \left(\frac{k}{k_*}\right)^{n_t},
\end{eqnarray}
where $A_{s}(k_*)$ is a normalization constant, and $k_*$ is a normalization point, which is often taken to be $k_* \sim 0.05$/Mpc. Here we ignored running of the indexes $n_{s}$ and $n_{t}$ since there is no observational evidence that it is significant.

One can also introduce the tensor/scalar ratio
$r$, the relative amplitude  of the tensor to  scalar modes,
\begin{equation}
 \label{eq:rdef} r \equiv \frac{A_{t}(k_*)}{A_{s}(k_*)}.
\end{equation}

There are two most important slow-roll parameters \cite{LL}
\begin{eqnarray}
 \label{eq:eps} \epsilon =
  \frac{1}{2}\left(\frac{V_{\phi}}{V}\right)^2, ~~
  \eta  =  \frac{V_{\phi\phi}}{V}, 
  \label{eq:xi} 
\end{eqnarray}
where prime denotes derivatives with respect to the field
$\phi$. All parameters must be
smaller than one for the slow-roll approximation to be valid. 

 A standard slow roll analysis gives observable
quantities in terms of the slow roll parameters to first order as
\begin{eqnarray}
 \label{eq:A} &&A_{s}  =  \frac{V}{24\pi^2\epsilon}  \ ,\\
    \label{eq:n_s} &&n_s =
1  -6\epsilon + 2\eta = 1 
  -{3}\left(\frac{V'}{V}\right)^2 + 2\frac{V''}{V} \ , \\
  \label{eq:r} &&r   =  16 \epsilon, \\
  \label{eq:n_t} &&n_t   =  -2\epsilon = -\frac{r}{8} \ .
\end{eqnarray}
 The equation
$n_t=-r/8$ is known as the consistency relation for single-field
inflation models; it becomes an inequality for multi-field
inflation models. If $V$ during inflation is sufficiently large, as in the simplest models of chaotic inflation, one may have a chance to find the tensor contribution to the CMB anisotropy. The most important information which can be obtained from the cosmological observations at present is related to Eqs. (\ref{eq:A}), (\ref{eq:n_s}) and  and (\ref{eq:r}).

According to \cite{Ade:2013uln},
\be
A_{s}(k_{*}) \approx 2.2 \times 10^{{-9}} \ ,
\ee
which corresponds to 
\begin{eqnarray}
 \label{eq:V} {V^{{3/2}}\over V'} \simeq 5.1\times 10^{-4} \ .
\end{eqnarray}
Here $V(\phi)$ should be evaluated for the value of the field $\phi$ which is determined by the condition that the perturbations produced at the moment when the field was equal $\phi$  evolve into the present time perturbations with momentum $k_{*} \sim 0.05$/Mpc. The results of the calculations corresponding to the perturbations on the present scale of the horizon slightly differ from these results because the spectrum is not exactly flat, but the difference is rather small, so one can use the results given above as a good first approximation for the amplitude of the perturbations on the scale of the horizon. In many inflationary models, these perturbations are produced at $N \approx 60$ e-foldings before the end of inflation. However, the number $N$ can be somewhat different, depending on details of the post-inflationary evolution. That is why when comparing expected results for various models with observations, cosmologists often make calculations for $N = 60$ and also for $N = 50$.

The number of e-foldings can be calculated in the slow roll approximation using the relation
\begin{eqnarray}
 \label{eq:N} {N} \simeq \int_{\phi_{\rm end}}^{\phi}{V\over V'} d\phi \ .
\end{eqnarray}
Equation (\ref{eq:V})  leads to the relation between $r$, $V$ and $H$, in Planck units:
\begin{eqnarray}
 \label{eq:rvh} {r} \approx 3 \times  10^{7}~V  \approx 10^{8}~ H^{2} \ .
\end{eqnarray}
 The latest Planck results, in combination with the results of WMAP and the results based on investigation of baryon acoustic oscillations (BAO), imply that
\be\label{r}
r \lesssim 0.11 
\ee
and 
\be \label{eq:ns}
n_{s} = 0.9607 \pm 0.0063 \ .
\ee
 These relations are very useful for comparing inflationary models with observations. A more detailed discussion of observational constraints can be found in Section \ref{observ}.
 
 Up to now, we discussed perturbations produced by  the simplest, standard mechanism described in \cite{Mukh,Hawk,Mukh2}. However, in models involving additional light scalars fields $\sigma$, other mechanisms of generation of perturbations are possible.

Let us assume, for example, that the products of the inflaton decay after inflation are ultra-relativistic and rapidly rapidly loose energy in an expanding universe, whereas the field $\sigma$ is heavy decay with a significant delay. In that case, the field $\sigma$ may dominate the energy density of the universe and perturbations of this field suddenly become important. When the field $\sigma$ decays, its perturbations under certain conditions can be  converted into the usual adiabatic perturbations of the metric. If this conversion is incomplete, one obtains a mixture of isocurvature and adiabatic perturbations  \cite{Ax,Mollerach:1989hu},  which should be very small in accordance with recent observational data  \cite{Ade:2013uln}. On the other hand, if the conversion is complete, one obtains a novel mechanism of generation of purely adiabatic density perturbations, which is called the curvaton mechanism \cite{LM,Enqvist:2001zp,LW,Moroi:2001ct}. Note that in many of the original versions of the curvaton scenario, it was assumed that at the epoch of the curvaton decay the universe was dominated by the classical curvaton field. In this case the curvaton decay produced significant amount of isocurvature perturbations, which strongly constrain such models  \cite{Ade:2013uln}. However, if one makes a natural assumption that a large number of curvaton {\it particles} are produced during the inflaton decay, this problem disappears \cite{Linde:2005yw,Demozzi:2010aj}. 
There are other closely related but different mechanisms of generation of adiabatic perturbations during inflation \cite{mod1}. 
All of these non-standard mechanisms are more complicated than the original one, but one should keep them in mind since they sometimes may work in the situations where the standard one does not. Therefore they can give us an additional freedom in finding realistic models of inflationary cosmology. 

\section{Universe or Multiverse?}

For most of the 20th century, scientific thought was dominated by the idea of uniformity of the universe and the uniqueness of laws of physics. Indeed, the cosmological observations indicated that the universe on the largest possible scales is almost exactly uniform, with the accuracy better than 1 in 10000. Uniformity of the universe was somewhat of a mystery, and instead of explaining it, scientists invoked the ``cosmological principle,'' which states that the universe must be uniform, because... well, because of the cosmological principle. 

A similar situation occurs with respect to the uniqueness of laws of physics. We knew, for example, that the electron mass is the same everywhere in the observable part of the universe, so the obvious assumption was that it must take the same value everywhere, it is just a physical constant. Therefore, for a long time, one of the great goals of physics was to find a single fundamental theory, which would unify all fundamental interactions and provide an unambiguous explanation for all known parameters of particle physics.

The strongest proponent of this idea was Einstein, who said ``I would like to state a theorem which at present can not be based upon anything more than a faith in the simplicity, i.e. intelligibility, of nature: There are no arbitrary constantsÉ that is to say, nature is so constituted that it is possible logically to lay down such strongly determined laws that within these laws only rationally completely determined constants occur (not constants, therefore, whose numerical value could be changed without destroying the theory).''

Intellectual honesty of Einstein shines through this statement of faith: ``a theorem which at present can not be based upon anything more than a faith of simplicity.'' However, we know that our world is anything but simple. Even the simplest theories have billions of different solutions. Simple laws of electromagnetism describe an infinite variety of colors in the spectrum. Biological laws allow an incredible variety of species. Even such simple substances as water can be either liquid, solid, or vapor. The same chemical composition - and yet a profound difference: fish can only live in liquid water. And our universe is not simple either: it contains many important deviations from its uniformity, such as galaxies, stars and planets. Thus the cosmological principle, which insists that the universe must be uniform, is not entirely correct, and if it is not entirely correct, it cannot be a general fundamental principle of science.

\begin{figure}
\centering{
\includegraphics[height=5.5cm]{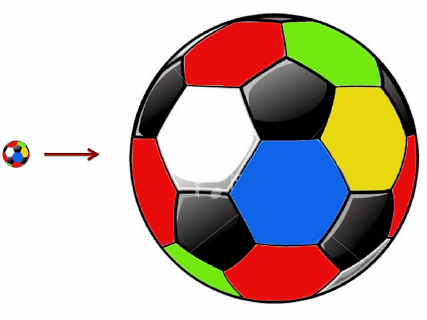}}
\caption{\footnotesize Inflation of a tiny universe consisting of many different parts with different properties makes each of these parts exponentially large and  uniform, while preserving distinct features of each of these parts. The universe  becomes a multiverse consisting of exponentially large parts with different properties.} \label{simplemulti}
\end{figure}

The best available explanation of the observed uniformity of the universe is provided by inflation. 
However, as soon as this mechanism was proposed, it was realized that inflation, while explaining why our part of the world is so uniform, does not predict that this uniformity must extend for the whole universe \cite{Linde:1982ur,nuff}. To give an analogy, suppose the universe is a surface of a big soccer ball consisting of multicolored hexagons, see Fig.  \ref{simplemulti}.    During inflation, the size of each hexagon becomes exponentially large. If inflation is powerful enough, those who live in a black part will never see parts of the universe of any different color, they will believe that the whole universe is black, and they will try to find a scientific explanation why the whole universe must be black. Those who live in a red universe will never see the black parts and therefore they will think that there is no other universe than the red universe, and everybody who says otherwise are heretics.

But what if the whole universe started in the red state? In the next section we will show how quantum fluctuations can lead to transitions between different colors and simultaneously make inflation eternal. This means that almost independently of the initial state of the universe, eventually it becomes a multicolored eternally growing fractal. 


\section{Eternal inflation}\label{eternalinfl}

One of the most unusual features of inflationary cosmology is the process of self-reproduction of inflationary
universe. This process was known to exist in old inflationary
theory \cite{Guth}, but there it was considered a big problem rather than an advantage. Then a similar process was found to exist in the new inflation scenario \cite{StLin,Linde:1982ur,Vilenkin:xq}. One of the most important features of this scenario was that the universe during this process was divided into many different exponentially large parts with different properties, which provided the first scientific justification of the cosmological anthropic principle in this context. The possibility of this combination of eternal inflation with the anthropic principle was first noticed at the famous Nuffield Symposium on inflationary cosmology, \cite{Linde:1982ur,nuff}. 
Duly impressed by free lunches provided to the participants, I summarized the main consequence  of this scenario as follows \cite{nuff}: ``As was claimed by Guth (1982), the inflationary universe is the only example  of a free lunch (all matter in this scenario is created from the unstable vacuum). Now we can add that the inflationary universe is the only lunch at which all possible dishes are available.'' 

However, the significance of the self-reproduction of the universe was fully recognized only after the discovery of the
regime of self-reproduction of inflationary
universe in the chaotic inflation scenario, which was called ``eternal inflation'' \cite{Eternal,Linde:1993xx}. It appears that in many inflationary models large quantum fluctuations produced during inflation
may significantly increase the value of the energy density in some
parts of the universe. These regions expand at a greater rate than
their parent domains, and quantum fluctuations inside them lead to
production of new inflationary domains which expand even faster.
This  leads to an eternal process of self-reproduction of the
universe. Most importantly, this process may divide the universe into exponentially many exponentially large parts with different laws of low-energy physics operating in each of them. The universe becomes an {\it inflationary multiverse} \cite{Eternal,Linde:1993xx} (see also \cite{Linde:1982ur,nuff}).

To understand the mechanism of self-reproduction one should
remember that the processes separated by distances $l$ greater
than $H^{-1}$ proceed independently of one another. This is so
because during exponential expansion the distance between any two
objects separated by more than $H^{-1}$ is growing with a speed
exceeding the speed of light. As a result, an observer in the
inflationary universe can see only the processes occurring inside
the horizon of the radius  $H^{-1}$. An important consequence of
this general result is that the process of inflation in any
spatial domain of radius $H^{-1}$ occurs independently of any
events outside it. In this sense any inflationary domain of
initial radius exceeding $H^{-1}$ can be considered as a separate
mini-universe.

To investigate the behavior of such a mini-universe, with an
account taken of quantum fluctuations, let us follow \cite{Eternal} and consider an
inflationary domain of initial radius $H^{-1}$ containing
sufficiently homogeneous field with initial value $\phi \gg M_p$.
Equation (\ref{E04}) implies that during a typical time interval
$\Delta t=H^{-1}$ the field inside this domain will be reduced by
$\Delta\phi = \frac{2}{\phi}$. By comparison this expression with
$|\delta\phi(x)| \approx \frac{H}{2\pi} =  {m\phi\over 2\pi\sqrt
6}$ one can easily see that if $\phi$ is much less than $\phi^*
\sim {5\over  \sqrt{ m}} $,
 then the decrease of the field $\phi$
due to its classical motion is much greater than the average
amplitude of the quantum fluctuations $\delta\phi$ generated
during the same time. But for   $\phi \gg \phi^*$ one has
$\delta\phi (x) \gg \Delta\phi$. Because the typical wavelength of
the fluctuations $\delta\phi (x)$ generated during the time is
$H^{-1}$, the whole domain after $\Delta t = H^{-1}$ effectively
becomes divided into $e^3 \sim 20$ separate domains
(mini-universes) of radius $H^{-1}$, each containing almost
homogeneous field $\phi - \Delta\phi+\delta\phi$.   In almost a
half of these domains the field $\phi$ grows by
$|\delta\phi(x)|-\Delta\phi \approx |\delta\phi (x)| = H/2\pi$,
rather than decreases. This means that the total volume of the
universe containing {\it growing} field $\phi$ increases 10 times.
During the next time interval $\Delta t = H^{-1}$ this process
repeats. Thus, after the two time  intervals $H^{-1}$ the total
volume of the universe containing the growing scalar field
increases 100 times, etc. The universe enters eternal process of
self-reproduction.

The existence of this process implies that the  universe will never disappear as a whole. Some of its parts may collapse,  life in our part of the universe may perish, but there always will be some other parts of the universe where life will appear again and again, in all of its possible forms.

One should be careful, however, with the interpretation of these results. There is still an ongoing debate of whether eternal inflation is eternal only in the future or also in the past. In order to understand what is going on, let us consider any particular time-like geodesic line at the stage of inflation. One can show that for any given observer following this geodesic, the duration $t_{i}$ of the stage of inflation on this geodesic will be finite. One the other hand, eternal inflation implies that if one takes all such geodesics and calculate the time $t_{i}$ for each of them, then there will be no upper bound for $t_{i}$, i.e. for each time $T$ there will be such geodesic which experience inflation for the time $t_{i} >T$.  Even though the relative number of long geodesics can be very small, exponential expansion of space surrounding them will lead to an eternal exponential growth of the total volume of inflationary parts of the universe.

Similarly, if one concentrates on any particular geodesic in the past time direction, one can prove that it has finite length \cite{Borde:2001nh}, i.e. inflation in  any particular point of the universe should have a beginning at some time $\tau_{i}$. However, there is no reason to expect that there is an upper bound for all $\tau_{i}$ on all geodesics. If this upper bound does not exist, then eternal inflation is eternal not only in the future but also in the past.

In other words, there was a  beginning for each part of the universe, and there will be an end for inflation at any particular point. But there will be no end for the evolution of the universe {\it as a whole} in the eternal inflation scenario, and at present we do not know whether there was a single beginning of the evolution of the whole universe at some moment $t = 0$, which was traditionally associated with the Big Bang.

\begin{figure}[h!]
\centering\leavevmode\epsfysize=10cm \epsfbox{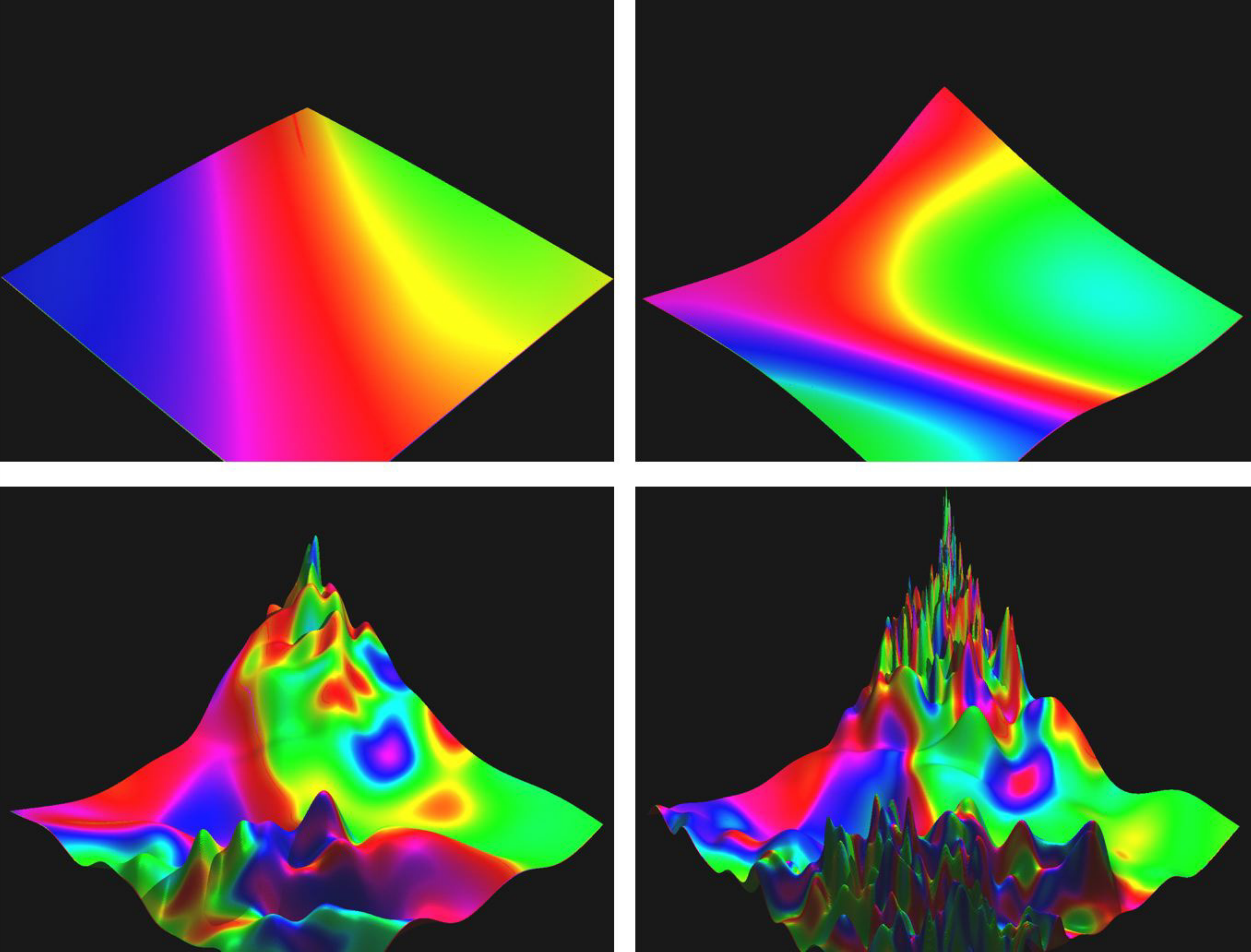}  
\caption{\footnotesize Evolution of scalar fields  $\phi$ and $\Phi$ during the process of self-reproduction of the universe.   The height of the distribution shows the value of the field $\phi$ which drives inflation. The surface is painted red, green or blue corresponding to three different minima of the potential of the field $\Phi$.   Laws of low-energy physics are different in the regions of different color. The peaks of the ``mountains'' correspond  to places where quantum fluctuations bring the scalar fields back to the Planck density. Each of such places in a certain sense can be considered as a beginning of a new Big Bang. At the end of inflation, each such part becomes exponentially large. The universe becomes a multiverse,   a huge eternally growing fractal  consisting of different exponentially large  locally homogeneous parts with different laws of low-energy physics operating in each of them.}
\label{fig:Fig0}
\end{figure}

To illustrate the process of eternal inflation, we present here the results of computer simulations of evolution of a system of two scalar fields during inflation. The field $\phi$ is the inflaton field driving inflation; it is shown by the height of the distribution of the field $\phi(x,y)$ in a two-dimensional slice of the universe. The second field, $\Phi$, determines the type of spontaneous symmetry breaking which may occur in the theory. We paint the surface  red, green or blue corresponding to three different minima of the potential of the field $\Phi$. Different colors correspond to different types of spontaneous symmetry breaking, and therefore to different sets of laws of low-energy physics in different exponentially large parts of the universe.

In the beginning of the process the whole inflationary domain is red, and the distribution of both fields is very homogeneous. Then the domain became exponentially large (but it has the same size in comoving coordinates, as shown in Fig. \ref{fig:Fig0}).  Each peak of the mountains corresponds to nearly Planckian density and can be interpreted as a beginning of a new ``Big Bang.'' The laws of physics are rapidly changing there, as indicated by changing colors, but they become fixed in the parts of the universe where the field $\phi$ becomes small. These parts correspond to valleys in Fig. \ref{fig:Fig0}. Thus quantum fluctuations of the scalar fields divide the universe into exponentially large domains with different laws of low-energy physics, and with different values of energy density. This makes our universe look as a multiverse, a collection of different exponentially large inflationary universes \cite{Eternal,Linde:1993xx}. 

Eternal inflation scenario was extensively studied during the last 30 years. I should mention, in particular, the discovery of the topological eternal inflation \cite{TopInf}  and the calculation of the fractal dimension of the universe \cite{Aryal:1987vn,Linde:1993xx}. But the most interesting recent developments of the theory of eternal inflation are related to the theory of inflationary multiverse and string theory landscape \cite{Eternal,Lerche:1986cx,Bousso:2000xa,Str,Kachru:2003aw,Douglas,Susskind:2003kw}.
These developments can be traced back to the very first paper on eternal inflation in the chaotic inflation scenario  \cite{Eternal}. It contained the following statements, which later became the manifesto of the string landscape scenario:

``{\sl As a result, our universe at present should contain an exponentially large number of mini-universes with {\it all}\, possible types of compactification and in {\it all}\, possible (metastable) vacuum states consistent with the existence of the earlier stage of inflation. If our universe would consist of one domain only (as it was believed several years ago), it would be necessary to understand why Nature has chosen just this one type of compactification, just this type of symmetry breaking, etc. At present it seems absolutely improbable that all domains contained in our exponentially large universe are {\it of the same type}. On the contrary, {\it all}\, types of mini-universes in which inflation is possible should be produced during the expansion of the universe, and it is unreasonable to expect that our domain is the only possible one or the best one. 

From this point of view, an enormously large number of possible types of compactification which exist e.g. in the theories of superstrings should be considered not as a difficulty but as a virtue of these theories, since it increases the probability of the existence of mini-universes in which life of our type may appear. The old question why our universe is the only possible one is now replaced by the question in which theories the existence of mini-universes of our type is possible. This question is still very difficult, but it is much easier than the previous one. In our opinion, the modification of the point of view on the global structure of the universe and on our place in the world is one of the most important consequences of the development of the inflationary universe scenario.'}'

At that time, the anthropic considerations have been extremely unpopular, and string theory seemed to suffer from the excessive abundance of possible types of compactification \cite{Lerche:1986cx,Bousso:2000xa}. Fortunately,  in the eternal inflation scenario one could make sense out of this complicated and controversial picture, and use it constructively. However, at that time it was not quite clear whether any of the string vacua are actually stable. It seemed especially difficult to propose a string theory realization of dS vacua. A possible solution of this problem was proposed only much later; an often discussed possibility is associated with the KKLT scenario \cite{Kachru:2003aw}. Subsequent investigations found \cite{Douglas}, in agreement with the earlier expectations \cite{Lerche:1986cx,Bousso:2000xa}, that the total number of different stable string theory compactifications in the KKLT construction can be as large as $10^{500}$, or maybe even greater. In the context of the inflationary cosmology, this corresponds to the universe divided into exponentially many exponentially large parts  with $10^{500}$ different laws of the low-energy physics operating in each of them \cite{Susskind:2003kw}, just as it was anticipated in  \cite{Eternal}.  We will return to a more detailed discussion of the new cosmological paradigm in Section \ref{land}.


\section{Initial conditions for low-scale inflation}\label{torus}

\subsection{Low-scale inflation and topology of the universe}

One of the advantages of the simplest versions of the chaotic
inflation scenario is that inflation may begin in the universe immediately after its creation at the largest
possible energy density $O(1)$, of a smallest possible size (Planck
length), with the smallest possible mass $M = O(1)$ and with the
smallest possible entropy $S = O(1)$. This provides a true
solution to the flatness, horizon, homogeneity, mass and entropy
problems \cite{Linde:2005ht}. 

However,  one may wonder whether  it  possible to solve the problem of initial conditions for the low scale inflation, if it occurs only for $V\ll 1$?
The answer to this question is positive though perhaps somewhat unexpected: The simplest way to solve the problem of initial conditions for the low scale inflation is to consider a compact flat  or open universe with nontrivial topology (usual flat or open universes are infinite). The universe may initially look like a nearly homogeneous torus of a Planckian size containing just one or two photons or gravitons. It can be shown that such a universe continues expanding and remains homogeneous until the onset of inflation, even if inflation occurs only on a very low energy scale \cite{ZelStar,chaotmix,topol4,Coule,Linde:2004nz}. 

Consider, e.g. a flat compact universe having the topology of a
torus, $S_1^3$,
\begin{equation}\label{2t}
ds^2 = dt^2 -a_i^2(t)\,dx_i^2
\end{equation}
with identification $x_i+1 = x_i$ for each of the three dimensions. Suppose
for  simplicity that $a_1 = a_2 = a_3 = a(t)$. In this case the curvature
of the universe and the Einstein equations written in terms of $a(t)$ will be
the same as in the infinite flat Friedmann universe with metric $ds^2 = dt^2
-a^2(t)\,d{\bf x^2}$. In our notation, the scale factor $a(t)$ is equal to the
size of the universe in Planck units $M_{p}^{{-1}} = 1$.

Let us assume,  that at the
Planck time $t_p \sim M_p^{-1}=1$ the universe was radiation dominated, $V\ll
 T^4 = O(1)$. Let us also assume that at the Planck time the total size of the
box was Planckian, $a(t_p) = O(1)$. In such case the whole universe initially
contained only $O(1)$ relativistic particles such as photons or gravitons, so
that the total entropy of the whole universe was O(1).

 The size of the universe dominated by relativistic particles was growing as
$a(t) \sim \sqrt t$, whereas the mean free path of the gravitons was growing as
$H^{-1}\sim t$. If the initial size of the universe was $O(1)$, then at the
time  $t \gg 1$ each particle (or a gravitational perturbation of metric)
within one cosmological time would run all over the torus many times, appearing
in all of its parts with nearly equal probability. This effect, called
``chaotic mixing,'' should lead to a rapid homogenization of the universe
\cite{chaotmix,topol4}. Note, that to achieve a modest degree of homogeneity
required for inflation to start when the density of ordinary matter drops down below $V(\phi)$,
we do not even need chaotic mixing. Indeed, density perturbations do not grow
in a universe dominated by ultra-relativistic particles if  the size of the
universe is smaller than $H^{-1}$. This is exactly what
happens in our model. Therefore the universe should remain relatively
homogeneous until the thermal energy density drops below $V$ and inflation
begins. And once it happens, the universe rapidly becomes exponentially large and homogeneous due to inflation.

Thus we see that in this scenario, just as in the simplest chaotic inflation scenario, inflation begins if we had a sufficiently homogeneous domain of a smallest possible size (Planck scale), with the smallest possible mass (Planck mass), and with the total entropy O(1). The only additional requirement is that this domain should have identified sides, in order to make a flat or open universe compact. We see no reason to expect that the probability of formation of such domains is strongly suppressed.

One can come to a similar conclusion from a completely different point of view. Investigation of the quantum creation of a closed or  an infinite open inflationary universe with $V\ll 1$ shows that this process is forbidden at the classical level, and therefore it occurs only due to tunneling. The main reason for this result is that a closed de Sitter space always has size greater than $H^{{-1}} \sim 1/\sqrt V$, and the total energy is greater than $H^{-3}V \sim 1/\sqrt V$. Than is why the universe with $V\ll 1$ is large, heavy and difficult to create.  As a result, the probability of this process is exponentially suppressed \cite{Linde:1983mx,Vilenkin:1984wp,Open}. Meanwhile,  creation of the flat or open universe  is possible without any need for the tunneling, and therefore there is no exponential suppression for the probability of quantum creation of a topologically nontrivial compact flat or open inflationary universe \cite{ZelStar,Coule,Linde:2004nz}.

These results suggest that the problem of initial conditions for low energy scale inflation can be easily solved if one considers topologically nontrivial compact universes.  If inflation can occur only  much below the Planck density, then the compact topologically nontrivial flat or open universes should be much more probable than the standard Friedmann universes described in every textbook on cosmology. This possibility is quite natural in the context of string theory, where all internal dimensions are supposed to be compact. Note, however, that if the stage of inflation  is sufficiently long, it makes the observable part of the universe so large that its topology does not affect observational data.

\subsection{Initial conditions in models with several non-interacting scalars}
In the context of more complicated inflationary model containing more than one scalar field, the problem of initial conditions for inflation can be solved in many other ways. 
To give some examples, let us start with a simple theory describing two non-interacting scar fields, $\phi$ and $\sigma$:
\be
U(\phi,\sigma) = V(\phi) + W(\chi)
\ee
We will assume that the potential $W(\chi)$ contains a local minimum $\chi_{0}$ with very high energy density, and $W$ vanishes after the tunneling to the true vacuum. (More exactly, $W+V$ become vanishingly small, about $10^{-120}$, after the tunneling and the end of inflation, to account for the present tiny cosmological constant.) This is a standard feature of the string theory landscape scenario, which describes a huge number of such metastable states at all energy densities, many of which approach the Planck energy density \cite{Kachru:2003aw,Douglas,Susskind:2003kw}. According to our arguments in Section \ref{ini}, inflation in such metastable vacua is quite probable. Of course, this will be inflation of the old inflation type, so it does not solve any cosmological problems by itself. That is where the inflaton field with the slow roll potential $V(\phi)$ comes handy, even if  $V(\phi)$ by itself can support slow roll inflation only for $V(\phi) \ll 1$.

Indeed, if the Hubble constant squared $H$ during the first stage of inflation supported  by the potential $W(\chi_{0})$ is greater than $V''$, fluctuations of the field $\phi$ are generated during this time, and after a while the universe becomes divided into exponentially large regions where the field $\phi$ takes all of its possible values. According to  \cite{Starobinsky:1986fx,Goncharov:1987ir,Linde:2005ht,Linde:1993xx,Linde:2006nw}, the probability to find a given value of the field $\phi$ at any given point in this regime is described by the probability distribution 
\be\label{distrib}
P(\phi) \sim \exp\left(-{24\pi^{2}\over W(\sigma_{0})}+{24\pi^{2}\over W(\chi_{0})+V(\phi)}\right) \approx \exp\left(-{24\pi^{2}V(\phi)\over W^{2}(\chi_{0})}\right)
\ee 
Therefore \rf{distrib} implies that at the moment prior to the decay of the metastable vacuum state in this model all values of the field $\phi$ such that $V(\phi)\ll 10^{{-3}} W^{2}(\chi_{0})$ will be equally probable. Note that in accordance with the Planck constraints on $r$ \rf{eq:rvh}, the value of the inflationary potential $V(\phi)$ during the last 60 e-foldings of inflation was smaller than $3\times 10^{-9}$. Therefore \rf{distrib} implies that initial conditions for the last 60 e-foldings of inflation in all theories $V(\phi)$ consistent with observations are quite  probable for $W^{2}(\chi_{0}) \gtrsim 10^{{-3}}$. Since the natural energy scale for the metastable vacua in the landscape can be almost as large as $O(1)$ in Planck units, one concludes that the initial conditions for the slow roll inflation in this scenario can naturally emerge after the stage of the false vacuum inflation in the landscape \cite{Linde:2005dd}.

One should add that the same scenario works even without any assumptions about the false vacuum inflation and landscape if one takes into account the possibility of the slow-roll eternal inflation in a theory of superheavy scalars $\chi$ and the light inflaton field $\phi$, see \cite{Linde:1987yb}.

\subsection{Initial conditions in models with several interacting scalars}

Consider now a theory with an effective potential 
\be
V(\phi,\chi) =  V(\phi) + W(\chi) +{g^2\over 2} \phi^2\chi^2
\ee
Here we assume that $V(\phi), W(\chi) \lesssim 10^{-9}$ are some potentials which cannot reach the Planckian values, which is the essence of the problem that we are trying to address. However, their interaction term ${g^2\over 2} \phi^2\chi^2$ can become $O(1)$, which would correspond to the Planck boundary. In that case the Planck boundary is defined by the condition
\begin{equation}\label{1}
{g^2\over 2} \phi^2\chi^2 \sim 1 \ ,
\end{equation}
which is represented by the set of four hyperbolas
\begin{equation}\label{2g}
 g |\phi ||\chi| \sim 1 \ .
\end{equation}
At larger values of $\phi$ and $\chi$ the density is greater than
the Planck density, so the standard classical description of
space-time is impossible there. In addition, the effective
masses of the fields should be smaller than $1$, and
consequently the curvature of the effective potential cannot be
greater than $1$. This leads to two additional conditions:
\begin{equation}\label{3}
  |\phi | \lesssim g^{-1}, ~~~~~~|\chi| \lesssim g^{-1}.
\end{equation}
We assume that $g \ll 1$. On the main part of the hyperbola \rf{2g} one either has $|\phi | \sim g^{-1} \gg |\chi \gg 1$, or $|\chi | \sim g^{-1} \gg |\phi \gg 1$. Consider for definiteness the first possibility. In this case, the effective mass of the field $\chi$, which is proportional to $g\phi$, is much greater than the effective mass of the field $\phi$, which is proportional to $g\chi$. Therefore in the beginning the field $\phi$ will move extremely slowly, whereas the field $\chi$ will move towards its small values much faster. Since its initial value on the Planck boundary is greater than $1$, the universe will experience a short stage of chaotic inflation determined by the potential ${g^2\over 2} \phi^2\chi^2$ with a nearly constant field $\phi$. After that, the first stage of inflation will be over, the field $\chi$ will oscillate, and within few oscillations it will give its energy away in the process of preheating \cite{KLS}. As a result, the classical field $\chi$ rapidly becomes equal to zero, the term ${g^2\over 2} \phi^2\chi^2$ disappears, and the potential energy reduces to $V(\phi)$. Soon after that, the second stage of inflation begins driven by the scalar field. As we already mentioned, it initial value can be $|\phi | \sim g^{-1} \gg |\chi \gg 1$. Thus the first stage of oscillations of the field $\phi$ provides good initial conditions for chaotic inflation driven by the scalar field $\phi$   \cite{Felder:1999pv}. Note that this effect is very general; it may occur for potentials $V(\phi)$ and $W(\chi)$ of {\it any} shape, either convex or concave, as long as they are small and at least one of them can support inflation. 

In conclusion, in this section we have shown that initial conditions for the slow-roll inflation may occur very naturally in a broad class of models, including the models where inflation is possible only for $V(\phi) \ll 1$. In Section \ref{land} we will discuss some new ideas related to initial conditions for inflation in the context of the string theory landscape.

\section{Inflation and observations}\label{observ}

Inflation is not just an interesting theory that can resolve many
difficult problems of the standard  Big Bang cosmology. This
theory made several   predictions which can be tested by
cosmological observations. Here are the most important
predictions:

1) The universe must be flat. In most models $\Omega_{total} = 1
\pm 10^{-4}$.

2) Perturbations of the metric produced during inflation are
adiabatic.

3) These perturbations are gaussian. In non-inflationary models, the parameter $f_{NL}^{\rm local}$ describing the level of the so-called local non-Gaussianity can be as large as $O(10^{4})$, but it is predicted to be $O(1)$ in all single-field inflationary models. Prior to the Planck data release, there were rumors that $f_{NL}^{\rm local} \sim 30$, which would rule out all or nearly all single field inflationary models.

4)  Inflationary perturbations generated during a  slow-roll regime with $\epsilon,\eta \ll 1$   have a nearly flat spectrum with $n_{s}$ close to 1.

5) On the other hand, the spectrum of inflationary perturbations usually is slightly non-flat. A small deviation of $n_{s}$ from $1$ is one of the distinguishing features of inflation. It is as significant for inflationary theory as the asymptotic freedom for the theory of strong interactions.

6) perturbations of the metric could be scalar, vector or tensor.
Inflation mostly produces scalar  perturbations, but it also
produces tensor perturbations with nearly flat spectrum, and it
does {\it not} produce vector perturbations. There are certain
relations between the properties of  scalar and tensor
perturbations produced by inflation.

7) Inflationary perturbations produce specific peaks in the
spectrum of CMB radiation. (For a simple pedagogical
interpretation of this effect see  e.g. \cite{Dodelson:2003ip}; a
detailed theoretical description can be found in
\cite{Mukhanov:2003xr}.)

It is possible to violate each of these predictions if one makes
inflationary theory sufficiently complicated. For example, it is possible
to produce vector perturbations of the metric in the models where
cosmic strings are produced at the end of inflation, which is the
case in some versions of hybrid inflation. It is possible to have
an open or closed inflationary universe, it is possible to have models with
nongaussian isocurvature fluctuations with a non-flat spectrum.
However, it is difficult to do so, and most of the
inflationary models obey the simple rules given above.

It is possible to test many of these predictions. The major breakthrough in this direction was achieved  due to the recent
measurements of the CMB anisotropy. The latest Planck results show that this anisotropy can be described by the angular distribution shown in the Figure \ref{cmb2}. For small $l$, comparison of the theory and experiment is complicated because of the cosmic variance,  for large multipoles $l$, where the effects of cosmic variance are very small, an agreement between the  observational data and the predictions of the simplest inflationary models is quite spectacular. 

\begin{figure}
\centering\leavevmode\epsfysize=7cm \epsfbox{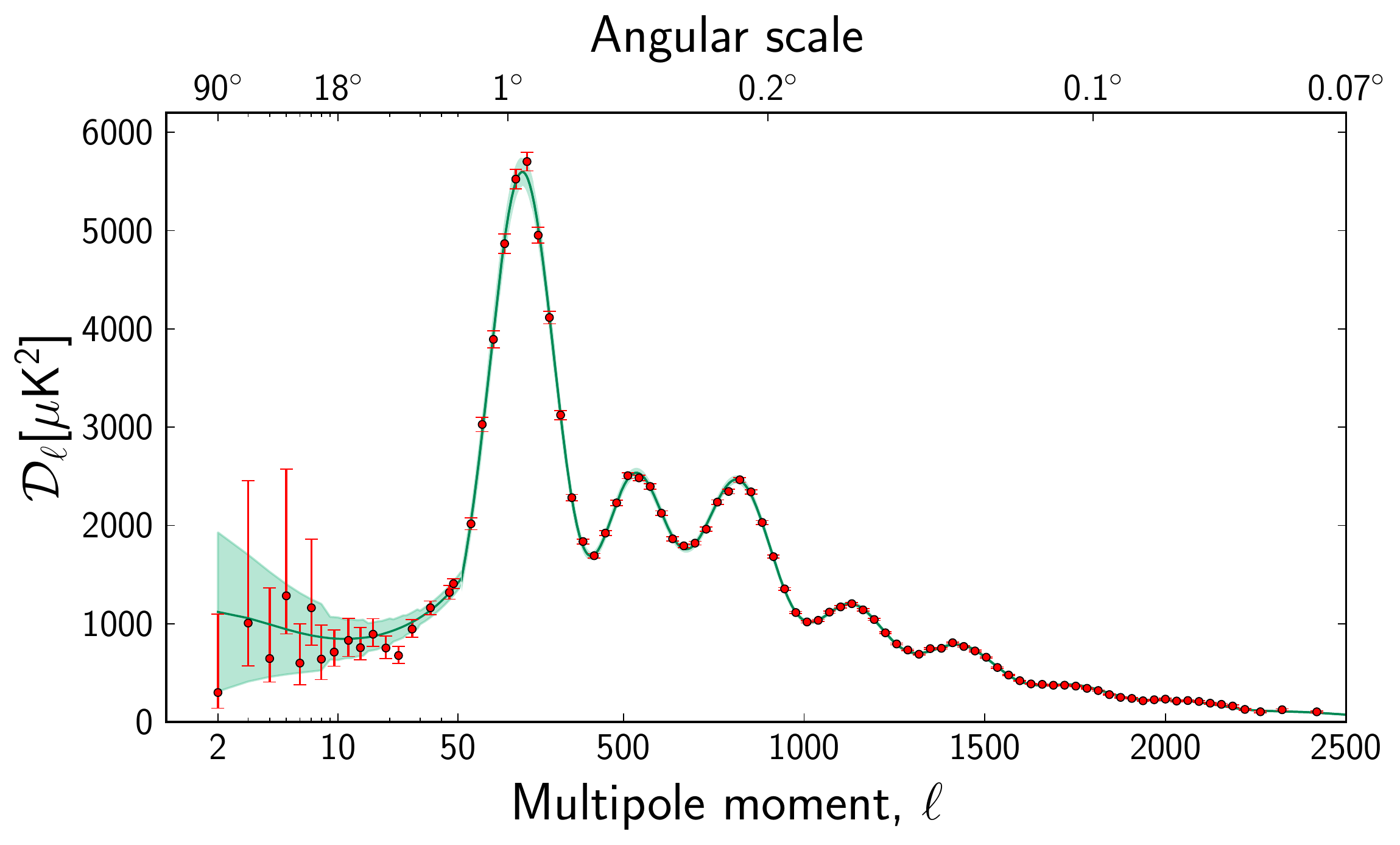}  
 \caption{\footnotesize CMB data (Planck 2013) versus the predictions of one of the simplest inflationary models with $\Omega = 1$ (green line).}
\label{cmb}
\end{figure}

The results of Planck 2013, in combination with the results of WMAP, high $l$ perturbations measurement by other observations  and the results based on investigation of baryon acoustic oscillations (BAO), imply that
\be
\Omega = 1.0005 \pm 0.0066\qquad  (\rm 95\%~confidence) ,
\ee
and 
\be
n_{s} = 0.96 \pm 0.007\qquad  (\rm 68\%~confidence).
\ee

These results are very significant. Before the discovery of dark energy, many astronomers believed that $\Omega \sim 0.3$. At that time, experts in inflationary theory made many attempts to save inflation by inventing inflationary models compatible with this value of $\Omega$, and none of these models worked except one of two models which were extremely artificial. Now one of the main predictions of inflationary cosmology is confirmed with accuracy better than 1\%.  Many years ago, critics of inflation argued that the prediction $n_{s} = 1$ is trivial because it was made by Harrison and Zeldovich (albeit without any real physical motivation) long before inflation.  Now the Harrison-Zeldovich spectrum is ruled out at the $5\sigma$ level, and inflationary expectations are confirmed.

One of the most important new results is the constraint on $f_{NL}^{\rm local}$  \cite{Ade:2013uln},
\be\label{fnlloc}
f_{NL}^{\rm local} = 2.7 \pm 5.8 \ .
\ee
This is a confirmation of one of the main predictions of the simplest versions of inflationary cosmology at the level $O(10^{{-4}})$. This result implies that even if the error bars in the measurement of $f_{NL}^{\rm local}$ will become two times smaller (which is very difficult to achieve in practice), the result will remain compatible with the predictions of the simplest single-field inflationary models at the $1 \sigma$ level. In practical terms this means that we may return to investigation of the simplest inflationary models, instead of concentrating on exceedingly complicated models which may produce local non-Gaussianity. 

There are still some question marks to be examined, such as a rather controversial issue of the low multipoles anomalies, see e.g. \cite{Bennett:2012zja, Ade:2013nlj}. The observational status and interpretation of these effects is still uncertain, hopefully some of these issues will be clarified in the next Planck data release, but if one takes these effects seriously one may try to look for some theoretical explanations. For example, there are several ways to suppress the large angle anisotropy, see e.g. \cite{Contaldi}. The situation with correlations between low multipoles requires more work, see e.g. \cite{LM,Erickcek:2008sm,Lyth:2013vha}.  One way or another,  it is quite significant that all proposed explanations of these anomalies are based on inflationary cosmology. 

\begin{figure}
\centering\leavevmode\epsfysize=7cm \epsfbox{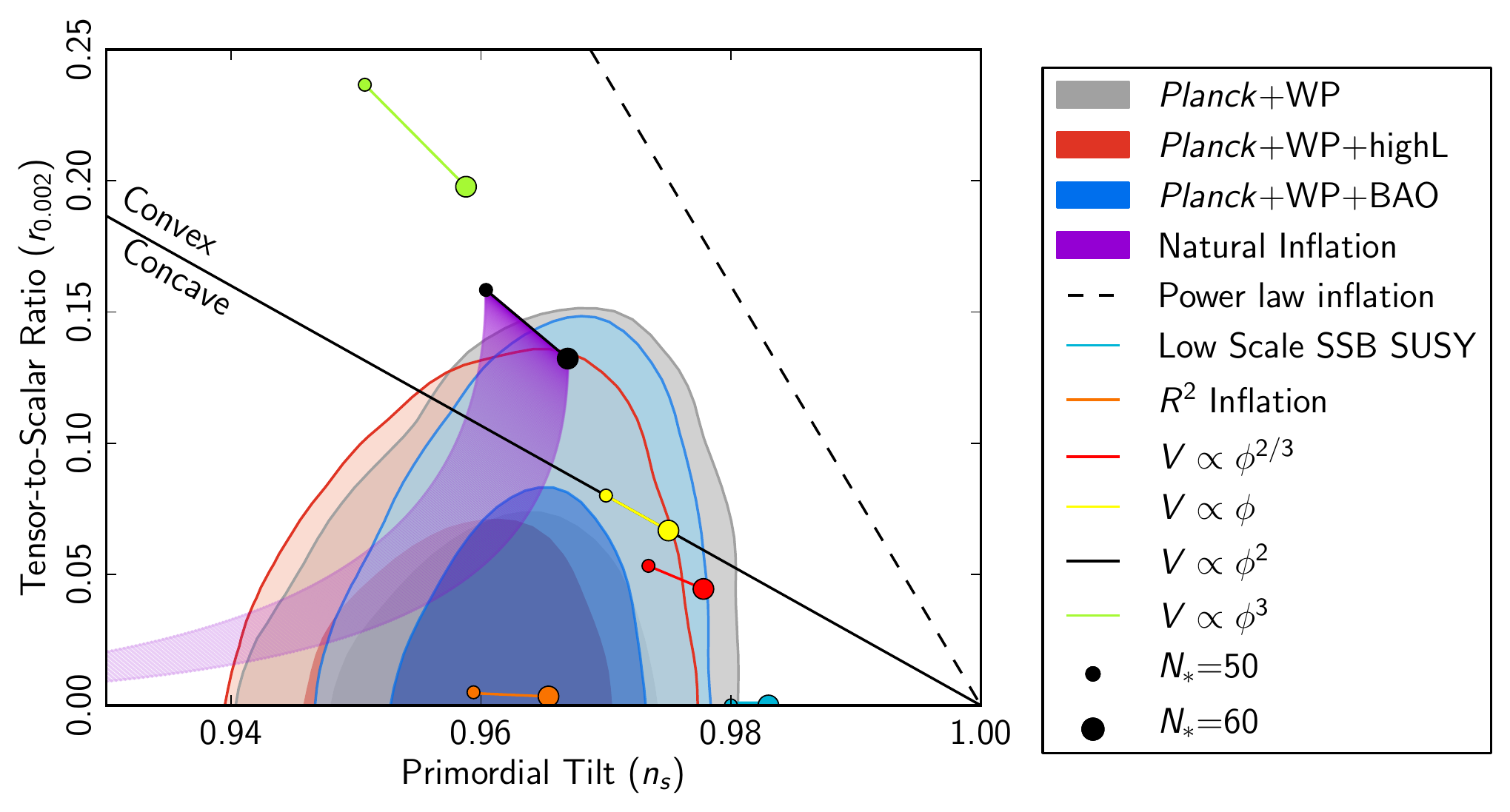}  
 \caption{\footnotesize Constraints on $n_{s}$ and $r$ according to Planck 2013.}
\label{cmb2}
\end{figure}

One of the interesting issues to be probed by the future observations is the possible existence of tensor perturbations, gravitational waves produced during inflation. Planck results did not reveal the existence of these perturbations, but considerably strengthened the upper bound on the tensor to scalar ratio: $r < 0.11$ at the $2\sigma$ level. These result is strong enough to disfavor (though not rule out) many previously popular inflationary models, including the simplest version of chaotic inflation with the quadratic potential. 

To understand this statement, let us look at one of the most famous figures from the Planck data release, Fig. \ref{cmb2}.
It shows predictions of the simplest versions of chaotic inflation with $V \sim \phi^{n}$ for $n =3$ , 2, 1, and 2/3 (green, black, yellow and red circles), as well as the predictions of the Starobinsky model and the Higgs inflation model, which we will describe shortly (orange circles). As we can see, the simplest versions of chaotic inflation are somewhat disfavored, whereas the Starobinsky model and the Higgs inflation model occupy the sweet spot at the center of the part of the $n_{s}-r$ plain preferred by the Planck data. 
A long list of various models satisfying all Planck data can be found in \cite{Ade:2013uln,Martin:2013tda,Martin:2013nzq}.

\begin{figure}
\centering\leavevmode\epsfysize=7cm \epsfbox{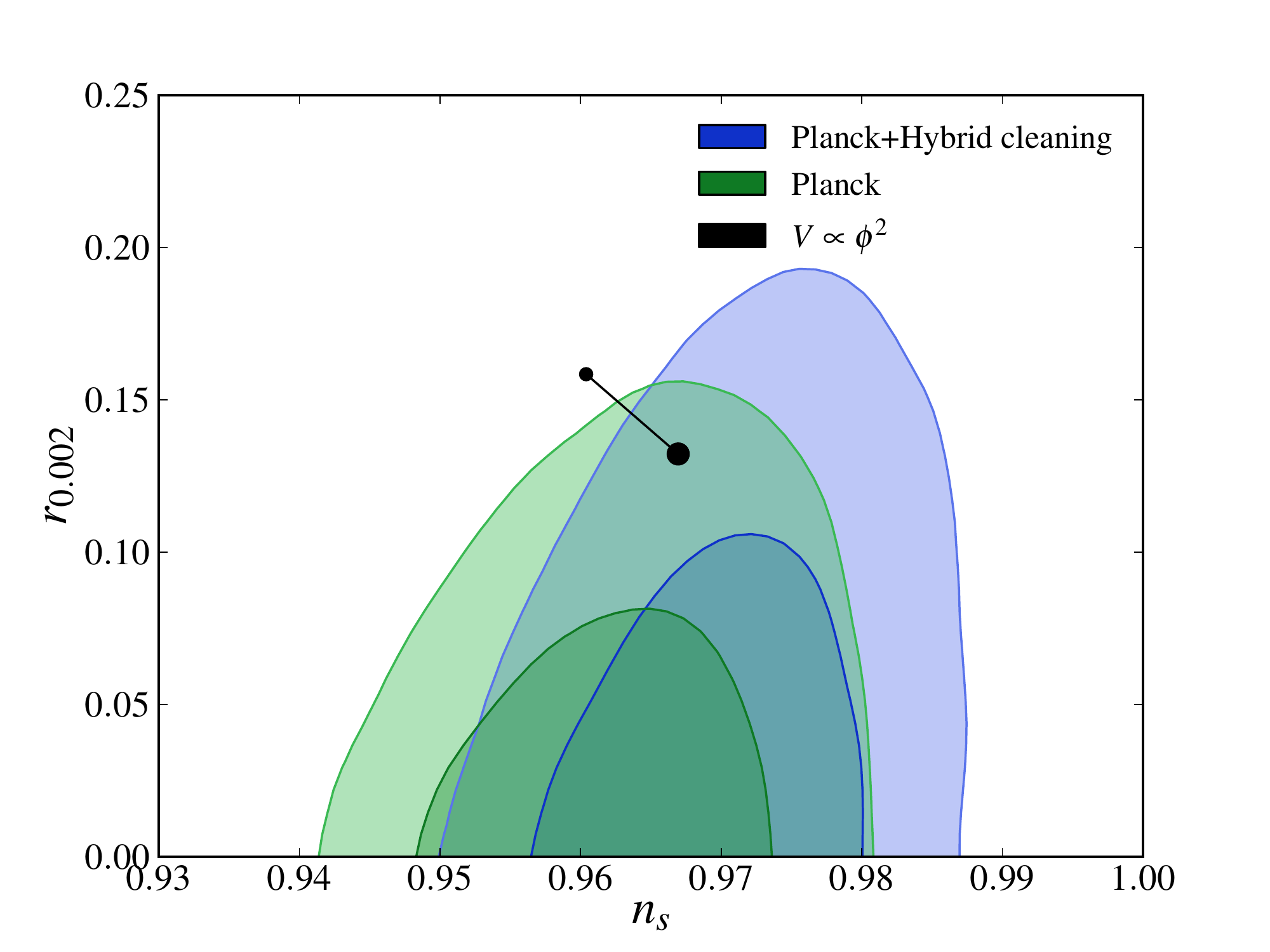}  
 \caption{\footnotesize  Modified constraints on $n_{s}$ and $r$ according to \cite{Spergel:2013rxa}.}
\label{cmb3}
\end{figure}

While we are waiting for the next data release by Planck, a recent alternative interpretation of their results  \cite{Spergel:2013rxa} suggests that some of their conclusions may require further attention and justification. In particular, the authors of \cite{Spergel:2013rxa}, using a different foreground cleaning procedure for the Planck data, find slightly different constraints on $n_{s}$ and $r$, see Fig. \ref{cmb3}.

The predictions of inflationary models with $V(\phi) \sim \phi$ and $V(\phi) \sim \phi^{2/3}$ are consistent with the results of \cite{Spergel:2013rxa} as shown in Fig.  \ref{cmb3}, and even the simplest quadratic model $V={m^{2}\over 2} \phi^{2}$ is marginally  consistent with the results shown in Figs. \ref{cmb2}, \ref{cmb3}.  It will be fantastic if further investigations confirm one of the simplest  models of chaotic inflation. Hopefully we will know much more about it soon. In the meantime we will concentrate on the gradually shrinking area in the $n_{s}$ - $r$ plane favored by the Planck data, with $r \lesssim 0.1$ and $0.95\lesssim  n_{s} \lesssim 0.98$. Is it possible to find simple models of inflation with predictions belonging to this area? Can we derive such models in the context of supergravity? Is there anything special in the models which fit the data? We will try to answer these questions in the next sections.

\section{Chaotic inflation in supergravity}\label{chinflsugra}

From a purely mathematical point of view, finding a realization of the chaotic inflation in the theory of a single scalar field $\phi$ is a nearly trivial exercise. One simply  finds a function $V(\phi)$ which is sufficiently flat in some interval of the values of the inflaton field  \cite{Linde:1983gd}. The simplest example of such potential is $m^{2}\phi^{2}/2$, or one can take any other function which approaches $\lambda_{n}\phi^{n}$ for $n>0$, $\phi > 1$, where $M_p=1$. If one wants to have inflation at $\phi < 1$, one can consider a function with a sufficiently flat maximum or an inflection point.

For many years, finding a proper inflaton potential in supergravity was much more difficult. 
The scalar potential in supergravity is a complicated function of the superpotential ${ W}$ and the \K\, potential. Usually the potential depends on several complex scalar fields. Therefore one has to investigate dynamical evolution in a multidimensional moduli space and verify stability of the inflationary trajectory. The main problem is related to the \K\, potential ${\cal K}$. The simplest  \K\, potential contains terms proportional to $\Phi\bar\Phi$. The F-term part of the potential is proportional to $e^{\cal K}\sim e^{|\Phi|^2}$ and is therefore much too steep for chaotic inflation at $\Phi \gg 1$. Moreover, the presence of the terms like $e^{|\Phi|^2}$ implies that the typical scalar masses are $\mathcal{O}(H)$, too large to support inflation even at $\Phi < 1$. For that reason, many attempts towards inflation in supergravity have been devoted to various versions of hybrid inflation, where inflation was possible even for $\Phi \ll1$ \cite{F,D}. 

From the point of view of an inflationary model builder, there is also an additional problem. One can always find $V(\phi)$ from ${W}$ and ${\cal K}$, but until the calculations are finished, one does not know exactly what kind of potential we are going to get. As a result, it is very difficult to solve the inverse problem: to find ${W}$ and ${\cal K}$ which would produce a desirable inflationary potential $V(\phi)$, which would fit the observational data. 

A significant step in this direction was made back in 2000 in \cite{Kawasaki:2000yn}. The basic idea is that instead of considering a minimal \K\, potential containing $\Phi\bar\Phi$, one may consider the   potential $(\Phi+\bar\Phi)^2/2$. This \K\,  potential has shift symmetry: It does not depend on the field combination $\Phi-\bar\Phi$. Therefore the dangerous term $e^{ K}$, which often makes the potential too steep, is also independent of $\Phi-\bar\Phi$. This makes the potential flat and suitable for chaotic inflation, with the field $\Phi-\bar\Phi$ playing the role of the inflaton. The flatness of the potential is broken only by the superpotential $mS\Phi$, where $S$ is an additional scalar field, which vanishes along the inflationary trajectory. As a result, the potential in the direction $\Phi-\bar\Phi$ becomes quadratic, as in the simplest version of chaotic inflation. Similarly, one can use the \K\, potential $(\Phi-\bar\Phi)^2/2$, with the field $\Phi+\bar\Phi$ playing the role of the inflaton.

This scenario was substantially generalized in \cite{Kallosh:2010ug,Kallosh:2010xz}.  The generalized scenario describes two scalar fields, $S$ and $\Phi$, with the superpotential 
 \be
{W}= Sf(\Phi) \ ,
\label{cond}
\ee
where $f(\Phi)$ is a real holomorphic function such that $\bar f(\bar \Phi) = f(\Phi)$. Any function which can be represented by Taylor series with real coefficients has this property. The \K \, potential can be  chosen to have functional form
\be\label{Kminus}
K= K((\Phi-\bar\Phi)^2,S\bar S).
\ee
In this case, the \K\, potential does not depend on ${\rm Re}\, \Phi$. Under certain conditions on the \K\, potential, inflation occurs along the direction $S = {\rm Im}\, \Phi = 0$. For $\Phi = (\phi+i\chi)/\sqrt 2$,  the field $\phi$ plays the role of the canonically normalized inflaton field with the potential 
\be
V(\phi) = |f(\phi/\sqrt 2)|^{2}.
\ee
 All scalar fields have canonical kinetic terms along the inflationary trajectory $S = {\rm Im}\, \Phi = 0$.

An alternative formulation of this class of models has
the \K\, potential \be\label{Kplus}
K= K((\Phi+\bar\Phi)^2,S\bar S).
\ee
In this class of models, the \K\, potential does not depend on ${\rm Im }\, \Phi$. The role of the inflaton field is played by the canonically normalized field $\chi$ with the potential
\be
V(\chi) = |f(\chi/\sqrt 2)|^{2}.
\ee
One should also make sure that the real part of this field vanishes during inflation. The simplest way to find a class of functions $f(\Phi)$ which lead to the desirable result is to consider any real holomorphic function $f(\Phi) = \sum_{n} c_{n} \Phi^{n}$, and then make the change of variables $\Phi \to -i\Phi$ there. 

Obviously, in the theory with $K= K((\Phi-\bar\Phi)^2,S\bar S)$ it is easier to formulate the required conditions for the function $f$.
However, as long as we do not consider interactions of the field $\Phi$ to vector fields, which are different for scalars and pseudo-scalars, the two approaches give identical results. For example, the theory with $K= -(\Phi-\bar\Phi)^2/2+S\bar S)$ and $f = \Phi + c\Phi^{2}$ leads to the same inflationary scenario as the theory with $K= (\Phi+\bar\Phi)^2/2+S\bar S)$ and $f = -i \Phi - c\Phi^{2}$. Alternatively, one may consider the function  $f = \Phi - i c\Phi^{2}$, obtained from the previous one by multiplication by $i$.

The generality of the functional form of the inflationary potential $V(\phi)$ allows one to describe {\it any} combination of the parameters $n_{s}$ and $r$. Indeed, the potential depends only on the function $f(\Phi)$. One can always Taylor expand it, with real coefficients, in a vicinity of the point corresponding to $N \sim 60$ of e-folds, so that the square of this function will fit any desirable function $V(\phi)$ with an arbitrary accuracy. In fact, one can show that there are {\it many} different choices of $f(\Phi)$ which lead to the same values of $n_{s}$ and $r$. Thus, this rather simple class of models may describe {\it any} set of observational data which can be expressed in terms of these two parameters by an appropriate choice of the function $f(\Phi)$ in the superpotential.  

The simplest example of such theory has $f(\Phi) = m\Phi$, which leads, in the context of the theory with the \K\ potential $K= -(\Phi-\bar\Phi)^2/2+S\bar S$, to the simplest parabolic potential  ${m^{2}\over 2} \phi^{2}$ \cite{Kawasaki:2000yn}. It is interesting to analyze various generalizations of this model.

\begin{figure}[t!]
\begin{center}
\hskip -0.76 cm \includegraphics[scale=0.25]{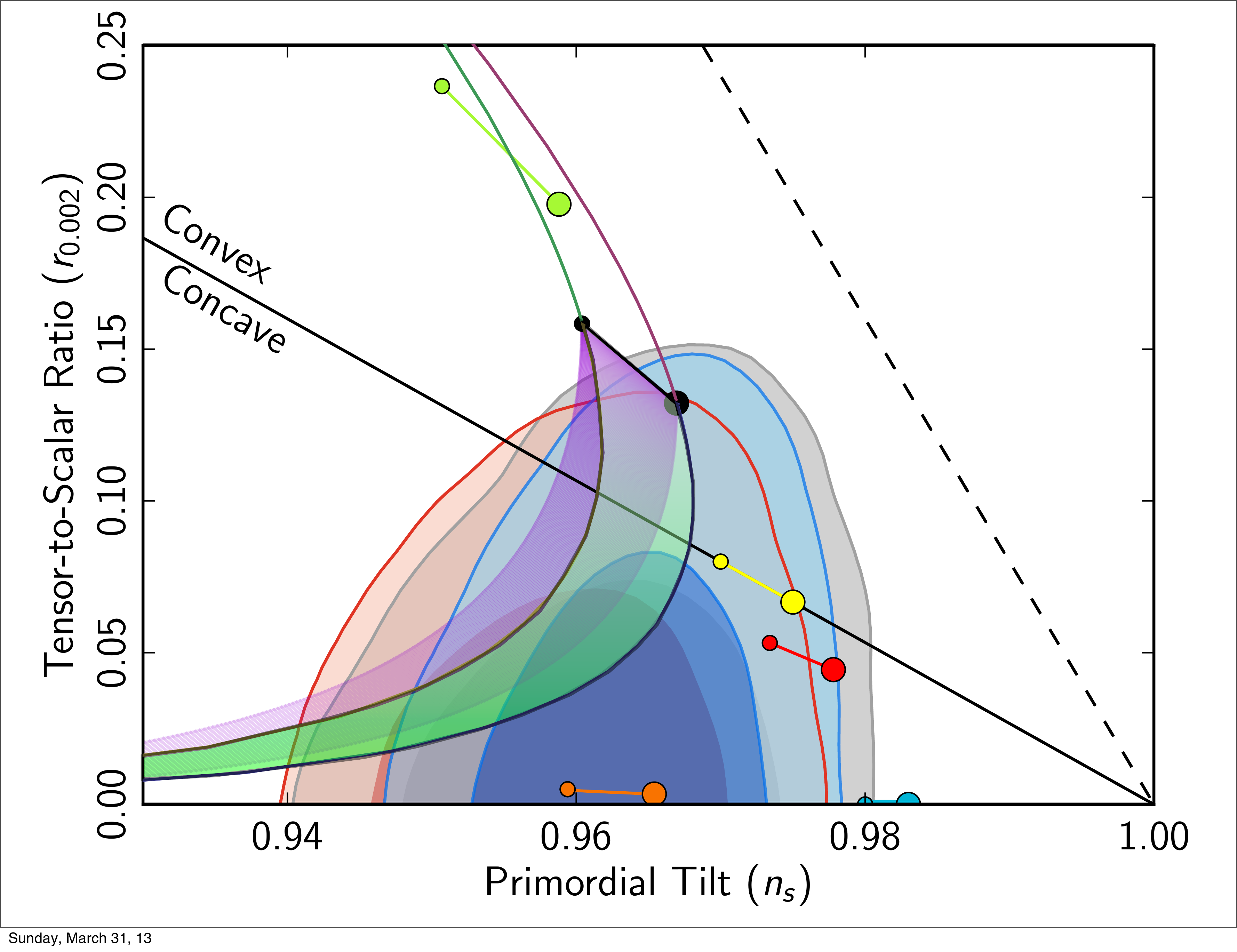}
\end{center}
\caption{\footnotesize The green area describes observational consequences of inflation in the Higgs model \rf{minhiggs} with  $v \gg 1$ ($a \ll 1$). In the limit $v\to \infty$, which corresponds to $a\to 0$, the predictions coincide with the predictions of the simplest chaotic inflation model with a quadratic potential ${m^{2}\over 2} \phi^{2}$, but for a wide range of values of the parameter $v > 1$ the predictions agree with the Planck results.}
\label{chi0}
\end{figure}

As a first step, one may add to the function $f(\Phi) = m\Phi$ a small higher order correction,
\be\label{corr}
f(\Phi) = m\Phi(1-a \Phi)
\ee
with $a \ll 1$. This function, upon the change of variables $\Phi \to \tilde\Phi +{1\over 2a}$, is equivalent to the function $f = -m a (\Phi^{2}-{1\over (2a)^{2}})$ used previously in \cite{Kallosh:2010ug,Linde:2011nh}. Representing $\tilde\Phi$ as $(\phi+i\chi)/\sqrt 2$, one finally obtains the Higgs-type inflationary potential
\be\label{minhiggs}
V(\phi) = {\lambda\over 4}(\phi^{2}-v^{2})^{2}\ , 
\ee
where $\lambda = m^{2} a^{2}$ and $v = {1\over\sqrt 2\, a}$. For $v > 1$, there is an inflationary regime when the field $\phi$ rolls from the maximum of the potential at $\phi = 0$, as in new inflation scenario. Initial conditions for inflation in this model are discussed in Section \ref{torus}.  The results of investigation of the observational consequences of this model \cite{Kallosh:2010ug,Kallosh:2007wm,Linde:2011nh,WestphalLinde} are described by the green area in Figure \ref{chi0}. Predictions of this model are in good agreement with observational data  for a certain range of values of the parameter $a \ll 1$.

However, this does not mean that absolutely any potential $V(\phi)$ can be obtained in this simple context, or that one has a full freedom of choice of the functions $f(\Phi)$. It is important to understand the significance of the restrictions on the form of the \K\ potential and superpotential described above. According to  \cite{Kallosh:2010xz}, in the theory with the \K\ potential $K= K((\Phi-\bar\Phi)^2,S\bar S)$ the symmetry of the \K\ potential $\Phi\to \pm \bar\Phi$, as well as the condition that $f(\Phi)$ is a real holomorphic function are required to ensure that the inflationary trajectory, along which the \K\ potential vanishes is an extremum of the potential in the direction orthogonal to the in\-fla\-ti\-o\-na\-ry trajectory $S = {\rm Im} \Phi = 0$. After that, the proper choice of the \K\, potential can make it not only an extremum, but a minimum,  thus stabilizing the inflationary regime \cite{Kallosh:2010ug,Kallosh:2010xz,Kallosh:2011qk}.

The requirement that $f(\Phi)$ is a real holomorphic function does not affect much the flexibility of choice of the inflaton potential: One can take any positively defined potential $V(\phi)$, take a square root of it, make its Taylor series expansion and thus construct a real holomorphic function which approximate $V(\phi)$ with great accuracy.  However, one should be careful to obey the rules of the game as formulated above.

For example, suppose one wants to obtain a fourth degree polynomial potential of the type of $V(\phi) = {m^{2}\phi^{2}\over 2}(1 + a \phi+ b\phi^{2})$ in supergravity. One may try to do it by taking $K= (\Phi+\bar\Phi)^2/2+S\bar S)$ and $f(\Phi) = m \Phi(1+ c e^{i\theta} \Phi)$ \cite{Nakayama:2013jka}. For general $\theta$, this choice violates our conditions for $f(\Phi)$. In this case, the potential will be a fourth degree polynomial with respect to  $\rm Im\,\Phi$ if $\rm Re\,\Phi = 0$. However, in this model the flat direction of the potential $V(\Phi)$ (and, correspondingly, the inflationary trajectory)   deviate from   $\rm Re\,\Phi = 0$, so we will no longer have the simple single-field inflation. (Also, in addition to the minimum at $\Phi = 0$, the potential will develop an extra minimum at $\Phi = - c^{{-1}} e^{-i\theta}$.) As a result, the potential along the inflationary trajectory is not exactly polynomial, contrary to the expectations of  \cite{Nakayama:2013jka}. Moreover, the kinetic terms of the fields will be non-canonical and non-diagonal.

This may not be a big problem, since the potential in the direction orthogonal to the inflationary trajectory is exponentially steep. Therefore the deviation of this field from $\rm Re\,\Phi = 0$ typically should be smaller than $O(1)$ in Planck units, and for sufficiently large values of the inflaton field $\chi \gg 1$ the potential will be approximately given by the simple polynomial expression $|f(\chi/\sqrt 2)|^{2}$. But in order to make a full investigation of inflation in such models one would need to study evolution of all fields numerically, and make sure that all stability conditions are satisfied. An advantage of the methods developed in \cite{Kallosh:2010ug,Kallosh:2010xz} is that all fields but one vanish during inflation, all kinetic terms are canonical and diagonal along the inflationary trajectory, and investigation of stability is straightforward.

\begin{figure}[t!]
\begin{center}
\hskip -0.76 cm \includegraphics[scale=0.42]{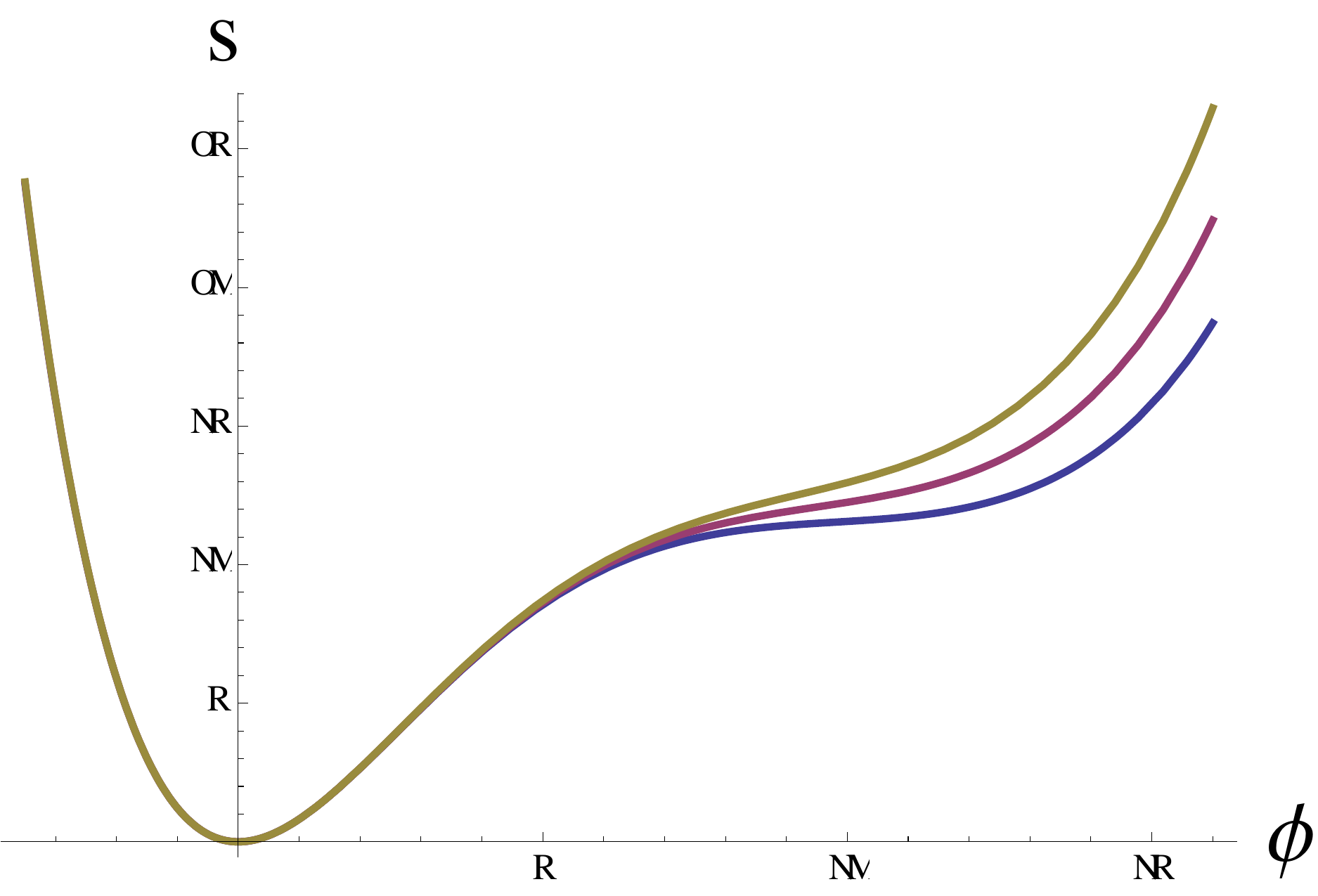}
\end{center}
\caption{\footnotesize The potential $V(\phi) = {m^{2}\phi^{2}\over 2}\,\bigl(1-a\phi +a^{2}b\,\phi^{2})\bigr)^{2}$ for $a = 0.1$ and $b = 0.36$ (upper curve), 0.35 (middle) and 0.34 (lower curve). The potential is shown in units of $m$, with $\phi$ in Planckian units. For each of these potentials, there is a range of values of the parameter $a$ such that the observational predictions of the model are in the region of $n_{s}$ and $r$ preferred by Planck 2013. For $b = 0.34$, the value of the field $\phi$ at the moment corresponding to 60 e-foldings from the end of inflation is $\phi \approx 8.2$. Change of the parameter $a$ stretches the potentials horizontally without changing their shape.}
\label{chi}
\end{figure}

Fortunately, one can obtain an exactly polynomial potential $V(\phi)$ and single-field inflation  in the theories with $K= K((\Phi-\bar\Phi)^2,S\bar S)$ using the methods of \cite{Kallosh:2010ug,Kallosh:2010xz}, if the polynomial can be represented as a square of a polynomial function $f(\phi)$ with real coefficients. As a simplest example, one may consider $f(\Phi) = m\Phi\bigl(1-c\Phi +d\Phi^{2}\bigr)$. The resulting potential of the inflaton field can be represented as
\be\label{polyn}
V(\phi) = {m^{2}\phi^{2}\over 2}\,\bigl(1-a\phi +a^{2}b\,\phi^{2})\bigr)^{2} \ .
\ee 
Here $a = c/\sqrt 2$ and $a^{2}b = d/2$. We use the parametrization in terms of $a$ and $b$ because it allows us to see what happens with the potential if one changes $a$: If one decreases $a$, the overall shape of the potential does not change, but it becomes stretched. 
The same potential can be also obtained in supergravity with vector or tensor multiplets \cite{Ferrara:2013rsa}.

\begin{figure}[t!]
\begin{center}
\hskip -0.76 cm \includegraphics[scale=0.16]{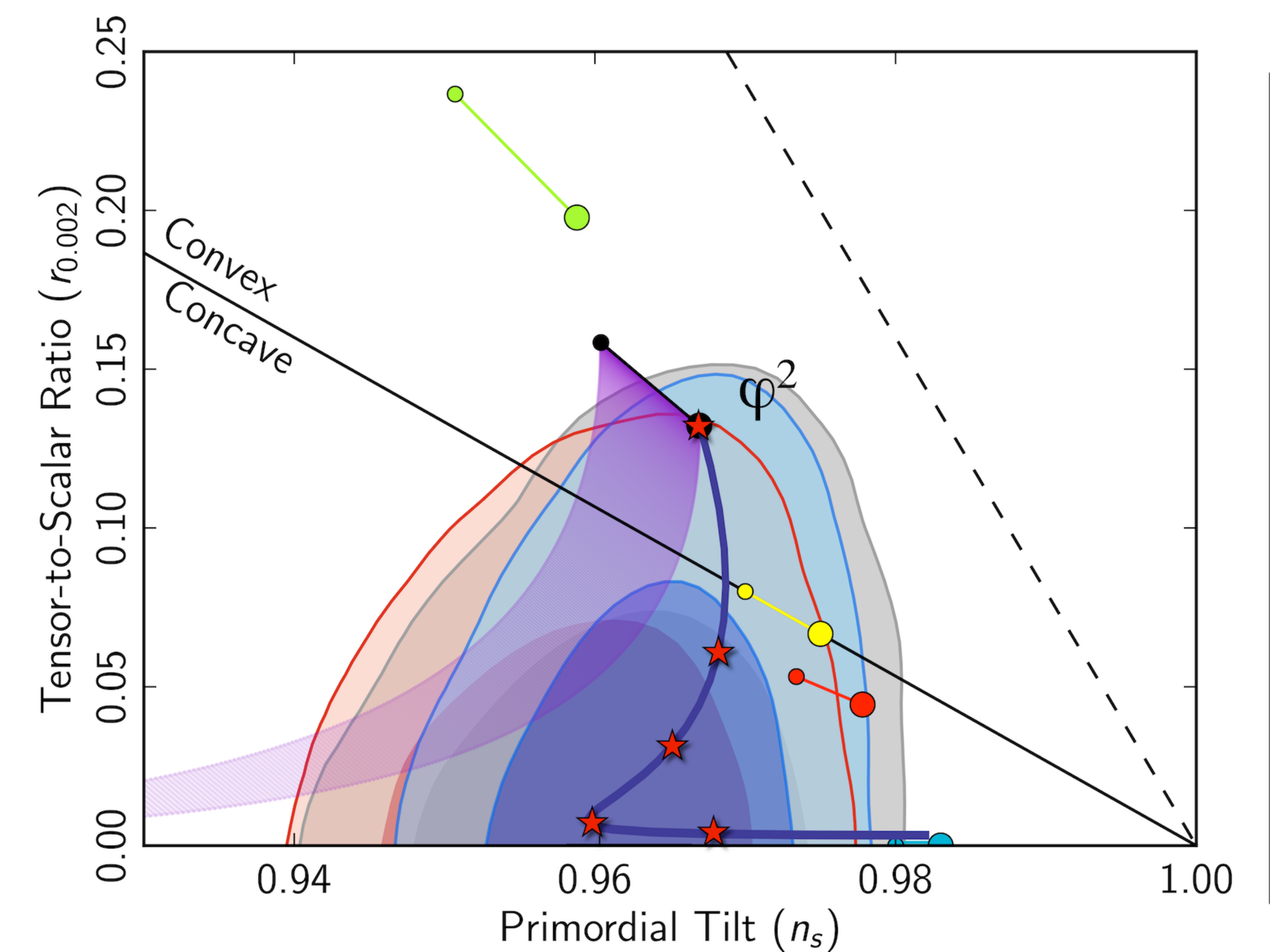}
\end{center}
\caption{\footnotesize Predictions for $n_{s}$ and $r$ in the model with $V(\phi) = {m^{2}\phi^{2}\over 2}\,\bigl(1-a\phi +a^{2}b\,\phi^{2})\bigr)^{2}$ for $b = 0.34$. The stars, from the top down, correspond to $ a = 0$, 0.03, 0.05, 0.1, 0.13. For $a = 0$ one recovers the predictions for the simplest chaotic inflation model with a quadratic potential,  for $a = 0.13$ the predictions almost exactly coincide with the predictions of the Starobinsky model and the Higgs inflation model.}
\label{chi1}
\end{figure}

Inflation in this theory may begin under the same initial conditions as in the simplest large field chaotic inflation models $\phi^{n}$. The difference is that in the small $a$ limit, the last 60 e-foldings of inflation are described by the theory $\phi^{2}$. Meanwhile for large $a$ one has the same regime as in the theory $\phi^{6}$, but at some intermediate values of $a$ the last 60 e-foldings of inflation occurs near the point where the potential bends and becomes concave, see Fig. \ref{chi}.  As a result, for $b = 0.34$ and $0.03\lesssim a \lesssim 0.13$ the observational predictions of this model are in perfect agreement with the Planck data, see Fig. \ref{chi1}. Agreement with the Planck data can be achieved, for a certain range of $a$, for each of the potentials shown in Fig. \ref{chi}.

As we see, a slight modification of the simplest chaotic inflation model with a quadratic potential leads to a model consistent with the results of Planck 2013. These results provide us with 3 main data points: The amplitude of the perturbations $A_{s}$, the slope of the spectrum $n_{s}$ and the ratio of tensor to scalar perturbations $r$. (Tensor perturbations have not been found yet, so following the set of all presently available data, we are talking about the upper bound $r \lesssim 0.1$.) 
The potential of the model (\ref{polyn}) also depends on 3 parameters which are required to fit the data. Thus we are not talking about fine-tuning where a special combination of many parameters is required to account for a small number of data points; we are trying to fit 3 data points, $A_{s}$,  $n_{s}$ and $r$ by a proper choice of 3 parameters, $m$, $a$ and $b$. The values of $n_{s}$ and $r$ do not depend on the overall scale of $V$; they are fully controlled by the parameters $a$ and $b$. One can show that by fixing a proper combination of $a$ and $b$  with a few percent accuracy, one can cover the main part of the area in the $n_{s} - r$ plane allowed by observations. After fixing these two parameters, one can determine the value of $m \sim 10^{{-5}}$ which is required to fit the observed value of $A_{s} \sim 2.2\times 10^{-9}$. 

This is very similar to what happens in the standard model of electroweak interactions, which requires about 20 parameters, which differ from each other substantially. For example, the Higgs coupling to the electron is about $2\times 10^{-6}$. This smallness is required to account for the anomalously small mass of the electron. Meanwhile the Higgs coupling to W and Z bosons and to the top quark are $O(1)$. By looking at this complexity, one could argue that the standard model of elementary particles is unnatural or `unlikely.' But it is extremely successful. The cosmological models discussed above are much simpler than the theory of elementary particles. Nevertheless, it would be very nice to simplify these models and identify some possible reasons why the data by WMAP and Planck gradually zoom to some particular area of $n_{s}$ and $r$. 

In particular, it is quite intriguing that the Starobinsky model and the Higgs inflation model, which are very different from each other, lead to nearly identical predictions for $n_{s}$ and $r$ favored by the Planck data. In what follows we will present some recent results which may help us to understand what is going on there.

\section{Cosmological attractors}

\subsection{Starobinsky model}

\begin{figure}[h!t!]
\begin{center}
\includegraphics[scale=0.41]{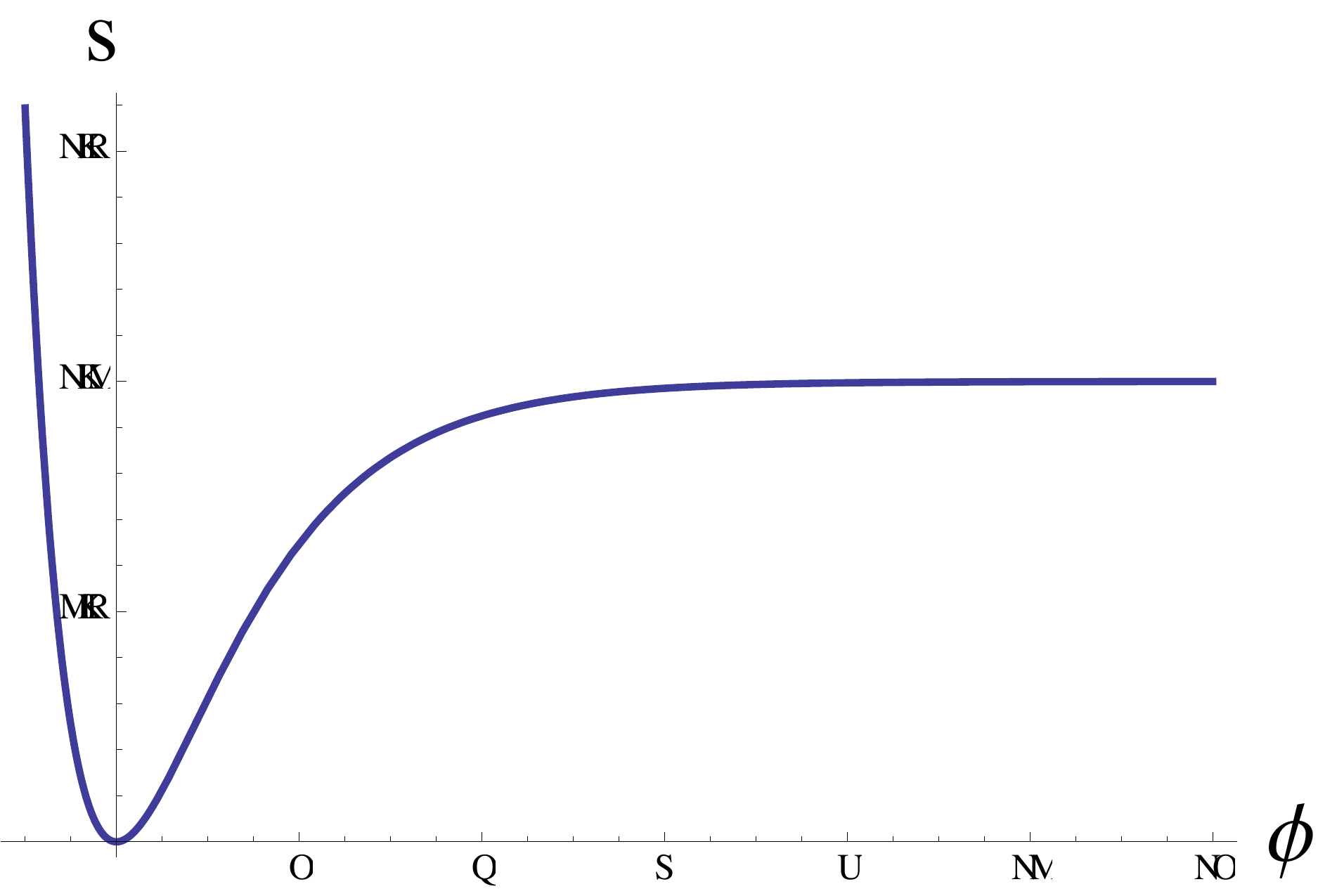}
\end{center}
\caption{\footnotesize The Starobinsky-Whitt potential in units $M=1$.}
\label{stwh}
\end{figure}

The original version of the Starobinsky model  \cite{Starobinsky:1980te} was based on the investigation of quantum effects in GR. These effects could become sufficiently large only if produced by enormously large number of elementary particles. The corresponding effective action consisted of several terms of different nature. Few years later, a streamlined version of this model was proposed \cite{Starobinsky:1983zz}, with the Lagrangian which contained the term $\sim R^{2}$ already at the classical level,
\be\label{star}
L={ \sqrt{-g}} \left({1\over 2} R+{R^2\over 12M^2}\right) \, ,
\ee
where $M \ll M_{p}$ is some mass scale; we keep $M_{p}= 1$ in our paper. Shortly after that, Whitt pointed out that at the classical level this theory  is conformally equivalent to canonical gravity plus a scalar field $\phi$ \cite{Whitt:1984pd}.
Indeed, by making the transformation $(1 + \phi/3M^2) g_{\mu\nu} \to{g}_{\mu\nu}$
and the field redefinition $\varphi = \sqrt{\frac{3}{2}} \ln \left( 1+ \frac{\phi}{3 M^2} \right)$,
one can find the equivalent Lagrangian 
\be\label{whitt}
L=\sqrt{-{g}}\left[{1\over 2}{R} - {1\over 2}\partial_{\mu} \varphi\partial^{\mu} \varphi - \frac{3}{4} M^2 \left(1- e^{-\sqrt{2/3}\,\varphi}\right)^2 \right] \, .
\ee
In the recent literature, this model is often called the Starobinsky model, even though it may not be fully equivalent to the original Starobinsky model at the quantum level.  Its potential is shown in Fig. \ref{stwh}.
The predictions of this model for $n_{s}$ and $r$ can be represented as a function of the number of e-foldings $N$. In the limit of large $N$, one finds
\be  
n_s  = 1-2/N \,, \quad 
  r =12/N^2 \, . \label{observables}
\ee
For $N \sim 60$, these predictions $n_s   \sim 0.967$, $r  \sim 0.003$ ($n_s   \sim 0.964$, $r  \sim 0.004$ for $N \sim 55$) are in the sweet spot of the WMAP9 and Planck2013 data.

In discussion of various supersymmetric generalizations of the Starobinsky model one should distinguish between supergravity extensions of the $R + R^{2}$ model  \rf{star} and the supergravity extensions of its dual version developed by Whitt  \rf{whitt}, where the term $R^2$ is absent and a scalar coupling is present. Supersymmetric generalization of the $R + R^{2}$ Starobinsky model \rf{star} was performed in old minimal supergravity in \cite{Cecotti:1987sa}. Its supersymmetric dual version without higher derivative $R^2$ term involves 2 chiral multiplets, see \cite{Cecotti:1987sa} and a more recent discussion in \cite{Ferrara:2013wka}.  It has therefore 4 scalars. The corresponding consistent supersymmetric cosmological model requiring stabilization of 3 extra moduli, which appear in the theory in addition to the inflaton, was presented in \cite{Kallosh:2013lkr}. In new minimal supergravity,  $R + R^{2}$ models and their dual were given in \cite{Cecotti:1987qe,Farakos:2013cqa,Ferrara:2013rsa,Ferrara:2013kca}. These dual supergravity models with vector or tensor multiplets have only one scalar and do not require extra moduli stabilization. Other supersymmetric extensions of the Starobinsky-Whitt model \rf{whitt} have been developed in \cite{Kallosh:2013lkr,Kallosh:2013hoa}. Supergravity models with  potentials closely approximating \rf{whitt} have been proposed in \cite{Ellis:2013xoa,Buchmuller:2013zfa}.

\subsection{Chaotic inflation in the theories with non-minimal coupling to gravity}

\subsubsection {Universal cosmological attractors}

An amazing fact that we are going to see over and over again in the emergence of the potentials very similar to the potential shown in Fig. \ref{stwh} in the context of a broad class of completely different theories, all of which share the same prediction \rf{observables}, or at least zooming towards it in a certain limit. We will call such theories ``cosmological attractors.'' We will discuss some of such models in this section.

\begin{figure}[h!t!]
\centering
\vskip 0.2cm \includegraphics[scale=0.45]{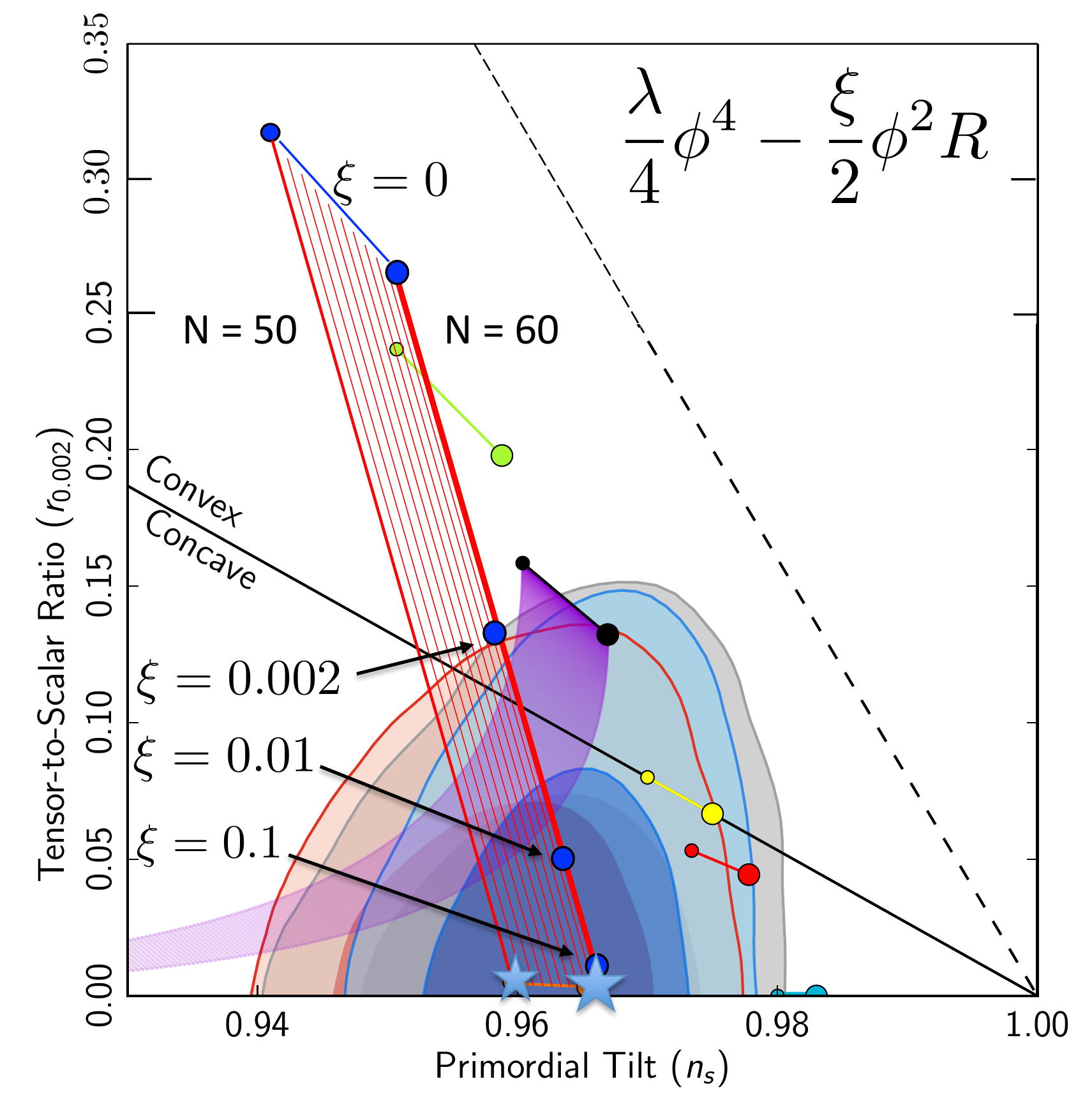}
\caption{\footnotesize Red stripes show inflation parameters $n_{s}$ and $r$ for the model $\lambda\phi^{4}/4$ for different values of $\xi$ in the  term $\xi\phi^{2} R/2$. The top range corresponds to $\xi = 0$. Different stripes correspond to different number of e-foldings $N$. The right one corresponds to $N = 60$, the left one corresponds to $N = 50$. The stars show the predictions of the Starobinsky model, which coincide with the predictions of the model $\lambda\phi^{4}/4$ with $\xi \gg 0.1$.}
\label{potential}
\end{figure}

The simple chaotic inflation model with potential $\lambda\phi^{4}$ in the Einstein frame \cite{Linde:1983gd} is conclusively ruled out by the data due to the high level of the tensor-to-scalar ratio $r$ which it predicts. Meanwhile the same model $\lambda \phi^4$, which includes the non-minimal gravitational coupling ${\xi\over 2} \phi^2 R/2$  in the Jordan frame, makes a dramatic comeback and is in perfect agreement with the Planck 2013  data for ${\xi\over 2} \gtrsim  10^{{-3}}$  \cite{Okada:2010jf,Bezrukov:2013fca}, see Fig. \ref{potential}. 
Effects of non-minimal coupling to gravity on inflation has received a lot of attention over the years \cite{Futamase:1987ua,Salopek:1988qh,Makino:1991sg,Sha-1}.  
The recent revival of interest to these models was related to the possibility that
the Higgs field of the standard model may play the role of the inflaton \cite{Salopek:1988qh,Sha-1}, which would require $\lambda = O(1)$ and  $\xi \sim 10^{5}$. A supersymmetric generalization of the Higgs inflation was found in \cite{Einhorn:2009bh,Ferrara:2010yw,Lee:2010hj,Ferrara:2010in,Linde:2011nh,Kallosh:2013pby}. However, the inflaton does not have to be a Higgs field, its quartic coupling $\lambda$ is not constrained by the standard model phenomenology. Therefore $\lambda$ can be small, which means in turn that the non-minimal coupling $\xi$ does not have to be large. It is amazing that adding to the Lagrangian the term $\xi \phi^2 R/2$ with a minuscule coefficient $\xi/2 >  10^{{-3}}$ is sufficient to make the simple chaotic inflation model $\lambda \phi^4$ viable.
Thus one may wonder whether there is something special about the model $\lambda\phi^{4}$ with nonminimal coupling to gravity, and would it be possible to find other models of chaotic where a similar effect is possible.

The answer to this question was found only recently \cite{Kallosh:2013tua}. Here we will describe the main results of  \cite{Kallosh:2013tua}, which allow to find a very large class of models which have a very similar property.

The starting point of the inflationary models to be discussed now is a Lagrangian 
  \begin{align}
 & \mathcal{L}_{\rm J} = \sqrt{-g} [ \tfrac12  \Omega(\phi) R  - \tfrac12 (\partial \phi)^2 -  V_J(\phi) ] \,, \label{Baction} 
\end{align}
with 
\begin{align}
& \Omega(\phi)  = 1 + \xi f(\phi)\, ,  \qquad V_J(\phi)=\lambda^2 f^2(\phi) \,.
\end{align}
Here we represented a positive inflationary potential as $\lambda^2 f^2(\phi)$, where $f(\phi)$ is the same function as in the expression for $\Omega$. Later we will show how one may relax this requirement. Due to the non-minimal coupling to gravity $\Omega(\phi) R$, we will refer to this form of the theory as Jordan frame. In order to transform to the canonical Einstein frame, one needs to redefine the metric $ g_{\mu \nu} \rightarrow \Omega(\phi)^{-1} g_{\mu \nu}$.
This brings the Lagrangian to the Einstein-frame form:
 \be\label{Einstein}
\mathcal{L}_{\rm E} = \sqrt{-g} [    \tfrac12   R - \tfrac12 \Big (\Omega(\phi)^{-1} + \tfrac32 (\log \Omega(\phi))'^2\Big ) (\partial \phi)^2  - V_E(\phi)]\ee
where
\be 
 {\rm with~~} V_E(\phi) =  \frac{V_J(\phi)}{\Omega(\phi)^2}  \,. 
 \ee
Note that in the absence of non-minimal coupling, $\xi = 0$, the distinction between Einstein and Jordan frame vanishes. In this case the inflationary dynamics is fully determined by the properties of the scalar potential $V_J(\phi)= V_E(\phi)$. In the presence of a non-minimal coupling, however, one has to analyze the interplay between the different contributions  to the inflationary dynamics due to $V_J(\phi)$ and $\xi$. 

Let us first concentrate on the large $\xi$ limit. In this limit, the two contributions to the kinetic terms in (\ref{Einstein})  scale differently under $\xi$. Retaining only the leading term, the Lagrangian becomes
  \begin{align}
  \mathcal{L}_{\rm E} = \sqrt{-g} \bigg[ & \tfrac12  R - \tfrac34 (\partial \log (\Omega(\phi)))^2 
  - \lambda^2 \frac{f(\phi)^2}{\Omega(\phi)^2}  \bigg] .
 \end{align}
The canonically normalized field $\varphi$ involves the function $\Omega(\phi)$,
 \begin{align}
  \varphi = \pm \sqrt{3/2} \log (\Omega(\phi)) \,. \label{canonical}
 \end{align}
Therefore, in terms of $\varphi$, the theory has lost all reference to the original scalar potential, it has the universal form. In case of odd $f(\phi)$ we choose the same sign in \eqref{canonical} for both signs   of $\varphi$ and find
  \begin{align}\label{potstar}
  \mathcal{L}_{\rm E} = \sqrt{-g} \left[ \tfrac12   R - \tfrac12 (\partial \varphi)^2 - \frac{ \lambda^2}{\xi^{2}} (1 - e^{-\sqrt{{2\over 3}} \, \varphi})^{2}  \right] \,,
 \end{align}
which is the scalar formulation of the Starobinsky-Whitt model \cite{Starobinsky:1980te}, with the potential shown in Fig. \ref{stwh}. 
If the function $f(\phi)$ is  even in $\phi$, we choose opposite signs and find the following attractor action
  \begin{align}\label{pothiggs}
  \mathcal{L}_{\rm E} = \sqrt{-g} \left[ \tfrac12   R - \tfrac12 (\partial \varphi)^2 - \frac{ \lambda^2}{\xi^{2}} \Big (1 - e^{-\sqrt{{2\over 3}  \varphi^2}} \, \, \Big )^{2}  \right] .
 \end{align}
In this case, the potential in the large $\xi$ limit coincides with the Starobinsky-Whitt potential for $\varphi > 0$, but, unlike that potential, it is symmetric under $\varphi \rightarrow -\varphi$. 
In both cases, these models in the large $\xi$ limit lead to the same observational predictions  (\ref{observables}), {\it for any choice of the potential $V(\phi)$}.

One can generalize this class of models even further. Indeed, in this class of models one can show that the last N e-foldings of inflation occur when the field satisfies the condition
\be\label{N}
N = \tfrac34 \xi f(\phi_N) \,.
\ee
For example, if $f(\phi_N) = \phi$, the last $N$ e-foldings of inflation occur for $\phi <  \tfrac43 \xi^{-1}$. This means that for large $\xi$ the description of the last N e-foldings of inflation discussed above works for all theories where the structure of the theory is the same as in \rf{Baction} in a small vicinity of the minimum of the potential where $f(\phi_N) \lesssim N/\xi$.  

Let us use this fact to our advantage. In the previous investigation, we assumed that $\Omega(\phi) = 1 + \xi f(\phi)$, and  $V_J(\phi)=\lambda^2 f^2(\phi)$, but one may also consider a more general class of theories, where
\be
\Omega(\phi) = 1 + \xi g(\phi)\, ,  \qquad V_J(\phi)=\lambda^2 f^2(\phi) \, . 
\ee
Here we introduce an additional functional freedom in the definition of $\Omega(\phi)$, disconnecting it from $V_J(\phi)$. As we will see now, many of the results obtained above will work in this case as well.

Note that when the field rolls to the minimum of its potential, $f(\phi)$ is supposed to vanish, or at least to become incredibly small to account for the incredible smallness of the cosmological constant $\sim 10^{-120}$. As in the previous analysis, we will assume that the same is true for the function $g(\phi)$, simply because its constant part is already absorbed into the definition of the Planck mass. Therefore we will expand both functions in a Taylor series in $\phi$, assuming that they vanish at some point (which can be taken as $\phi =0$ by a field redefinition) and that they are differentiable at this point:
 \be
f(\phi) = \sum_{n=1}^{\infty} f_{n} \phi^{n} \ , \qquad g(\phi) = \sum_{n=1}^{\infty}  g_{n} \phi^{n} \ . 
 \ee
By rescaling $\lambda $ and $\xi$, one can always redefine $f_{1} = g_{1} = 1$ without loss of generality.

Let us first ignore all higher order corrections, i.e. take $f(\phi) = g(\phi) = \phi$. In this case our investigation is reduced to the one performed earlier, and equation (\ref{N}) yields $\phi_{\rm N} = {4N\over 3\xi}$. This result implies that for $\xi \gg N$ one has $\phi_{\rm N}\ll 1$.

If one now adds all higher order terms and makes an assumption that  the coefficients $f_{n}$ and $g_{n}$ are $O(1)$, one finds that in the large coupling limit $\xi \gg N$, these corrections are suppressed by the powers of ${4N\over 3\xi}$, so one can indeed ignore these terms. Note that the assumption that $f_{n}$ and $g_{n}$ do not blow up is just that, an assumption, but for a broad class of function for which this assumption is valid we have an important result: In the large $\xi$ limit, the potential $V(\varphi)$ in terms of the canonically normalized inflaton field $\varphi$ coincides with the potential (\ref{potstar}), and all observational predictions of this new broad class of theories coincide with (\ref{observables}). 

In this analysis we assumed that the Taylor series begins with the linear term. However, if the theory is symmetric with respect to the change $\phi \to -\phi$, as is the case e.g.~in the $\phi^4$ theory, then the expansion for $f(\phi)$ and $g(\phi)$ begins with the quadratic terms. The rest follows just as in the case discussed above: For  $\xi\gg N$, higher order corrections do not affect the description of the observable part of the universe, we have the same observational predictions (\ref{observables}) as before, but now the relevant part of the potential is even with respect to the field $\varphi$, and its large $\xi$ limit is given by (\ref{pothiggs}).

To conclude, instead of the two apparently unrelated models, the Starobinsky model and the Higgs inflation model, we now have a huge variety of different models which lead to identical observational predictions in the large $\xi$ limit.

\begin{figure}[t!]
\vspace{-.3cm}
\centerline{\includegraphics[scale=.39]{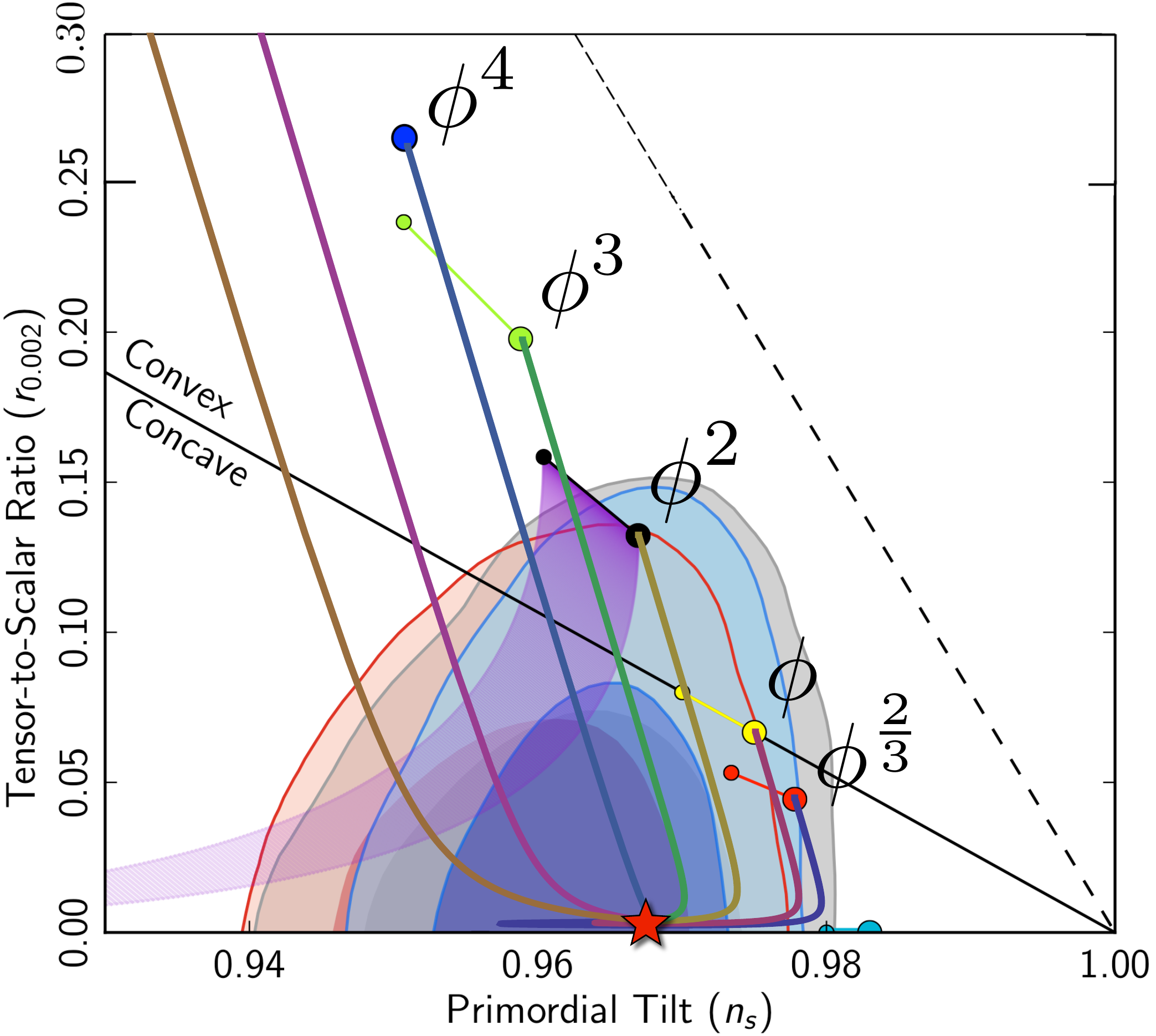}}
\caption{\footnotesize{The $\xi$-dependence of $(n_s,r)$ for different chaotic inflation models $\phi^{n}$ with $n=(2/3, 1, 2, 3, 4, 6, 8)$, from right to left, for 60 e-foldings. }}
\vspace{-.3cm}
\label{attrstr2}
\end{figure}

But what if we do not take this limit and follow the evolution of the predictions while $\xi$ grows from $0$? This issue was studied in Ref. \cite{Kallosh:2013tua},  Fig. \ref{attrstr2} illustrates the main  results of this investigation. As we see, predictions of all chaotic inflation models  with potentials $\sim \phi^{n}$ continuously flow towards the attractor point (\ref{observables}) when the value of the nonminimal coupling $\xi$ increases. Importantly, for $n = 4$ the attractor is reached very early, for $\xi \gtrsim 10^{-1}$, and for $n > 4$ the convergence occurs even much faster, for much smaller values of $\xi$. In this sense, the words ``strong coupling limit'', or ``the large $\xi$ limit'' are not entirely adequate, because the ``strong coupling limit'' is reached in these models very early, at $\xi \ll 1$.

Meanwhile in the limit $\xi \to 0$ the prediction of this class of models approach the standard chaotic inflation result for the theories $V\sim \phi^{n}$:
\be
n_{s} = 1 -{2+n\over 2 N}\, , \qquad r = {4n\over N} \ .
\ee
As one can see from the Fig.  \ref{attrstr2}, the set of all available possibilities covers the broad range of values of $n_{s}$ and $r$ allowed by the WMAP9 and Planck results.

From the point of view of fundamental physics it may be important that all of these theories can be incorporated into the superconformal theory and supergravity. The way it can be done is explained in \cite{Kallosh:2013tua}. In Section \ref{susyattr} we will describe another broad class of superconformal attractors, which is considerably different from the one discussed above, and yet it leads to the same observational consequences.

\subsubsection{Unitarity bound?}

The models discussed above significantly generalize the chaotic inflation model $\lambda\phi^{4}/4$ with non-minimal coupling to gravity ${\xi\over 2}\phi^{2} R$, which was used in \cite{Salopek:1988qh} for the discussion of the Higgs inflation scenario with $\lambda = O(1)$, $\xi \sim 10^{5}$. In this respect we should mention that during the discussion of the Higgs inflation models, several authors claimed that Higgs inflation suffers from the unitarity problem \cite{Burgess:2009ea,Barbon:2009ya,Hertzberg:2010dc,Germani:2010gm}, whereas some others argued that this problem does not exist  \cite{Ferrara:2010in,Bezrukov:2010jz}. 

Here we will briefly discuss this issue and check whether a similar issue arises in the new class of models as well \cite{Kehagias:2013mya,Giudice:2014toa}. 
A more precise statement of the result obtained in \cite{Burgess:2009ea,Barbon:2009ya,Hertzberg:2010dc,Germani:2010gm}  is that in the vicinity of the Higgs minimum, at $\vp \ll 1/\xi \sim O(10^{-5})$, higher order quantum corrections to scattering amplitudes become greater than the lower order effects for energies greater than $O(1/\xi)$. The energy scale $O(1/\xi)$ above which the perturbation theory fails at $\vp \ll 1/\xi$ was called the unitarity bound. Since the Hubble constant during inflation is $O(\sqrt\lambda/\xi)$, and $\lambda = O(1)$ in Higgs inflation, it was conjectured in  \cite{Burgess:2009ea,Barbon:2009ya,Hertzberg:2010dc,Germani:2010gm} that the description of the Higgs inflation using perturbation theory is unreliable. However, inflation happens at $\vp \geq O(1)$, which is $10^{5}$ times greater than the range of the values of the field where the existence of the problem was established. An investigation performed in \cite{Ferrara:2010in,Bezrukov:2010jz} demonstrated that the higher order corrections are negligible during inflation because at  large $\vp$ the potential \rf{pothiggs} is exponentially flat and the effective coupling constant $\lambda(\vp)$ is exponentially small.

After inflation, when the field $\vp$ becomes very small, one may encounter problems in describing reheating by perturbation theory. But is it a real problem? It is well known that the perturbative approach to reheating fails in many inflationary models anyway, which requires using non-perturbative methods developed in  \cite{KLS,latticeold}. This does not affect inflationary predictions and does not lead to any problems with the inflationary scenario. 

This does not mean that the unitarity issue is entirely inconsequential.  The problem may re-appear if one attempts to develop LHC-related particle phenomenology models with the Higgs field playing the role of the  inflation. This may require solving RG equations up to the Planckian energies, which is problematic for $\xi \gg 1$. But this is not a problem of consistency of inflationary models but rather a specific problem of particle phenomenology beyond the Standard Model. An interesting way to avoid this problem was recently proposed in \cite{Giudice:2014toa}. However, the mass of the inflaton field in the class of the universal attractor models developed in  \cite{Giudice:2014toa} is 10 orders of magnitude greater the Higgs mass, so this approach is not directly related to Higgs inflation and particle phenomenology beyond the Standard Model.

In these lectures, we do not make any attempts to relate the inflaton field to the Higgs field discovered at LHC. In particular, the coupling constant $\lambda$ in the generic chaotic inflation models $\lambda\phi^{4}/4$ with non-minimal coupling to gravity ${\xi\over 2}\phi^{2} R$ can be extremely small. This makes the inflationary energy scale $H \sim \sqrt\lambda/\xi$ much smaller than $O(1/\xi)$ and alleviates the problems discussed in \cite{Burgess:2009ea,Barbon:2009ya,Hertzberg:2010dc,Germani:2010gm}. Moreover, in the theories $\phi^{n}$ with $n\lesssim 1$ the unitarity bound is much higher than the Planck mass even for large $\xi$ \cite{Kehagias:2013mya}. For $n\geq 2$, our results show that the attractor behavior occurs starting from $\xi <1$, in which case the unitarity bound is also super-planckian. Therefore we believe that the perturbative unitarity problem does not affect the main results obtained above.

\subsection{Superconformal attractors}\label{susyattr}

The standard approach to cosmological evolution is based on the Einstein theory of gravity. The gravitational constant in this theory is indeed a constant, $G = (8\pi M_{p})^{-2}$. Since it is a constant, it is customary to simply take $M_{p} = 1$ in all equations. In the standard approach to supergravity, one can also take $M_{p} = 1$. However, in the superconformal formulation of supergravity, which is one of the most powerful tools used since the very early days of this theory  \cite{Siegel:1977hn,Kaku:1978ea,Cremmer:1978hn,Freedman:2012zz}, the theory possesses an additional set of fields and symmetries. Planck mass becomes a constant only after gauge fixing of these extra symmetries. Just like in the theory of spontaneous symmetry breaking in the Higgs model, the original symmetry does not disappear after the symmetry breaking and/or gauge fixing: One can either use the unitary gauge, where the physical contents of the theory are manifest, or other gauges where the calculations can be easier; all physical results do not depend on this choice. Similarly, the original symmetries of the superconformal theory are still present in the standard formulation of supergravity with $M_{p} = 1$, but they are well hidden and therefore often forgotten and rarely used. We believe that this is going to change, as the superconformal approach becomes very useful in the cosmological context \cite{Kallosh:2000ve,Kallosh:2013oma}.

Since these lectures are cosmology-oriented, we will mostly discuss non-supersymmetric models based on a local conformal symmetry, instead of the superconformal one. Generalization to superconformal theory and supergravity is presented in the original papers to which we will refer, and also in the lecture by Renata Kallosh at this school \cite{RenataLecture}. 

\subsubsection{The simplest conformally invariant models of dS/AdS space}

As a first step towards the development of the new class of inflationary models based on spontaneously broken conformal symmetry, consider a simple  conformally invariant model of gravity 
of a scalar field $\chi$ with the following Lagrangian
\begin{equation}
\mathcal{L} = \sqrt{-{g}}\left[{1\over 2}\partial_{\mu}\chi \partial_{\nu}\chi \, g^{\mu\nu}  +{ \chi^2\over 12}  R({g}) -{\lambda\over 4}\chi^4\right]\,.
\label{toy2aa}
\end{equation}
This
theory is locally conformal invariant under the following
transformations: 
\be \tilde g_{\mu\nu} = e^{-2\sigma(x)} g_{\mu\nu}\,
,\qquad \tilde\chi =  e^{\sigma(x)} \chi\ . \label{conf2aaa}
\ee
The field $\chi(x)$ is referred to as  a conformal compensator, which we will call `conformon.'   It has negative sign kinetic term, but this is not a problem because it
can be removed from the theory by fixing the gauge symmetry
(\ref{conf}), for example by taking a gauge $\chi =\sqrt 6$. This gauge fixing can be interpreted as a spontaneous breaking of conformal invariance due to existence of a classical field $\chi =\sqrt 6$.

After fixing $\chi = \sqrt 6$ the interaction term $-{\lambda\over 4}\chi^4$  becomes a cosmological constant $\Lambda = {9}\lambda$
\be
\mathcal{L}=  \sqrt{-{g}}\,\left[{  R({g})\over 2}-{9\lambda}\right]\, .
\label{toy2b}
\ee
For $\lambda > 0$, this theory has a simple de Sitter solution 
\be\label{dssol1}
a(t) = e^{Ht}  = e^{\sqrt{\Lambda/3}\ t}  = e^{\sqrt{3\lambda}\ t}.
\ee
Meanwhile for $\lambda < 0$ it becomes the AdS universe with a negative cosmological constant.

As a next step, consider the model of two real scalar fields, $\phi$ and $\chi$, which has also an $SO(1,1)$ symmetry:
\begin{equation}
\mathcal{L}_{\rm toy} = \sqrt{-{g}}\left[{1\over 2}\partial_{\mu}\chi \partial^{\mu}\chi  +{ \chi^2\over 12}  R({g})- {1\over 2}\partial_{\mu} \phi\partial^{\mu} \phi   -{\phi^2\over 12}  R({g}) -{\lambda\over 4}(\phi^2-\chi^2)^{2}\right]\,.
\label{toy}
\end{equation}
This
theory is locally conformal invariant under the following
transformations: 
\be \tilde g_{\mu\nu} = e^{-2\sigma(x)} g_{\mu\nu}\,
,\qquad \tilde \chi =  e^{\sigma(x)} \chi\, ,\qquad \tilde \phi =  e^{\sigma(x)}
\phi\ . \label{conf}\ee 
The global $SO(1,1)$ symmetry is a boost between these two fields. Because of this symmetry,   it is very convenient to choose an  $SO(1,1)$ invariant conformal gauge,
\be
\chi^2-\phi^2=6 \ ,
\ee
which reflects the $SO(1,1)$ invariance of our model. This gauge condition (we will refer to it as a `rapidity' gauge) represents a hyperbola which can be parametrized by a canonically normalized field $\varphi$ 
\be
\chi=\sqrt 6 \cosh  {\varphi\over \sqrt 6}\, , \qquad \phi= \sqrt 6 \sinh {\varphi\over \sqrt 6} \ .
\ee
This allows us to modify the kinetic term into a canonical one for the independent field $\varphi$
\begin{equation}
{1\over 2}\partial_{\mu}\chi \partial^{\mu}\chi  - {1\over 2}\partial_{\mu} \phi\partial^{\mu} \phi   \,  \quad \Rightarrow \quad {1\over 2}\partial_{\mu}\varphi \partial^{\mu}\varphi
\label{kin}
\end{equation}
and the non-minimal curvature coupling into a minimal one
\begin{equation}
{ \chi^2\over 12}  R({g})   -{\phi^2\over 12}  R({g})   \,  \quad \Rightarrow \quad { 1\over 2}  R({g})
\,.
\label{R}
\end{equation}

In this gauge the Higgs-type potential ${\lambda\over 4}(\phi^2-\chi^2)^{2}$ turns out to be a cosmological constant $9\lambda$,
and our action \rf{toy} becomes
\begin{equation}\label{LE}
L = \sqrt{-g} \left[  \frac{1}{2}R - \frac{1}{2}\partial_\mu \varphi \partial^{{\mu}} \varphi -   9 \lambda  \right].
\end{equation}

This theory has the constant potential $V = 9\lambda$. Therefore this model can describe de Sitter expansion with the Hubble constant $H^{2} = 3\lambda$, for positive $\lambda$. For negative $\lambda$ in \rf{toy} the model can describe an AdS space. But unlike the one-field model, this model also describes a massless scalar field $\varphi$, which may have its kinetic and gradient energy. If one takes this energy into account, the universe approaches dS regime as soon as expansion of the universe makes the kinetic and gradient energy of the field $\varphi$ sufficiently small.

\subsubsection{Chaotic inflation from conformal theory: T-Model}\label{tmod}

Now we will consider a  conformally invariant class of models
\begin{equation}
\mathcal{L} = \sqrt{-{g}}\left[{1\over 2}\partial_{\mu}\chi \partial^{\mu}\chi  +{ \chi^2\over 12}  R({g})- {1\over 2}\partial_{\mu} \phi\partial^{\mu} \phi   -{\phi^2\over 12}  R({g}) -{1\over 36} F\left({\phi/\chi}\right)(\phi^{2}-\chi^{2})^{2}\right]\,.
\label{chaotic}
\end{equation}
where $F$ is an arbitrary function in term of the homogeneous variable $z = \phi/\chi$, which is the natural variable for this class of theories. When this function is present, it breaks the $SO(1,1)$ symmetry of the de Sitter model \rf{toy}. Note that this is the only possibility to keep local conformal symmetry and to deform the $SO(1,1)$ symmetry:  an arbitrary function $F(z)$ has to deviate from the critical value $F(z)=\rm const$,  where $SO(1,1)$ symmetry is restored.

This theory has the same conformal invariance as the theories considered earlier. As before, we may use the gauge $\chi^2-\phi^2=6$ and resolve this constraint in terms of the fields 
$\chi=\sqrt 6 \cosh  {\varphi\over \sqrt 6}$,  $ \phi= \sqrt 6 \sinh {\varphi\over \sqrt 6} $ 
and the  canonically normalized field $\varphi$:~ ${\phi\over\chi} = \tanh{\varphi\over \sqrt 6}$.
Our action \rf{chaotic} becomes
\begin{equation}\label{chaotmodel}
L = \sqrt{-g} \left[  \frac{1}{2}R - \frac{1}{2}\partial_\mu \varphi \partial^{\mu} \varphi -   F(\tanh{\varphi\over \sqrt 6}) \right].
\end{equation}
Note that asymptotically $\tanh\varphi\rightarrow \pm 1$ and therefore $F(\tanh{\varphi\over \sqrt 6})  \rightarrow \rm const$, the system  in large $\varphi$ limit evolves asymptotically  towards its critical point  where the $SO(1,1)$ symmetry is restored.

Since $F(z)$ is an {\it arbitrary} function, by a proper choice of this function one can reproduce an {\it arbitrary} chaotic inflation potential $V(\varphi)$ in terms of a conformal theory with spontaneously broken conformal invariance. But this would look rather artificial. For example, to find a theory which has potential $m^{2}\varphi^{2}/2$ in terms of the canonically normalized field $\varphi$, one would need to use $F(z) \sim (\tanh^{{-1}}z)^{2}$, which is a possible but rather peculiar choice.

Alternatively, one may think about the function $F\left({\phi/\chi}\right)$ as describing a deviation of inflationary theory from the pure cosmological constant potential which emerges in the theory (\ref{toy}) with the coupling $(\phi^{2}-\chi^{2})^{2}$. Therefore it is interesting to study what will happen if one takes the simplest set of functions $F\left({\phi/\chi}\right) = \lambda \left({\phi/\chi}\right)^{2n}$ as we did in the standard approach to chaotic inflation.
In this case one finds 
\begin{equation}\label{TModel}
V(\varphi) = \lambda\ {\tanh}^{2n}(\varphi/\sqrt6) .
\end{equation}
This is a basic representative of the universality class of models depending on $\tanh(\varphi/\sqrt6) $.  We will call it a T-Model, because it originates from different powers of $\tanh(\varphi/\sqrt6) $, and also because its basic representative $ \lambda\ {\tanh}^{2}(\varphi/\sqrt6)$ is the simplest version of the class of conformal chaotic inflation models considered in this paper. Moreover, as we will see, observational predictions of a very broad class of models of this type, including their significantly modified and deformed cousins, have nearly identical observational consequences, thus belonging to the same universality class. 

\begin{figure}[h!t!]
\centering
\includegraphics[scale=0.9]{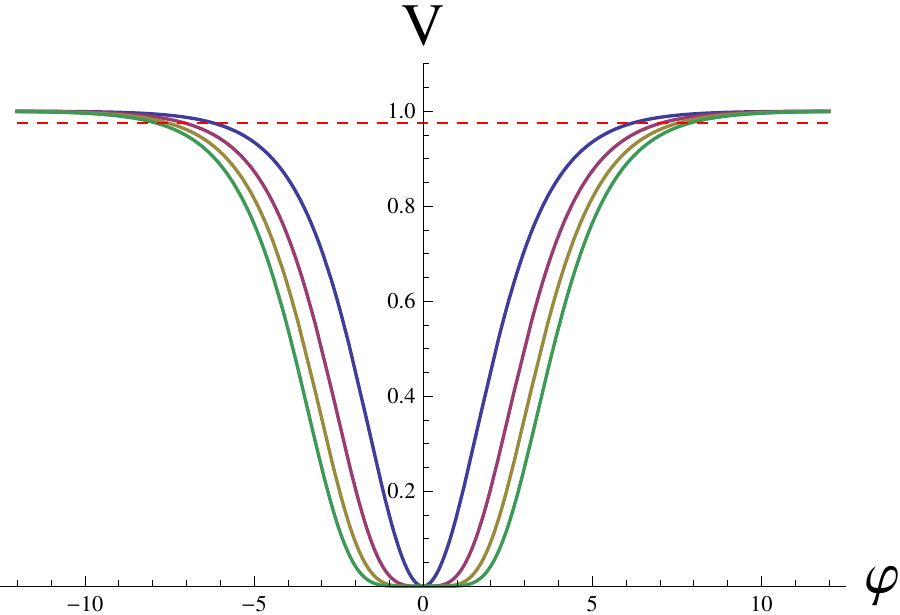}
\caption{\footnotesize Potentials for the T-Model inflation ${\tanh}^{2n}(\varphi/\sqrt6)$ for $n = 1,2,3,4$ (blue, red, brown and green,  corresponding to increasingly wider potentials).  We took $\lambda = 1$ for each of the potentials for convenience of comparison. All of these models predict the same values $n_{s} =1-2/N$, $r = 12/N^{2}$, in the leading approximation in $1/N$, where $N\sim 60$ is the number of e-foldings. The points where each of these potentials cross the red dashed line $V = 1-3/2N = 0.975$ correspond to the points where the perturbations are produced in these models on scale corresponding to $N = 60$. Asymptotic height of the potential is the same for all models of this class, see (\ref{energylevel}).}
\label{tmodelfig}
\end{figure}

Functions ${\tanh}^{2n}(\varphi/\sqrt6)$ are symmetric with respect to $\varphi \to - \varphi$. To study inflationary regime in this model at $\varphi \gg 1$, it is convenient to represent them as follows:
\begin{equation}\label{eframe111a}
V(\varphi) = \lambda \left({1- e^{-\sqrt{2/3}\, \varphi}\over 1+ e^{-\sqrt{2/3}\, \varphi}}\right)^{2n}= \lambda \left(1- 4n\, e^{-\sqrt{2/3}\, \varphi} + O\left(n^{2}\, e^{-2\sqrt{2/3}\, \varphi}\right)\right).
\end{equation}
One can easily check that in all models with potentials of this type, independently on $n$, one has the same ``attractor'' result \rf{observables}:
\be
1 -n_{s} =2/N\, , \qquad r = 12/N^{2} \ ,
\ee
 in the leading approximation in $1/N$.

Interestingly, Starobinsky-Whitt potential \rf{whitt} also appears the class of chaotic conformal inflation models discussed above, with $F({\phi/\chi}) \sim {\phi^{2}\over (\phi+\chi)^{2}}$, which leads to 
\be
V(\varphi)\sim   \Big [{ \tanh (\varphi/\sqrt6)) \over 1+ \tanh (\varphi/\sqrt6) ) }\Big ]^2 \sim  \left(1- e^{-\sqrt{2/3}\, \varphi}\right)^{2} \ .
\ee

\subsubsection{Universality of conformal inflation}
In this section we will describe the roots of the universality of predictions of conformal inflation in a more general way. But first of all, we will consider some instructive examples, which will help to explain the main idea of our approach.

\begin{figure}[ht!]
\centering
\includegraphics[scale=0.9]{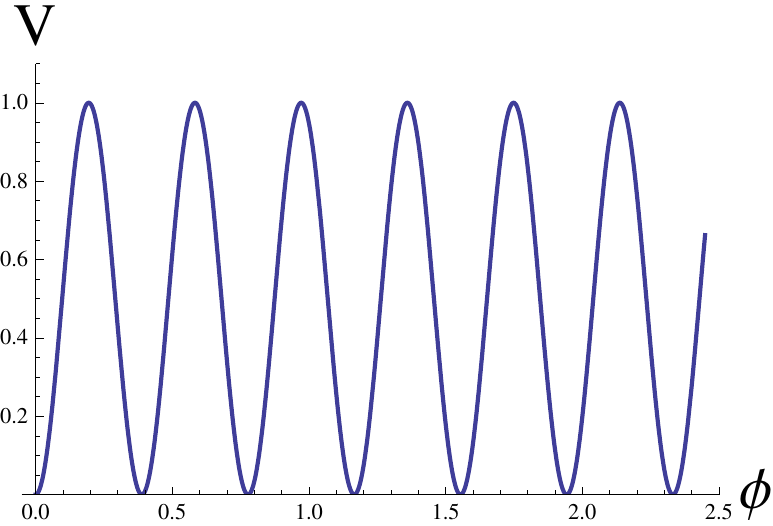}\hskip 2 cm \includegraphics[scale=0.9]{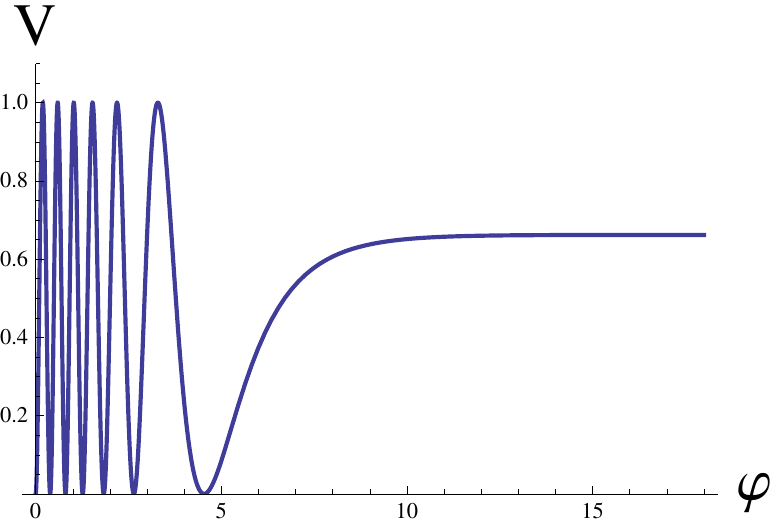}
\caption{\footnotesize  Flattening of the sinusoidal potential $V(\phi)$ near the boundary of the moduli space $\phi = \sqrt 6$ by boost  in the moduli space, $V(\phi) \to V(\sqrt 6 \, \tanh {\varphi\over \sqrt{6}})$. Inflationary plateau of the function $V(\phi)$ appears because of the exponential stretching of the last growing part of the sinusoidal function $V(\phi)$.}\label{stretchsin}
\end{figure}

Consider a sinusoidal function $F\left({\phi/\chi}\right) \sim \sin(a+ b \phi)$ and check what will happen to it after the boost $V(\phi) \to V(\sqrt 6 \, \tanh {\varphi\over \sqrt{6}})$. As we see from the Fig. \ref{stretchsin}, the main part of the stretch of the potential occurs very close to the boundary of moduli space, near $\phi = \sqrt 6$. The rising segment of the sinusoidal function bends and forms a plateau, which has an ideal form for the slow-roll inflation.

\begin{figure}[ht!]
\centering
\includegraphics[scale=0.18]{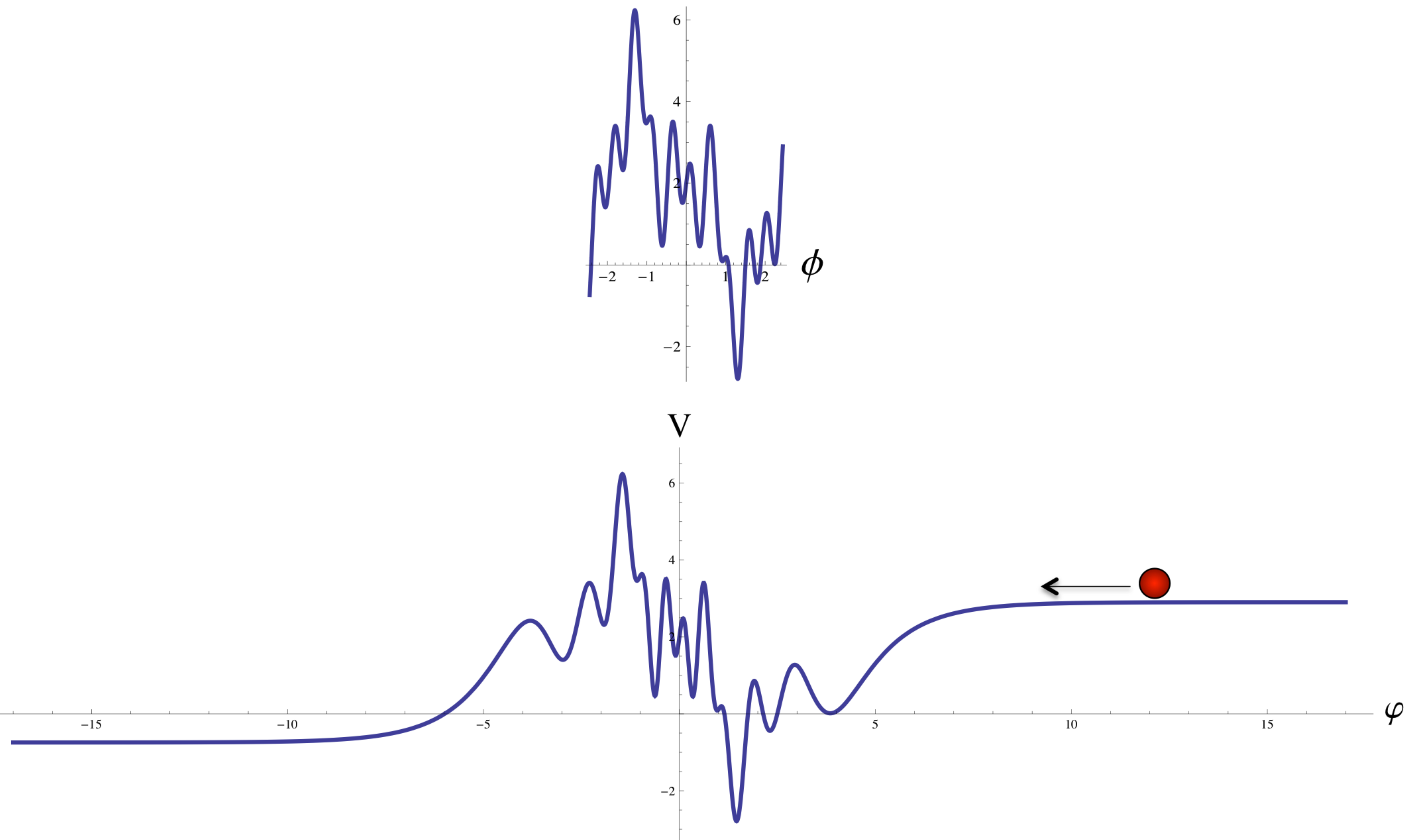}
\caption{\footnotesize Basic mechanism which leads to inflation in the theories with generic functions $F(z)$. The potential $V$ in the Einstein frame can have an arbitrary shape in terms of the original conformal variable $z$, which becomes $\phi/\sqrt 6$ in the gauge $\chi = \sqrt 6$; see e.g. the potential $V(\phi/\sqrt 6)$ in  the upper panel. If this potential is non-singular at the boundary of the moduli space $|z| = 1$ ($\phi= \sqrt 6$), it looks exponentially stretched and flat at large values of the canonically normalized field $\varphi$. This stretching  makes inflation very natural, and leads to universal observational predictions for a very broad class of such models \cite{Kallosh:2013hoa}. In essence, what we see is `inflation {\it of}\, the inflationary landscape' at the boundary of the moduli space, which solves the flatness problem of the inflationary potential required for inflation {\it in}\, the landscape. }
\label{stretch}
\end{figure}

Now let us study a more general and complicated potential  on the full interval $-\sqrt 6 < \phi< \sqrt 6$, as shown in Fig. \ref{stretch}. It shows the same effect as the one discussed above: The part of the landscape at $|\phi| \ll \sqrt 6$ does not experience any stretching. The inflationary plateau appears because of the exponential stretching of the growing branch of $V(\phi)$ near $\phi = \sqrt 6$.

In this scenario, inflationary regime emerges each time when $V(\phi)$ grows at the boundary of the moduli space, which is a rather generic possibility, requiring no exponential fine-tuning.\footnote{As usual, there is an unavoidable fine-tuning of the cosmological constant, which is the value of the potential at the local minimum corresponding to our part of the universe. This issue can be taken care off by the usual considerations involving inflationary multiverse and string theory landscape.}
 
By expanding the potential in powers of the distance from the boundary of the moduli space, one can show that it is given by 
\be
V(\varphi) = V_{*}\ \left(1 - c\  e^{-\sqrt{2/3}\, \varphi} \right) \ ,
\ee
up to exponentially suppressed higher order corrections. Just as in different versions of the T-model, the resulting inflationary predictions in the large $N$ limit are given by \rf{observables}.

From these results, one can derive an additional universal parameter, the energy scale of inflation, $V \approx V_{*}$, which takes the same value for all models described above, in the leading order in $1/N$ 
\be\label{energylevel}
V(\varphi_{N}) \approx  4 \times 10^{-7} N^{-2} \sim  10^{-10}  
\ee
in Planck units, for all models in this universality class  \cite{Kallosh:2013hoa}.

\subsubsection{Multifield conformal attractors}

\begin{figure}[h!t!]
\centering
\includegraphics[scale=0.17]{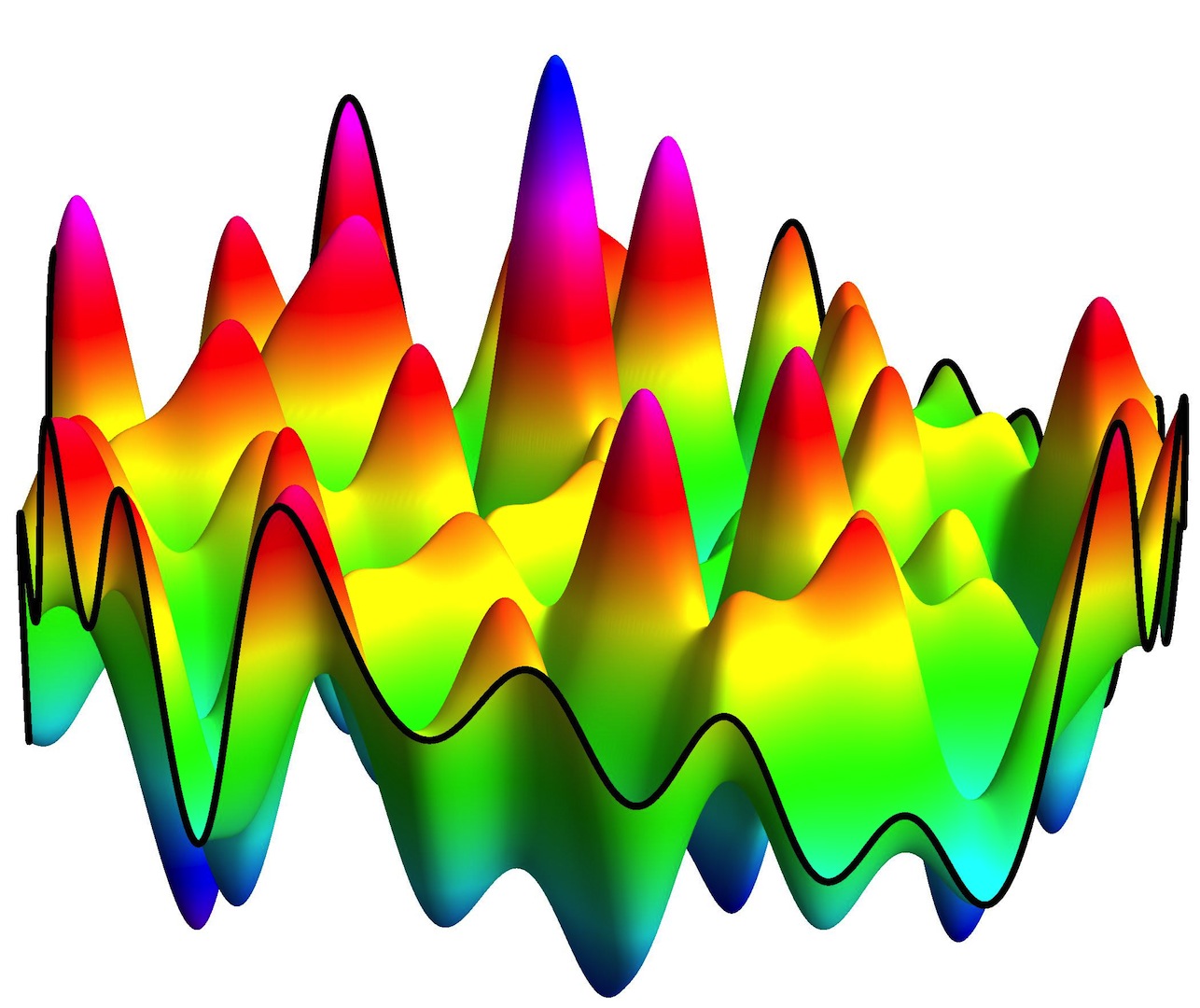}
\caption{\footnotesize Random function $F\left(z_{i}\right)$ in terms of the conformal variables $z_{i} = {\phi_{i}\over \chi}$. The black  line shows the boundary of the moduli space $|z_{i}|^{2} = 1$. This boundary moves to infinity upon switching to the canonically normalized radial variable $\vp$ in the Einstein frame. This is the main reason of the stretching of the potential in the radial direction, as shown in Figures \ref{stretch} and \ref{tmodelfig4}.}
\label{tmodelfig3}
\end{figure}

This mechanism can be easily generalized to the theory describing many fields $\phi_{i}$ \cite{Kallosh:2013daa}. Here we will only show the chaotic potential of the uniform variables $z_{i}$, Fig. \ref{tmodelfig3}, which generalizes the potential shown in the top panel of Fig. \ref{stretch}.  

\begin{figure}[ht!]
\centering{\hskip1cm \vskip0.5cm
\includegraphics[scale=0.31]{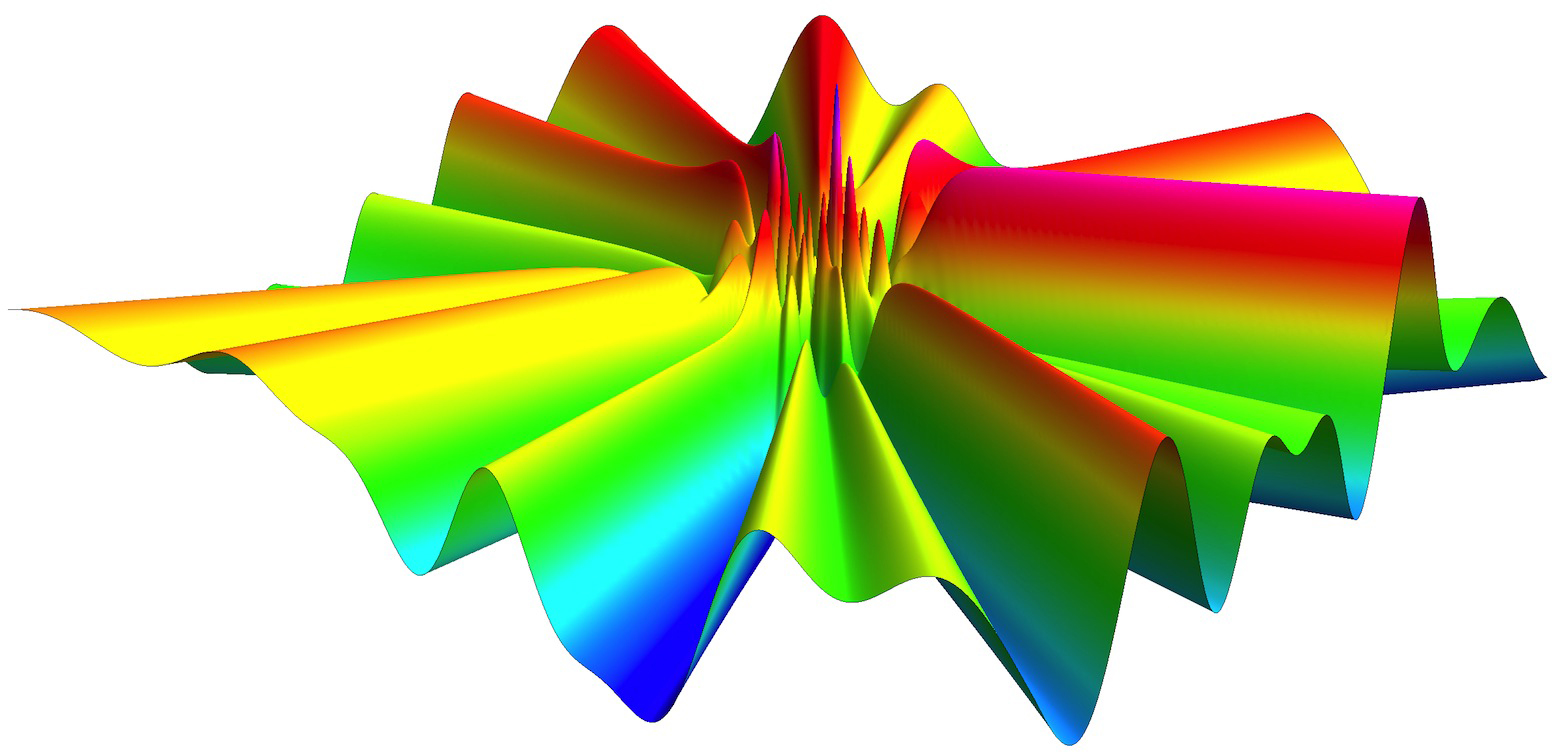}}
\caption{\footnotesize The Einstein frame potential corresponding to the random function $F\left(z_{i}\right)$ shown in Figure \ref{tmodelfig3}. Yellow and light green valleys are inflationary directions with $V> 0$, deep blue valleys correspond to $V<0$.}
\label{tmodelfig4}
\end{figure}

The potential looks very steep and disorderly, it does not contain any nicely looking flat regions, so naively one would not expect it to lead to a slow-roll inflation. Indeed, upon switching to the canonically normalized field $\vp$ in the Einstein frame, the central part of this distribution stretches by the factor of $\sqrt 6$, but it remains chaotic and mostly unsuitable for inflation. However, just as in the one-field case, the boundary of the moduli space experiences an infinitely large stretching, forming infinitely long and straight ridges and valleys. What looked like sharp Minkowski or dS minima near the boundary become the final destinations for inflationary evolution of the field rolling towards smaller values of $\vp$ along de Sitter valleys, see Figure \ref{tmodelfig4}. Meanwhile the blue valleys correspond to negative cosmological constant; we do not want to go there because the corresponding parts of the universe rapidly collapse.

What we have is a partially traversable landscape, divided into semi-infinite inflationary areas with positive potential energy, separated by the deep valleys with negative potential energy, corresponding to collapsing parts of the multiverse.
The structure of the inflationary valleys is determined by the properties of the function $F\left(z_{i}\right)$ in the close vicinity of the boundary of the moduli space. The sharper are the minima of the function $F\left(z_{i}\right)$, the more of these minima fit near the boundary, the greater is the variety of inflationary valleys we are going to obtain. If a typical size of such sharp minima is equal to $\Delta z$, the number of different inflationary valleys slowly bending towards different Minkowski or dS minima should be proportional to $(\Delta z)^{n-1}$, where $n$ is the total number of the different moduli. For $\Delta z \ll 1$ and $n\gg 1$, one can get an exponentially large variety of different possibilities, reminiscent of the string theory landscape.

It is instructive to compare this scenario with the more conventional multi-field scenario. If one assumes that from the very beginning we deal with a random Einstein frame potential without any symmetries protecting its flat directions, one may conclude that such flat directions are rather unlikely, see e.g. \cite{Marsh:2013qca}. On the other hand, the more chaotic is the function $F$ in terms of the original conformal variables $z_{i}$, and the greater number such fields we have, the greater is the variety of inflationary valleys which naturally and nearly unavoidably emerge in the scenario outlined in our paper. 

According to  \cite{Kallosh:2013daa}, most of the inflationary trajectories in this landscape correspond to the inflationary regime with the universal observational predictions $1 -n_{s} =2/N$, $r = 12/N^{2}$, just as for the single-field attractors studied in \cite{Kallosh:2013hoa}.

\subsubsection{$\alpha$-attractors}

\begin{figure}[h!t!]\centering
\includegraphics[scale=.4]{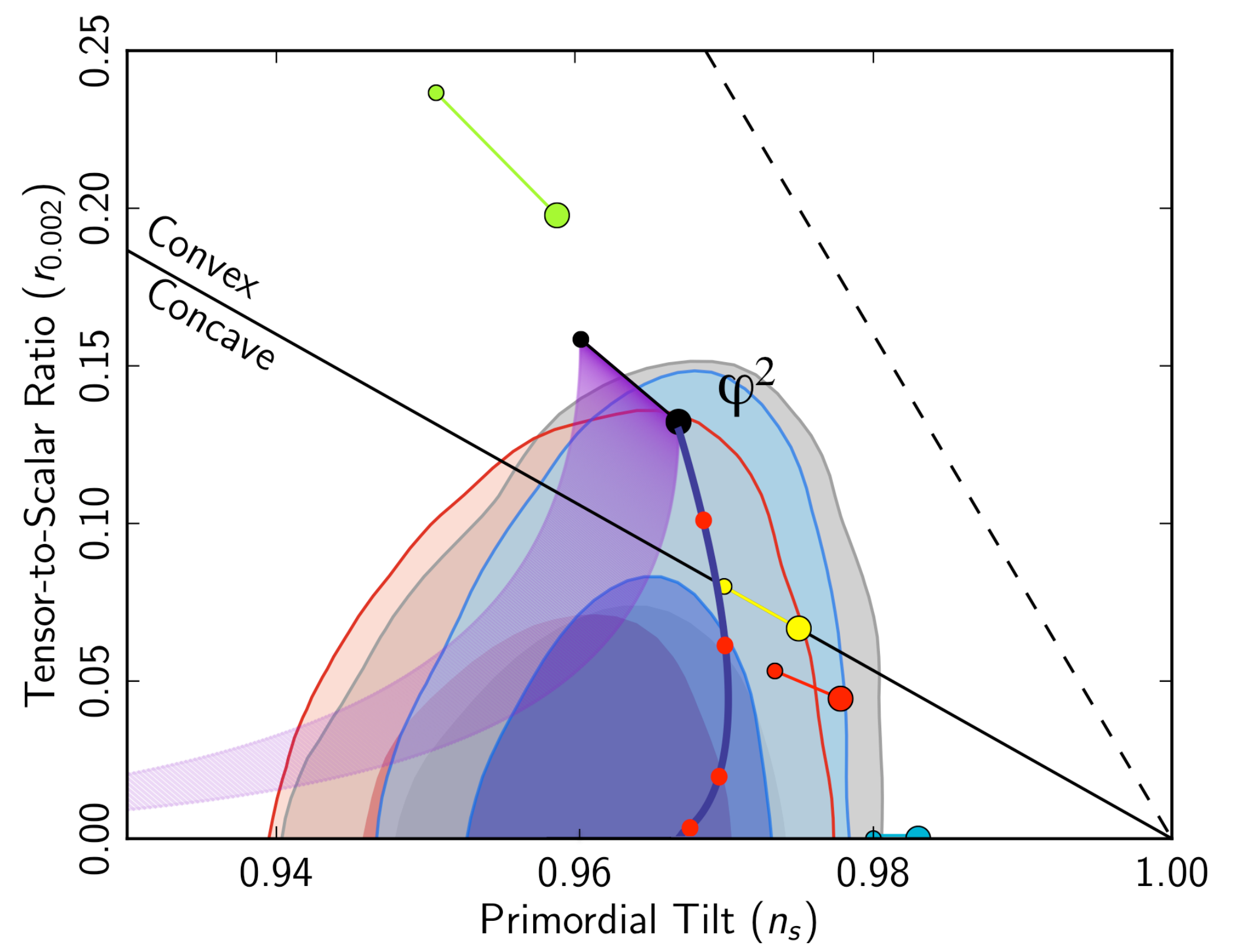} 
\caption{ \footnotesize{The cosmological observables $n_s$ and $r$ for the theory with a potential $V_{0} \bigl(1- e^{-\sqrt {2\over 3\alpha} \varphi}\bigr)^2$ for $N=60$. As shown by the thick blue line, $n_s$ and $r$ for this model depend on $\alpha$ and continuously interpolate between the prediction of the chaotic inflation with $V \sim \varphi^{2}$ for $\alpha \rightarrow \infty$, the prediction of the Starobinsky model for $\alpha = 1$ (the lowest red circle), and the prediction $n_{s} =1-\frac{2}{N}$, $r = 0$ for $\alpha \to 0$. The red dots on the thick blue line correspond to $\alpha = 10^{3},\, 10^{2},\, 10,\, 1$, from the top down.}}
\label{fig:interpol}
\vspace{-.3cm}
\end{figure}

During  investigation of supersymmetric generalizations of the Starobinsky model in \cite{Ferrara:2013rsa} we found a model with the potential
\be\label{alphanew1}
V= V_{0}\big(1- e^{-\sqrt {2\over 3\alpha} \varphi}\big)^2 \ ,
\ee
which coincides with the Starobinsky-Whitt potential \rf{whitt} for $\alpha = 1$. On the other hand, for $\varphi \ll \sqrt{\alpha}$, near the minimum of the potential \rf{alphanew1},  this potential is quadratic. The number of e-foldings in the purely quadratic potential is $N =\varphi^{2}/4$, see Eq. \rf{E04aa}. Thus, for $N \ll {\alpha/4}$, the last $N$ e-foldings of inflation in this theory are described by the simplest chaotic inflation model with a quadratic potential. In other words, for $\alpha \gtrsim 10^{3}$ all observational predictions of this model for $N \lesssim 60$ are expected to be the same as in the simplest chaotic inflation with a quadratic potential. Meanwhile for $\alpha < 1$, one has  \cite{Ferrara:2013rsa}
 \begin{align}
  n_s  = 1-\frac{2}{N}\,, \qquad 
  r  = \alpha \frac{12 }{N^2} \, .
\label{attr} \end{align}
Indeed, this is what we have found in the investigation of this model. The full set of predictions for this model is shown in Fig. \ref{fig:interpol} \cite{Kallosh:2013yoa}. We see here two limiting attractor points: In the limit $\alpha \to 0$ one has an attractor at 
 \begin{align}
  n_s  = 1-\frac{2}{N}\,, \qquad   r  = 0 
\label{attr2} \end{align}
Meanwhile in the limit $\alpha \to \infty$ the attractor point corresponds to the predictions of the simplest chaotic inflation model $m^{2}\phi^{2}/2$:
 \begin{align}
  n_s  = 1-\frac{2}{N}\,, \qquad 
  r  = \frac{8}{N} \, .
\label{attrquard} \end{align}

\ 

\begin{figure}[t!h!]
\centering
\includegraphics[scale=.4]{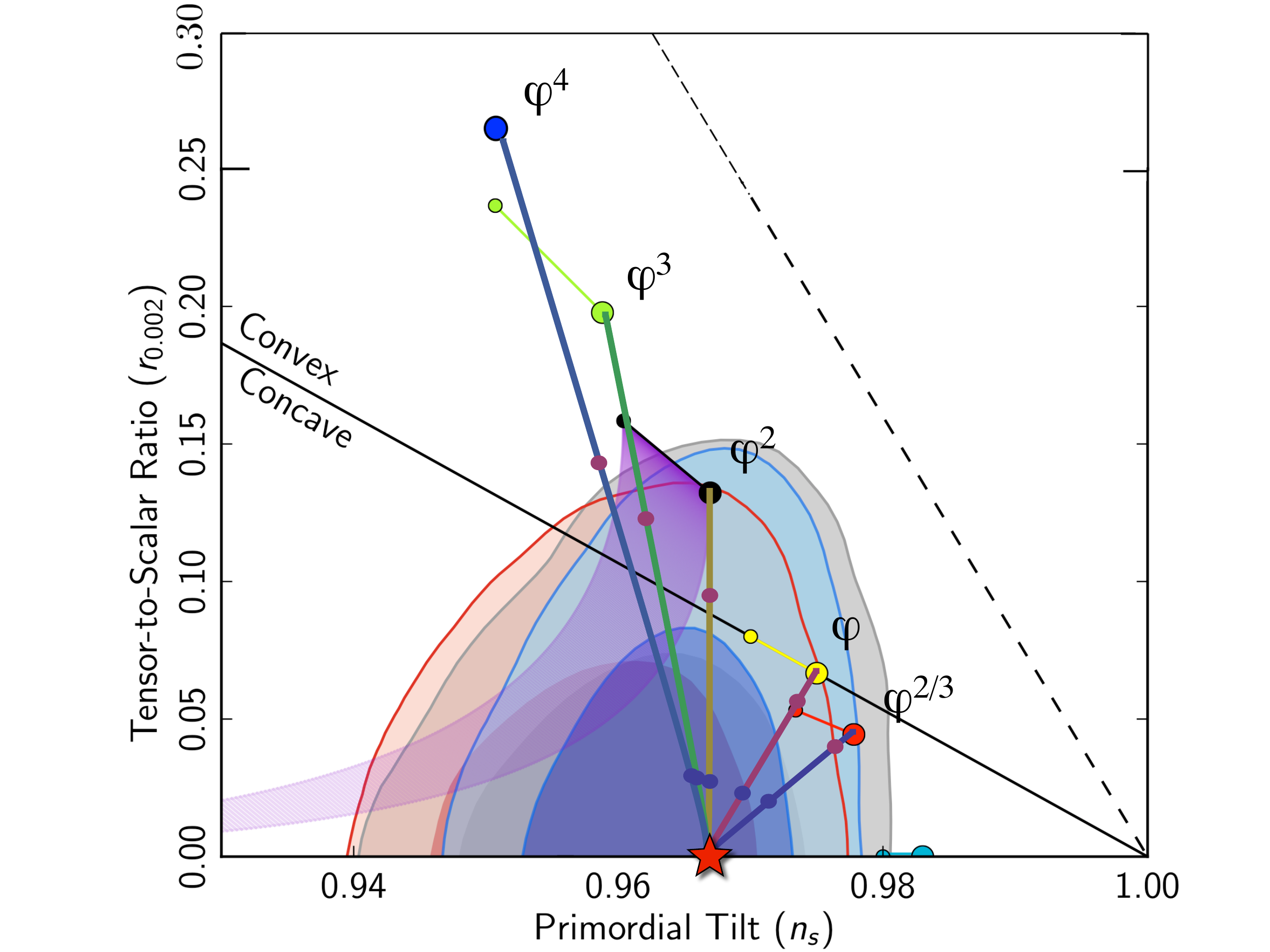}
\caption{\footnotesize{ The cosmological observables $(n_s,r)$  for different scalar potentials $\tanh^{2n} ({\vp \over \sqrt{6 \alpha}})$ with $2n = (2/3, 1, 2, 3, 4)$  for $N=60$. These continuously interpolate between the predictions of the simplest inflationary models with the monomial potentials $\varphi^{2n}$ for $\alpha \rightarrow \infty$, and the attractor point $n_{s} =1-2/N$, $r = 0$ for $\alpha \to 0$, shown by the red star. The different trajectories  form a fan-like structure for $\alpha \gg n^2$. The set of dark red dots at the upper parts of the interpolating straight lines corresponds to $\alpha = 100$. The set of dark blue dots corresponds to $\alpha = 10$. The lines gradually merge for $\alpha = O(1)$.}}
\label{fig:ClassI}
\end{figure}

Another class of models found in  \cite{Kallosh:2013yoa} generalizes superconformal attractors discussed in Section \rf{tmod}. In particular, generalized T-models have the potential
\begin{align}
  V=\tanh^{2n} (\varphi/\sqrt{6\alpha}) \,. \label{monomial-potential}
 \end{align}
Cosmological predictions in this class of theories continuously interpolate between the predictions of the simplest chaotic inflation models $\varphi^{n}$ and the 
prediction \rf{attr2}
for $\alpha \to 0$, see Fig. \ref{fig:ClassI}, which resembles Fig. \ref{attrstr2}. In all of these cases, as well as in some other models discussed in \cite{Kallosh:2013maa}, we observe attractor behavior, with an attractor point \rf{observables}, or its close neighbor \rf{attr2}.

\section{Inflation as a conformon instability} \label{confinstab}

As we already mentioned in section \ref{susyattr}, in theories with spontaneously broken superconformal invariance, such  as the standard supergravity models, one can directly go to the Einstein gauss where the Planck mass is fixed, but one may also use other gauges, and convert the results to the Einstein gauge only at the end of the calculations. The results should be gauge-independent, but some of the gauges may be more convenient for investigation. This new perspective was extremely helpful for the development of the models of superconformal attractors. But there is something else about it: Friedmann universe is conformally flat. So if we use superconformal symmetry as long as possible, it may simplify investigation of physical processes in the Friedmann universe by relating them to the processes in a static Minkowski space. A detailed discussion of related issues can be found in \cite{Kallosh:2013oma}; here I will only present some results using a toy model \rf{toy2aa} as a simple example.

This model describes gravity interacting with the conformon field $\chi$ with the Lagrangian \rf{toy2aa}: 
\begin{equation}
\mathcal{L} = \sqrt{-{g}}\left[{1\over 2}\partial_{\mu}\chi \partial_{\nu}\chi \, g^{\mu\nu}  +{ \chi^2\over 12}  R({g}) -{\lambda\over 4}\chi^4\right]\,.
\label{toy2aa11}
\end{equation}
This theory is locally conformal invariant under the 
transformations \rf{conf2aaa}: 
$\tilde g_{\mu\nu} = e^{-2\sigma(x)} g_{\mu\nu}$, 
$ \tilde\chi =  e^{\sigma(x)} \chi$.
We already discussed this model in section \ref{susyattr}, but here we would like to look at it from a different perspective, and compare its description in different conformal gauges. 

\subsection{$\chi = \sqrt 6$ conformal gauge}

In the gauge $\chi = \sqrt 6$, the kinetic term of the scalar field disappears, the term ${ \chi^2\over 12}  R({g})$ becomes the standard Einstein action, and the  term ${\lambda\over 4}\chi^4$ becomes a cosmological constant $\Lambda = {9}\lambda$, so the Lagrangian becomes  $\mathcal{L}=  \sqrt{-{g}}\,\left[{  R({g})\over 2}-{9\lambda}\right]$, see   \rf{toy2b}
This theory has a simple de Sitter solution 
with metric
\be \label{friedmann}
ds^{2} = -dt^{2} + a^{2}(t) d\vec x^{2} \ ,
\ee
where 
\be\label{dssol}
a(t) = e^{Ht}  = e^{\sqrt{\Lambda/3}\ t}  = e^{\sqrt{3\lambda}\ t} \ ,
\ee
and
\be
H = \sqrt{\Lambda/3} = \sqrt{3\lambda} \ .
\ee
One can also make a change of variables $d\eta = {dt}/a(t)$ and write the metric (\ref{friedmann}) in a conformally flat form,
\be \label{friedmann2}
ds^{2} =a^{2}(\eta)[-d\eta^{2} + d\vec x^{2}] \ .
\ee
For de Sitter space with $a(t) = e^{Ht}$ this yields
\be\label{eta22}
\eta = -H^{{-1}} e^{{-Ht}} = - {1\over \sqrt{3\lambda}}e^{-\sqrt{3\lambda}\ t} \ ,
\ee
and therefore
\be \label{friedmann22}
ds^{2} ={1\over H^{2}\eta^{2}}(-d\eta^{2} +  d\vec x^{2}) ={1\over 3\lambda\eta^{2}}(-d\eta^{2} +  d\vec x^{2}) \ .
\ee
Here we made a normalization $\eta = -H^{{-1}}$ for $t = 0$. Note that $\eta$ runs from $-\infty$ to $0$ when $t$ runs from $-\infty$ to $+\infty$.

\subsection{$a=1$ conformal gauge}

Instead of the gauge $\xi = \sqrt 6$, one may also use the gauge $a = 1$.
In this gauge,  the metric is flat in conformal time,
\be\label{confflat}
ds^{2} =-d\eta^{2} +  d\vec x^{2} \ ,
\ee
and the theory describes the scalar field $\xi$ in flat Minkowski space. The action becomes
\begin{equation}
\mathcal{L} = {1\over 2}\partial_{\mu}\chi \partial_{\nu}\chi \, \eta^{\mu\nu}   -{\lambda\over 4}\chi^4\,.
\label{toya=1}
\end{equation}
Equation of motion for the field $\chi$ in Minkowski space is
\be
 \chi'' =\lambda \chi^{3} \ .
\ee
Here $ \chi''   = {d^{2} \chi\over d \eta^{2}}$. Note that because of the ``wrong'' sign of the kinetic term of the curvaton field, its equation of motion is the same as of the normal field with a negative potential $-{\lambda\over 4} \chi^4$. Therefore the conformon field experiences an instability, falling down in its potential unbounded from below. This equation has a general solution (up to a time redefinition $\eta \to \eta -\eta_{0}$), such that $  \chi \to  +\infty$ for $\eta$ growing from $-\infty$ to $0$:
\be\label{chieq}
 \chi = -{\sqrt 2\over \sqrt \lambda \eta} \ .
\ee

\subsection{Relation between gauges} 
To compare our result (\ref{chieq})   to the results obtained  in the gauge $\chi=\sqrt 6$, one can  use the conformal transformation \rf{conf2aaa} with $e^\sigma= -\sqrt {3\lambda} \eta$:
\be\label{chiexpand}
 \chi = -{\sqrt 2\over \sqrt \lambda \eta}\, , \qquad \Rightarrow  \qquad  \chi =  {\sqrt 2\over \sqrt \lambda \eta} \,  \sqrt {3\lambda} \eta=\sqrt 6 \ .
\ee
The flat metric of the $a=1$ gauge becomes  $ e^{-2\sigma(x)} \eta_{\mu\nu}$,
\be\label{confflat3}
ds^{2} = {1\over 3\lambda \eta^{2}} (-d\eta^{2} + dx^{2}) \ ,
\ee
which coincides with \rf{friedmann22}, as it should. Finally, form this metric one can recover the usual Friedmann metric by requiring that $\eta < 0$, and therefore from $a(\eta) d\eta = {dt}$ one finds that 
\be\label{eta}
\eta = -H^{{-1}} e^{{-Ht}} = - {1\over \sqrt{3\lambda}}e^{-\sqrt{3\lambda}\ t} \ ,
\ee
which brings back the dS solution  (\ref{dssol}), (\ref{eta22}), which we earlier obtained by the standard method.

\subsection{Interpretation and consequences: Inflation as the conformon instability}

Let us say few words about interpretation of our result, which will turn out to be much more general than the simple model discussed so far. In order to do it, let us express the value of the conformon field $\chi$ in a non-expanding Minkowski space \rf{chiexpand} in terms of time $t$ in the Friedmann universe:
\be\label{expconformon}
\chi = -{\sqrt 2\over \sqrt \lambda \eta} = {\sqrt 6}\, e^{{H\, t}} = {\sqrt 6} M_{p}\, e^{{H\, t}} \ .
\ee
Note that since the theory is locally conformally invariant, one can always ``freeze'' the evolution of the conformon field at any moment $t$, and allow the scale factor to evolve starting from this moment, by making a proper conformal transformation, or choosing an appropriate gauge.  The corresponding wavelength, corresponding to the effective Planck length, decreases as $e^{{-H\, t}}$. Thus, Minkowski space does seem exponentially expanding if its size is measured in units of the exponentially contracting Planck length. This is a general result, which is applicable to any kind of uniform cosmological evolution, including inflation. In this context, exponential growth of space during inflation \rf{dssol} is directly related (equivalent) to the exponential growth of the conformon field in Minkowski space (\ref{expconformon}).

In order to understand this general result, which is going to be valid for all models studied in this paper, it is sufficient to look at the equation (\ref{conf2aaa}). In the standard investigation of the cosmological evolution, one goes to what can be called the Einstein frame gauge, fixes the conformon $\chi = \sqrt 6$ (or, more generally, the Planck mass), and investigates the evolution of the scale factor $a$, as measured in the Planck length units.  However, one can equally well work in the gauge where the scale factor is fixed. The transition from one gauge to another is achieved by conformal transformation (\ref{conf2aaa}), which absorbs expansion of the universe in terms of its scale factor $a(t)$ and converts it into the exactly equal time-dependent factor describing the growth of the conformon field.

In application to inflation, this means that one can equally well describe it as the exponentially fast expansion of the scale factor, or as the equally fast growth of the conformon field, obeying {\it the same Einstein equations as the scale factor, up to a trivial rescaling}. Alternatively, one can work in the original conformally invariant setting, without fixing the gauge, and study evolution of all fields while preserving the original conformal invariance and enjoying simplifications provided by conformal flatness of the Friedmann universe. Then in the end of the calculations one can re-formulate all results in terms of the Einstein frame gauge where the Planck mass is fixed.

In this section, we discussed the simplest applications of the hidden (super)conformal invariance of the theory in the context of a rather trivial model. However, the results discussed above are quite general, they are valid for all theories based on supergravity, which can be most conveniently formulated using the superconformal approach. This suggests that the superconformal invariance can be much more than just a useful tool for deriving the standard supergravity. The concept of the hidden superconformal symmetry was crucially important for finding the supersymmetric generalization of the scalar theories with nonminimal coupling to gravity, for the development of the cosmological attractor models, and for the re-interpretation of the cosmological evolution discussed in this section.  For a more detailed discussion of the superconformal approach to the cosmological evolution see \cite{Kallosh:2013oma,RenataLecture}.

\section{Towards Inflation in String Theory}

In the previous sections we discussed models of inflation which can be implemented in superconformal theory and supergravity. Implementation of the ideas discussed above in string theory requires several additional steps. Related ideas have been discussed in the lectures of Silverstein \cite{Silverstein:2013wua}  and also in the recent review by Burgess, Cicoli and Quevedo \cite{Burgess:2013sla}. Here we will  make some additional comments on this issue. They will be related to vacuum stabilization and its  phenomenological and cosmological implications, and also to the string landscape scenario. 

\subsection{de Sitter vacua in string theory}

For a long time, it seemed rather difficult to obtain inflation in
M/string theory, though some intriguing suggestions related to brane inflation have been made \cite{Dvali:1998pa}.
The main problem was the stability of
compactification of internal dimensions. For example, ignoring
non-perturbative effects to be discussed below, a typical
effective potential of the effective 4d theory obtained by
compactification in string theory of type IIB can be represented
in the following form:
\begin{equation}
V(\varphi,\rho,\phi) \sim e^{\sqrt 2\varphi -\sqrt6\rho}\ \tilde
V(\phi)
\end{equation}
Here $\varphi$ and $\rho$ are canonically normalized fields
representing the dilaton field and the volume of the compactified
space; $\phi$ stays for all other fields, including the inflaton field.

If $\varphi$ and $\rho$ were constant, then the potential $\tilde
V(\phi)$ could drive inflation.  However, this does not happen
because of the steep exponent $e^{\sqrt 2\varphi -\sqrt6\rho}$,
which rapidly pushes the dilaton field $\varphi$ to $-\infty$, and
the volume modulus $\rho$ to $+\infty$. As a result, the radius of
compactification becomes infinite; instead of inflating, 4d space
decompactifies and becomes 10d.

Thus in order to describe phenomenological consequences of string theory one should first learn how to
stabilize the dilaton, and the volume modulus. The dilaton
stabilization was achieved in \cite{GKP}. The most difficult
problem was to stabilize the volume. Here we will briefly describe a possible solution of this problem
found in \cite{Kachru:2003aw} (the KKLT construction). It consists of two
steps.

\begin{figure}[h!]
\centering\leavevmode\epsfysize=5.5cm \epsfbox{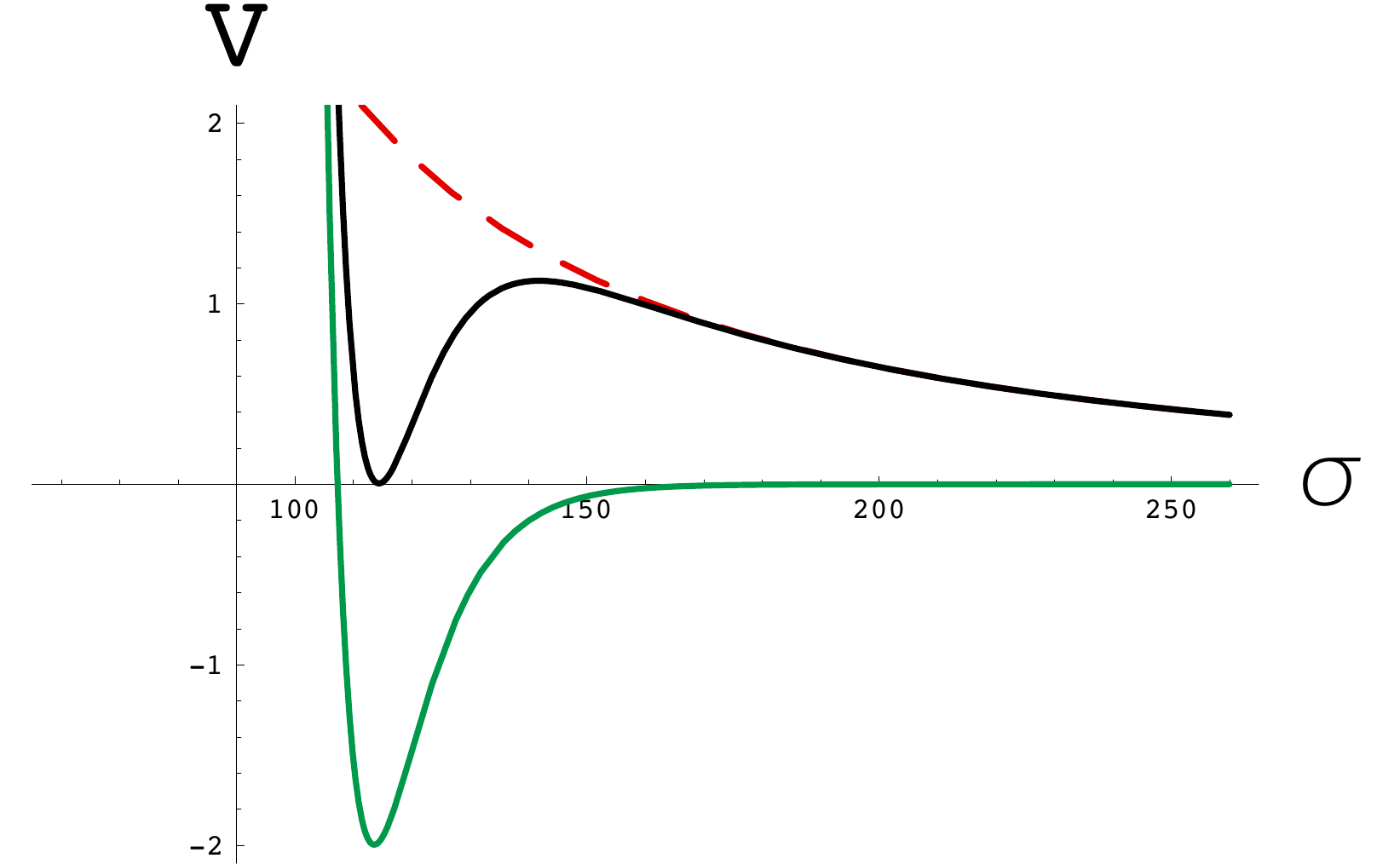} \caption[fig1]
{\footnotesize KKLT potential as a function of $\sigma = {\rm Re}\,\rho$. Thin green line corresponds to AdS stabilized potential for $W_0 =- 10^{{-4}}$,
$A=1$, $a =0.1$. Dashed line shows the additional term,
which appears either due to the contribution of a $\overline{D3}$ brane or of a
D7 brane. Thick black line shows the resulting potential  with a very small but positive value of $V$ in the minimum. The potential is shown multiplied by $10^{15}$.} \label{1str}
\end{figure}

First of all, due to a combination of effects related to warped
geometry of the compactified space and nonperturbative effects
calculated directly in 4d (instead of being obtained by
compactification), it was possible to obtain a supersymmetric AdS
minimum of the effective potential for $\rho$. In the original version of the KKLT scenario, it was done in the theory with the  K\"ahler potential
\begin{equation}
K = -3\log (\rho+\bar\rho)
, \end{equation} and with the nonperturbative superpotential of the form \begin{equation}\label{KKLTsp}
W=W_0+
Ae^{-a\rho}, 
\end{equation} 
with $a = 2\pi/N$.
The corresponding effective potential for the complex field $\rho =
\sigma +i\alpha$ had a minimum at finite, moderately large values of the volume modulus field $\sigma_{0}$, which  fixed the
volume modulus  in a state with a negative vacuum energy. Then
 an anti-${D3}$ brane with the positive energy $\sim
\sigma^{-2}$ was added. This addition uplifted the minimum of the potential to
the state with a positive vacuum energy, see Fig. \ref{1str}.

Instead of adding an anti-${D3}$ brane, which explicitly breaks
supersymmetry, one can add  a D7 brane with fluxes. This results
in the appearance of a D-term which has a similar dependence on
$\rho$, but leads to spontaneous supersymmetry breaking
\cite{Burgess:2003ic}. In these lectures we will discuss yet another way, related to F-term uplifting, following recent papers  \cite{Kallosh:2011qk,Linde:2011ja,Dudas:2012wi}.   In either case, one ends up with a
metastable dS state which can decay by tunneling and formation of
bubbles of 10d space with vanishing vacuum energy density. The
decay rate is extremely small \cite{Kachru:2003aw}, so for all practical
purposes, one obtains an exponentially expanding de Sitter space
with the stabilized volume of the internal space.

\subsection{Inflation, vacuum stabilization and the scale of SUSY breaking in string theory}

\begin{figure}[h!]
\centering\leavevmode\epsfysize=5.4cm \epsfbox{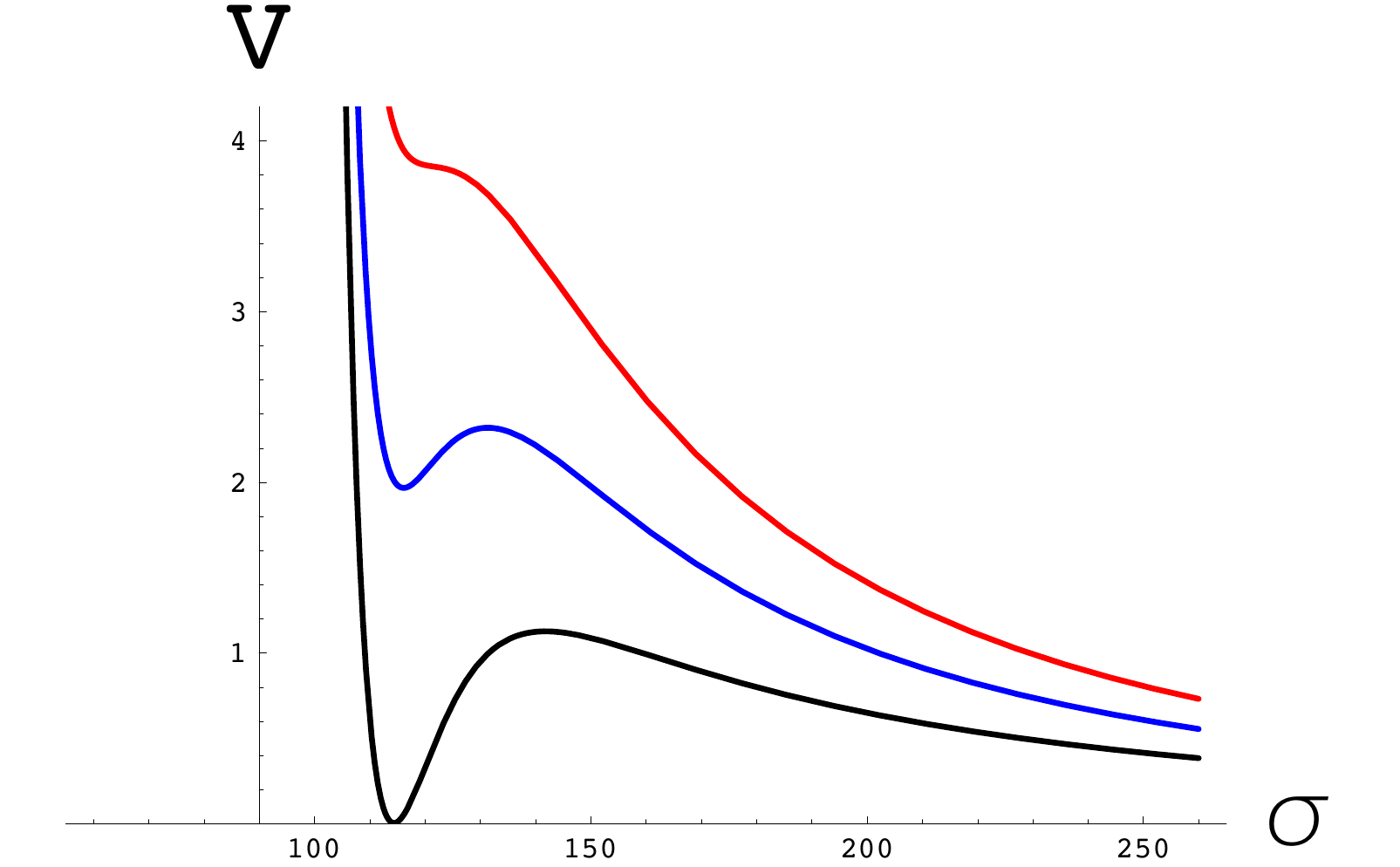} \caption[fig2]
{\footnotesize The lowest curve with dS minimum is the one from the KKLT model. The height of the barrier in this potential is of the order $m_{3/2}^{2}$.  The second line shows the $\sigma$-dependence of the inflaton potential. When one adds it to the theory, it always appears divided by $\sigma^{n}$, where in the simplest cases $n = 2$  or 3.  Therefore an addition of the inflationary potential lifts up the potential at small $\sigma$. The top curve shows that when the  inflation potential becomes too large, the barrier disappears, and the internal space decompactifies. This explains the origin of the constraint $H\lesssim   m_{3/2}$.  } \label{2}
\end{figure}

Once we learned how to stabilize extra dimensions, it became possible to study various inflationary models based on string theory. A detailed discussion of various models of inflation in string theory can be found in the lectures of Silverstein \cite{Silverstein:2013wua}  and also in the recent review by Burgess, Cicoli and Quevedo \cite{Burgess:2013sla}.
Here we will limit ourselves to a discussion of a rather peculiar relation between inflation in string theory, particle phenomenology, and  the possibility to observe the tensor modes in the CMB. This relation is rather unexpected and may impose strong constraints on particle phenomenology and on inflationary models:  in the simplest models based on the KKLT mechanism the Hubble constant $H$ and the inflaton mass $m_{\phi}$   are  smaller than the gravitino mass  \cite{Kallosh:2004yh},
\begin{equation}
 m_{\phi} \ll H \lesssim m_{{3/2}} \ .
\end{equation}
The reason for the constraint $H \lesssim m_{{3/2}}$ is that the height of the barrier stabilizing the KKLT minimum is $O(m_{3/2}^{2})$. Adding a large vacuum energy density  to the KKLT potential, which is required for inflation, may destabilize it, see Fig. \ref{2}. The constraint $ m_{\phi} \ll H$ is a consequence of the slow-roll conditions.

Therefore if one believes in the standard SUSY phenomenology with $m_{{3/2}} \lesssim O(1)$ TeV, one should find a realistic particle physics model where  inflation occurs  at a density at least 30 orders of magnitude below the Planck energy density. Such models are possible, but their parameters should be substantially different from the parameters used in all presently existing models of string theory inflation.

An interesting observational consequence of this result is that the amplitude of the gravitational waves in all string inflation models of this type should be extremely small. Indeed, according to Eq. (\ref{eq:rvh}), one has 
$ {r} \approx 3\times  10^{7}~V  \approx 10^{8}~ H^{2}$, which implies that 
\begin{equation}\label{bound}
r \lesssim 10^{8}~m_{{3/2}}^{2}  \ ,
\end{equation}
in Planck units. In particular, for $m_{{3/2}} \lesssim 1$ TeV $\sim 4 \times 10^{-16}~ M_{p}$, which is in the range most often discussed by SUSY phenomenology, one has \cite{Kallosh:2007wm}
\begin{equation}
r \lesssim 10^{-24} \ .
\end{equation}
If CMB experiments find that $r \gtrsim 10^{-2}$, then this will imply, in the class of theories described above, that 
\begin{equation}
m_{{3/2}} \gtrsim 10^{-5}~ M_{p} \sim 2.4 \times 10^{13}~{\rm GeV}  \ ,
\end{equation}
which is 10 orders of magnitude greater than the standard gravitino mass range discussed by particle phenomenologists.

These constraints appear in the original KKLT approach to the moduli stabilization, but they appear in other approaches as well. For example, in the string inflation scenario based on large volume stabilization \cite{Burgess:2013sla}, the constraints on the Hubble constant are even much stronger,
\be
H \lesssim m_{3/2}^{3/2} \ ,
\ee
see \cite{Conlon:2008cj}.

There are several different ways to address this problem. Here we will discuss the simplest solution proposed in  \cite{Kallosh:2004yh}. 
The basic idea is to use a slightly more complicated superpotential for the volume modulus, the so-called racetrack potential
\begin{equation}\label{race}
W=W_0+
Ae^{-a\rho} + B e^{-b\rho}.
 \end{equation}


\begin{figure}[h!]
\centering\leavevmode\epsfysize=5cm \epsfbox{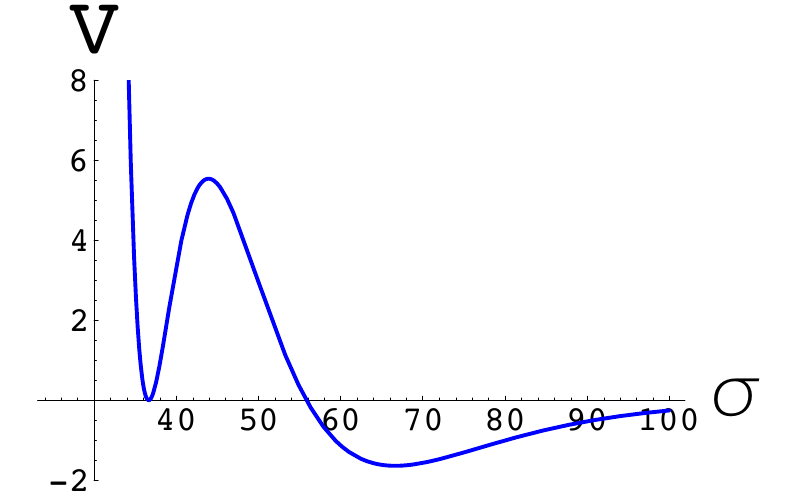} \caption[fig2]
{The potential in the theory (\ref{race}) for $A=1,\ B=-5,\ a=2\pi/100,\ b=2\pi/50,\ W_0= -0.05$. A Minkowski minimum at $V=0$ stabilizes the volume at $\sigma_{0}\approx 37$.  The height of the barrier in this model is not correlated with the gravitino mass, which   vanishes  if the system is trapped in Minkowski vacuum. Therefore in this model one can avoid the constraint $H \lesssim m_{{3/2}}$   \cite{Kallosh:2004yh}. } \label{3a}
\end{figure}

For the particular choice of $W_0$,
\be\label{w0}
W_0= -A \left({a\,A\over
b\,B}\right)^{a\over b-a} +B \left ({a\,A\over b\,B}\right) ^{b\over b-a} ,
\ee
the potential of the field $\sigma$ has a supersymmetric minimum with $W_{\rm KL} (\sigma_{0})=0$,  $D_\rho W_{\rm KL}(\sigma_{0}) = 0$,  $V(\sigma_{0})=0$, and $m_{3/2} = 0$, see Fig. \ref{3a}. Then, by adding a small quantity $\Delta W$ to $W_{0}$, such that $|\Delta W|\ll |W_{0}|$, one finds a very shallow AdS minimum without changing the overall shape of the potential. This minimum can be uplifted back to Minkowski or dS vacuum with a tiny cosmological constant by adding the standard Polonyi field to the model, as described in \cite{Dudas:2012wi}.  In this case, one finds $m_{3/2} = |\Delta W| (2\sigma_{0})^{{-3/2}}$. 

Thus by dialing $|\Delta W|$ one can have an arbitrarily small gravitino mass without changing the arbitrarily large height of the barrier. This stabilizes the vacuum and eliminates the constraint $H \lesssim m_{{3/2}}$. This version of the KKLT construction is often called the KL model  \cite{Kallosh:2004yh}. The only fine-tuning required there is the tuning of $ |\Delta W|$, which is necessary if one wants to make the gravitino mass $m_{3/2}$ small. In the original version of the KKLT model, the same kind of fine-tuning was imposed on $W_{0}$. The simpler Polonyi model of the supers symmetry breaking required a similarly small parameter, for the same purpose of making the gravitino mass small. Thus the degree of fine-tuning of parameters of the KL model is exactly the same as in the original KKLT model and in many other phenomenological models requiring smallness of the gravitino mass \cite{Kallosh:2011qk,Linde:2011ja,Dudas:2012wi}.

The use of the KL stabilization allows to construct string inflation models with sufficiently large values of $r$, such as the monodromy inflation models developed in \cite{Silverstein:2008sg}.

\section{Inflationary multiverse,  string theory landscape and the anthropic principle}\label{land}

For many decades people have tried to explain strange correlations
between the properties of our universe, the masses of elementary
particles, their coupling constants, and the fact of our existence.
We know that we could not live in a 5-dimensional universe, or in a
universe where the electromagnetic coupling constant, or the masses
of  electrons and protons would be just a few times greater or
smaller than their present values. These and other similar
observations have formed the basis for the anthropic principle.
However, for a long time many scientists  believed that the universe
was given to us as a single copy, and therefore speculations
about these magic coincidences could not have any scientific
meaning. Moreover, it would require a wild stretch of imagination and a certain degree of arrogance to assume that somebody was creating one universe after another, changing their parameters and fine-tuning their design, doing all of that for the sole purpose of making the universe suitable for our existence.

The situation changed dramatically with the invention of
inflationary cosmology. It was realized that inflation may divide
our universe into many exponentially large domains corresponding to
different metastable vacuum states, forming a huge inflationary
multiverse \cite{Linde:1982ur,nuff,Eternal}. 
The total number of
such vacuum states in string theory can be enormously large, in the range of $10^{100}$ or $10^{1000}$; the often cited number is $10^{500}$
\cite{Lerche:1986cx,Bousso:2000xa,Kachru:2003aw,Douglas}. A combination of these
two facts  gave rise to what the experts in inflation call `the inflationary multiverse,' \cite{Linde:2005ht,Linde:1993xx,Linde:2002gj} and string theorists call  
`the string theory landscape'  \cite{Susskind:2003kw}. 

This leads to an interesting twist in the theory of initial conditions. Suppose that we live in one of the many metastable de Sitter minima, say, $dS_{i}$. Eventually this dS state decays, and each of the {\it points} belonging to this initial state  jumps to another vacuum state, which may have either a smaller vacuum energy, or a greater vacuum energy (transitions of the second type are possible because of the gravitational effects). But if the decay probability   is not too large, then the total {\it volume} of the universe remaining in the  state $dS_{i}$  continues growing exponentially \cite{Guth:1982pn}. This is eternal inflation of the old inflation type. If the bubbles of the new phase correspond to another de Sitter space, $dS_{j}$, then some parts of the space $dS_{j}$ may jump back to the   state  $dS_{i}$. On the other hand, if the tunneling goes to a Minkowski vacuum, such as the non-compactified 10d vacuum corresponding to the state with $\sigma \to \infty$ in   Fig. \ref{1str}, the subsequent jumps to dS states no longer occur. Similarly, if the tunneling goes to the state with a negative vacuum energy, such as the AdS vacuum  in Fig. \ref{3a}, the interior of the bubble of the new vacuum rapidly collapses. Minkowski and AdS vacua of such type are called terminal vacua, or sinks.

If initial conditions in a certain part of the universe are such that it goes directly go to the sink, without an intermediate stage of inflation, then it will never return back, we will be unable to live there, so for all practical purposes such initial conditions (or such parts of the universe) can be discarded (ignoring for a moment the possibility of the resurrection of the universe after the collapse, to be discussed in the Appendix). On the other hand, if some other part of the universe goes to one of the dS states, the process of eternal inflation begins, which eventually produces an inflationary multiverse consisting of all possible dS states. This suggests that all initial conditions that allow life as we know it to exist, inevitably lead to formation of an eternally inflating multiverse.

The string theory landscape describes an incredibly large number of {\it discrete} parameters, which is often estimated by ${\cal M} \sim 10^{500}$ \cite{Douglas}. These parameters correspond to different string theory vacua.
However, the theory of inflationary multiverse goes even further. Some of the features of our world are determined not by the final values of the fields in the minima of their potential in the landscape, but by the dynamical, time-dependent values, which these fields were taking at different stages of the evolution of the inflationary universe. This introduces a large set of {\it continuous} parameters, which may take different values in different parts of the universe. For example, in the theory of dark energy, inflationary fluctuations may divide the universe into exponentially large parts with the effective value of the cosmological constant  taking a continuous range of values \cite{300}. In such models, the effective cosmological constant $\Lambda$ becomes a continuous parameter.  Similarly, inflationary fluctuations of the axion field make  the density of  dark matter  a continuous parameter, which takes  different values in different parts of the universe \cite{Linde:1987b,Rees}. Another example of a continuous parameter is the baryon asymmetry $n_{b}/n_{\gamma}$, which can  take different values in different parts of the universe in the Affleck-Dine scenario of baryogenesis \cite{Affleck:1984fy,Linde:gh,Bousso:2013rda}. Yet another important example is the large-scale structure of the universe, which emerges due to quantum fluctuations which lead to different galaxy distributions in different parts of the world.  As a result, the large-scale structure and the matter content in each of the locally Friedmann parts of the cosmic fractal may be quite different, which leads to an additional diversity of the observable outcomes for any particular vacuum in the landscape. One may wonder how many  different, observationally distinguishable locally Friedmann universes one may encounter in any particular part of the landscape  \cite{Linde:2009ah}. 

The meaning of these words can be explained as follows. Slow-roll inflation produces long-wavelength perturbations of the metric, which become imprinted on the cosmological background and determine the matter content and the large scale structure of the universe. Even though these perturbations are created from quantum fluctuations, they become essentially classical due to inflation. These perturbations provide different classical initial conditions in different parts of the universe. Since these classical initial conditions may change continuously, one could expect that this can lead to infinite number of different outcomes, but if these outcomes are too close to each other, one cannot distinguish between them due to quantum mechanical uncertainty. the number of distinctly different classical geometries which may appear as a result of this effect. According to  \cite{Linde:2009ah}, the number of distinctly different classical geometries which may appear as a result of inflationary perturbations is proportional to $e^{e^{3N}}$, where $N$ is the number of e-foldings of slow-roll inflation. 

Not all of these variations can be locally observed. The main bound on various potentially observable geometries follows from the fact that one can make any observations only inside the dS horizon. The size of the horizon may take different values in different parts of the inflationary multiverse, depending on the local value of the cosmological constant there. Combining all of these considerations suggests that the total number of locally distinguishable configurations in string theory landscape can be as large as $e^{{\cal M}^{3/4}}$, where ${\cal M}$ is the total number of vacua in string theory  \cite{Linde:2009ah}. In other words, the total number of locally distinguishable geometries is expected to be exponentially greater than the total number of vacua in the landscape. In particular, using the popular estimate ${\cal M} \sim 10^{500}$ \cite{Douglas}, one finds the total number of the potentially observable outcomes can be as large as $10^{10^{375}}$. 

This means that the same physical theory may yield an incredibly large number of exponentially large  parts of the universe that have diverse properties.  This generalizes and strengthens the statement \cite{Linde:1982ur,nuff} that inflation provides a scientific justification of the anthropic principle: we find  ourselves inside a  part of the universe with our kind of physical laws, matter abundance and the large-scale structure not because the parts with dramatically different properties are impossible or improbable, but simply because we cannot live there. 
 
This fact can help us understand many otherwise mysterious features of our world. The simplest example is the dimensionality of our universe. String theorists usually assume  that the universe is ten dimensional, so why do we live in the universe where only 4  dimensions of space-time are large? There were many attempts to address this question, but no convincing answer was found. This question became even more urgent after the development of the KKLT construction. Now we know that all de Sitter states, including the state in which we live now, are either unstable or   metastable. They tend to decay by producing bubbles of a collapsing space, or of a 10-dimensional Minkowski space. So what is wrong about the 10-dimensional universe if it is so naturally appears in string theory?

The answer to this question was given in 1917 by Paul Ehrenfest  \cite{ehr}: In space-time with dimensionality $d > 4$, gravitational forces between distant
bodies fall off faster than $r^{-2}$, and in space-time with $d<4$, the general theory of relativity tells us that such forces are absent altogether.
This rules out the existence of stable planetary systems for $d\not = 4$. 
A similar conclusion is valid for atoms:  stable atomic systems could not exist for $d> 4$. This means that we do not need to prove that the 4d space-time is a {\it necessary} outcome of string cosmology  (in fact, it does not seem to be the case). Instead of that, we only need to make sure that the 4d space-time is {\it possible}.

Anthropic considerations may help us to understand why the electron mass is 2000 times smaller than the proton mass, and why the proton mass is almost exactly equal to the neutron mass. They may even help us to understand  why the amount of dark matter is approximately 5 times greater than the amount of normal matter \cite{Linde:1987b,Rees} and why the baryon asymmetry is so small, $n_{b}/n_{\gamma} \sim 10^{{-10}}$  \cite{Linde:gh}. But perhaps the most famous example of this type is related to the cosmological constant problem. 

Naively, one could expect vacuum energy to be equal to the Planck density,  $\rho_{\Lambda} \sim 1$, whereas the recent observational data show that $\rho_{\Lambda} \sim 10^{-120}$, in Planck units, which is approximately 3 times greater than the density of matter in the universe. Why is it so small but nonzero? Why $\rho_{\Lambda}$  is about 3 times greater than the density of matter in the universe now, even though at the Planck time, the density of usual matter was $10^{120}$ times greater than $\rho_{\Lambda}$, and in the future it will be much smaller. What is so special about $\rho_{\Lambda} \sim 10^{-120}$? What is so special about the present time?

The first anthropic solution to the cosmological constant problem in the context of inflationary cosmology was proposed  in 1984
\cite{Linde:1984ir}. The basic assumption was that the vacuum energy density is a sum of the scalar field potential $V(\phi)$ and the energy of fluxes $V(F)$. According to  \cite{Linde:1983mx}, quantum creation of the universe is not suppressed if the universe is created at the Planck energy density, $V(\phi) +V(F) = O(1)$, in Planck units. Eventually the field $\phi$ rolls to its minimum at some value $\phi_0$, and the vacuum energy becomes $\Lambda = V(\phi_0) + V(F)$. Since initially $V(\phi)$ and $V(F)$  could take any values with nearly equal probability, under the condition $V(\phi) +V(F) = O(1)$, we get a flat probability distribution to find a universe with a given value of the cosmological constant after inflation, $\Lambda = V(\phi_0) + V(F)$, for $\Lambda \ll 1$. The flatness of this probability distribution is crucial, because it allows us to study the probability of emergence of life for different $\Lambda$. Finally, it was  argued in   \cite{Linde:1984ir}  that life as we know it is possible only for  $|\Lambda|  \lesssim \rho_{0}$, where $\rho_{0} \sim 10^{{-120}}$ is the present energy density of the universe.  This fact, in combination with inflation, which makes such universes exponentially large, provided a possible solution of the cosmological constant problem. 

The second mechanism was proposed by Sakharov \cite{Sakharov}. He mentioned, following \cite{nuff}, that   the universe may consist of many different parts with different types of compactification, and then argued that the number of different types of compactifications may be exponentially large. He emphasized that if this number is large enough, the typical energy gap between different levels can be extremely small, which will allow to explain the smallness of the cosmological constant by using anthropic considerations.

The third mechanism was proposed in \cite{300} (see also \cite{Banks}). It was based on a combination of eternal inflation driven by the inflaton field $\phi$ and a subsequent slow roll of what was later called `quintessence' field $\Phi$. The role of eternal inflation was to generate perturbations of the field $\Phi$, which then give this field different values in different exponentially large parts of the universe. As a result, the universe becomes divided into different parts with a flat probability distribution for different values of the effective cosmological constant. Once again, this provided a possibility to use anthropic considerations for solving the cosmological constant problem.

All of these proposals were based on the assumption that life as we know it is possible only for  $-\rho_{0} \lesssim \rho_{\Lambda} \lesssim \rho_{0}$. This bound seemed almost self-evident to many of us at that time, and therefore we concentrated on the development of the theoretical framework where the anthropic arguments could be applied  to the cosmological constant. However, it is important to understand how strong are the anthropic bounds discussed in  \cite{Linde:1984ir,Sakharov,300,Banks}.

The fact that $\rho_{\Lambda}$ could not be much smaller than $-\rho_{0}$ is indeed quite obvious, since such a universe would rapidly collapse \cite{DaviesUnwin,BarrowTipler,Linde:1984ir}. However, the origin of the constraint $\rho_{\Lambda} \lesssim \rho_{0}$ is less trivial. The first detailed derivation of this bound was made in 1987 in the famous paper by Weinberg \cite{Weinberg:1987dv}, but the constraint obtained there allowed the cosmological constant to be three orders of magnitude greater than its present value.

Since that time, the anthropic approach to the cosmological constant problem developed in two different directions. First of all, it became possible, under certain assumptions, to significantly strengthen the constraint on the positive cosmological constant, see e.g. \cite{Weinberg1998,Garriga:1999hu,Tegmark:2005dy,Bousso:2007kq}.  The final result of these investigations, $|\Lambda|  \lesssim O(10)\ \rho_{0} \sim 10^{-119}$, is very similar to the bound used in \cite{Linde:1984ir}.

Simultaneously, new models have been developed which may allow us to put the anthropic approach to the cosmological constant problem on a firm ground. In particular, the existence of a huge number of vacuum states in string theory  implies that in different parts of our universe, or in its different quantum states, the cosmological constant may take all of its possible values, from $-1$ to $+1$, with an increment which may be as small as $10^{-500}$. If  the prior probability to be in each of these vacua does not depend strongly on $\Lambda$, one can justify the anthropic bound on $\Lambda$ using the methods of \cite{Weinberg1998,Garriga:1999hu,Tegmark:2005dy,Bousso:2007kq}.

One should note, that the issue of probabilities in the inflationary multiverse is  very delicate, so one should approach numerical calculations related to anthropic arguments with some care. Let us explain the nature of this problem, and simultaneously separate it from the eternal inflation scenario. Consider the standard textbook theory of an infinite open or flat Friedmann universe, or, alternatively, a cyclic universe which is supposed to pass an infinite number of cycles of evolution, growing larger and larger with each new cycle (see a critical discussion of this scenario in the Appendix). The total volume of space-time in each of these cases is {\it infinite}. This means that in any of such universes every improbable event must happen infinitely many times. For example, in the cyclic universe, if the scenario actually works, there are infinitely many places where broken glass magically becomes unbroken. Does it mean that in an infinite universe one cannot make a statement that a broken glass typically remains broken?

This is called the measure problem. It was first recognized back in 1993 in \cite{Linde:1993xx}, where we proposed two different ways of describing an eternally inflating universe. It was found in  \cite{Linde:1993xx} that if we start with some small part of an eternally inflating universe, wait until it becomes divided into different parts with different properties and compare volumes of parts of each type, then the ratios of these volumes will gradually approach a stationary, time-independent limit. It would be very tempting to use these ratios for the calculation of the probabilities in an inflationary multiverse. However, as emphasized in \cite{Linde:1993xx}, this procedure is ambiguous because in general relativity one can use different space-time slicing. It was found in \cite{Linde:1993xx} that the results of the calculations do depend on this choice, so it was not clear which one of these procedures, if any, should be used in calculations of the probability distributions.

Later it was found that one of these two probability measures introduced in \cite{Linde:1993xx} leads to paradoxical conclusions, which are associated with the so-called ``youngness problem''  \cite{Tegmark:2004qd,Guth:2007ng}. The source of the paradox was identified  in \cite{Linde:2007nm}: In the implementation of the slicing of the multiverse in \cite{Linde:1993xx}, we did not take into account that the stationary regime was achieved at different times for different processes, so it was incorrect to use the equal time cut-off for comparison of different parts of the multiverse. A properly modified version of this volume-weighted measure does not lead to the youngness problem \cite{Linde:2007nm,Linde:2008xf}. The second of the two measures proposed in \cite{Linde:1993xx}, the scale factor cutoff measure, also does not suffer from this problem and remains quite popular \cite{DeSimone:2008if}. During the last few years, many other, more sophisticated probability measures have been introduced; for a recent discussion see e.g. \cite{Guth:2013sya} and references therein. We cannot say yet which one of these measures is the best, but none of the measures considered now lead to the paradoxes discussed in  \cite{Tegmark:2004qd,Guth:2007ng}.

Different approaches to the theory of inflationary multiverse lead to slightly different constraints on the possible values of the cosmological constant, but these differences do not change the main qualitative conclusion: Many parts of inflationary multiverse are expected to be in a state with a very large value of the cosmological constant, and we cannot live there. But there are many exponentially large parts of the multiverse with an extremely small value of the cosmological constant, compatible with our existence. Here the word ``our'' is used in a very narrow sense. We are talking not about ``life in general,'' but about the correlations between our own properties and the properties of the part of the multiverse where we can live. For example, different types of living organisms populate the Earth, but fish is typically surrounded by water and people typically live on dry land. Similarly, other types of life might exist in parts of the universe with much greater values of the cosmological constant, but these parts would be less hospitable for us. That is all that we can say with reasonable certainty, and this is already quite sufficient to make the cosmological problem if not completely and unambiguously solved then at least significantly ameliorated. 

We do not know yet which of the recently developed approaches to the  theory of the inflationary multiverse is going to be more fruitful, and how far we will be able to go in this direction.  One  way or another, it would be very difficult  to forget about what we just learned and return to our search for the theory which unambiguously explains all parameters of our world. Now we know that some features of our part of the universe may have an unambiguous explanation, whereas some others can be purely environmental and closely correlated with our own existence.

When inflationary theory was first proposed, its main goal was to address many problems which at that time could seem rather metaphysical: Why is our universe so big? Why is it so uniform? Why parallel lines do not intersect? It took some time before we got used to the idea that the large size, flatness and uniformity of the universe should not be dismissed as trivial facts of life. Instead of that, they should be considered  as  observational data requiring an explanation, which was provided with the invention of inflation.

Similarly,  the existence of an amazingly strong correlation between our own properties and the values of many parameters of our world, such as the masses and charges of electron and proton, the value of the gravitational constant, the amplitude of spontaneous symmetry breaking in the electroweak theory, 
the value of the vacuum energy, and the dimensionality of our world, is an experimental fact requiring an explanation. A combination of the theory of inflationary multiverse and the string theory landscape provides a unique framework where this explanation can be found.

\section{Conclusions}
 
Three decades ago, inflationary theory looked like an exotic product of vivid scientific imagination. Some of us believed that it possesses such a great explanatory potential that it must be correct; some others thought that it is too good to be true. Not many expected that it is possible to verify any of its predictions in our lifetime.  Thanks to the enthusiastic work of many scientists, inflationary theory is gradually becoming a broadly accepted cosmological paradigm, with many of its predictions being confirmed by observational data.  

The new data release by Planck 2013 stimulating the development of new cosmological theories, by changing the goal from finding various complicated models capable of describing large local non-Gaussiantiy to the development of new elegant models of inflation capable of explaining increasingly precise data in the $(n_{s},r)$ plane. The simplicity of the chaotic inflation scenario with potentials $\phi^{n}$ is unmatched, but what if the observational data force us to modify these models? Just few months ago, we did not know many good inflationary models which would naturally predict the data favored by Planck 2013. Now the situation has changed. The existence of the universal attractor regime for a large set of different inflationary models does not guarantee that we are on the right track, but it is hard to ignore that all of these cosmological attractors point in the same direction, and their predictions converge at the ``sweet spot'' in the $(n_{s},r)$ plane preferred by WMAP9 and Planck 2013. It may be equally important that if one follows the attractor trajectories in the opposite direction (i.e. takes the limit $\xi \to 0$ or $\alpha \to \infty$ in this class of models), one continuously interpolates back to the predictions of the simplest models of chaotic inflation.

 
I am grateful to the organizers of the Les Houches School ``Post-Planck Cosmology'' in July-August 2013 for 
their hospitality. I would like to thank my friends and collaborators who made my work on inflation so enjoyable, especially Richard Bond, Sergio Ferrara, George Efstathiou, Renata Kallosh, Slava Mukhanov,  Diederik Roest, Eva Silverstein and Alexander Westphal.  This work was supported by the NSF Grant PHY-1316699. 


\section{Appendix: Alternatives to inflation?}\label{alt}

There were many attempts to propose an alternative to inflation in recent years. In general, this could be a very healthy tendency. If one of these attempts  succeeds, it will be of great importance. If none of them are successful, it will be an additional demonstration of the advantages of inflationary cosmology. However, since the stakes are high, we are witnessing a growing number of premature announcements of success in developing an alternative cosmological theory.

Perhaps the most famous example is the ekpyrotic/cyclic scenario. I would not return to the rather uninspiring history of this scenario in this paper, but it will help us to understand the nature of the recent comments about inflation made by some of the authors of the ekpyrotic/cyclic scenario and their collaborators \cite{Steinhardt:2011zza,Ijjas:2013vea,Ijjas:2013sua}. 

The ekpyrotic/cyclic scenario   \cite{KOST} claims that it can solve all
cosmological problems without using the stage of inflation. However, the original version of the ekpyrotic scenario \cite{KOST} did not work. The number of incorrect statements made in  \cite{KOST}  was unusually high. Instead of solving the homogeneity problem, this scenario required the universe to be exactly homogeneous from the very beginning \cite{KKL}.  The authors claimed that their mechanism was based on string theory, but they never made any attempt to implement it in the modern versions of string theory with stabilized moduli. Most of the experts agreed that the mechanism of generation of the scalar perturbations of metric proposed in \cite{KOST} did not produce the desirable perturbations, see e.g. \cite{Lyth:2001pf}. After many years of fierce debates, the authors of the ekpyrotic scenario conceded and introduced another, much more complicated mechanism, which was borrowed from the curvaton mechanism used in the inflationary theory. Few years later, the authors made an attempt to return back to the original single-field mechanism of generation of perturbations \cite{Khoury:2009my}, but an investigation of this new proposal revealed that their work was based on investigation of processes at density 100 orders of magnitude higher than the Planck density \cite{Linde:2009mc}.   But the most significant problem of the ekpyrotic scenario proposed in \cite{KOST} was that instead of the big bang predicted in  \cite{KOST}, there was a big crunch  \cite{KKL,KKLTS}. This error completely invalidated the original scenario.

For that reason, the ekpyrotic scenario was replaced by the cyclic scenario, which postulated the existence of an infinite number of periods of expansion and contraction of the universe  \cite{cyclic}.  The origin of the required scalar field potential in this model remains unclear, and the very existence of the cycles postulated in \cite{cyclic} have never been demonstrated. When we analyzed this scenario  using the particular potential given in  \cite{cyclic}, and took into account the effect of particle production in the early universe, we found a very different cosmological regime \cite{Felder:2002jk,Linde:2002ws}.

Cyclic scenario relied on the existence of an infinite number of very long stages of ``superluminal expansion,'' i.e.
inflation, in order to solve the major cosmological problems. In particular, the recent version of this scenario discussed in \cite{Lehners:2008qe} requires at least 60 e-foldings of inflation of the universe (does this number look familiar?) in the dark energy state. In this sense,  cyclic scenario is not a true alternative to inflationary scenario, but its rather peculiar version. The main difference between the usual inflation and the cyclic inflation is the energy scale of inflation, and the mechanism of generation of density perturbations.
However, since the theory of density perturbations in cyclic
inflation requires a solution of the cosmological singularity
problem, it is difficult to say anything definite about it. 

Originally there was a hope that the problem of the cosmological singularity, which emerges during the ekpyrotic/cyclic collapse, will be solved in the context of string theory, but despite the attempts of the best experts in string theory, this problem remains unsolved  \cite{Liu:2002ft,Horowitz:2002mw,Berkooz}. 
Few years ago, there was an attempt to revive the original (non-cyclic) version of the ekpyrotic scenario  by involving  a nonsingular bounce. This regime requires violation of the null energy condition \cite{KKLTS}, which usually leads to a catastrophic vacuum instability and/or causality violation. One may hope to avoid these problems in the ghost condensate theory; see a series of papers on this subject \cite{Creminelli:2006xe,Buchbinder:2007ad,Creminelli:2007aq}. However, investigation of these models demonstrated that they contained ghosts resulting in a nearly instant vacuum decay in this theory \cite{Kallosh:2007ad}. Since that time, there were many additional modifications of this scenario;  it became even more complicated than before, consisting of different parts describing Galileons, ghost condensate, various higher order terms required for the cosmological bounce, additional higher order terms required for stabilization of the theory, etc. 

Another attempt to solve the singularity problem  was made by Bars, Steinhardt and Turok   \cite{Bars:2011th}. They suggested that one can alleviate this problem by passing through the antigravity regime first. A possibility of a smooth transition from gravity to antigravity was proposed in my own paper more than 30 years ago  \cite{Linde:1979kf}. However, the subsequent investigation has shown that during the transition from gravity to antigravity, the effective gravitational constant blows up and the cosmological singularity develops  \cite{Starobinskii1981}. 
Recently it was shown  \cite{Carrasco:2013hua} that a similar problem appears in the models studied by Bars, Steinhardt and Turok   \cite{Bars:2011th}: At the transition from gravity to antigravity and back, the universe passes through the singularities where the curvature invariants become infinitely large. In the vicinity of each of these two singularities, the higher order corrections blow up, and the standard classical GR methods employed in  \cite{Bars:2011th} become inapplicable \cite{Kallosh:2013oma,Carrasco:2013hua}.  Bars, Steinhardt and Turok agreed with the existence of the singularities found in \cite{Kallosh:2013oma,Carrasco:2013hua}, but responded that it does not change their conclusions ``one iota'' \cite{Bars:2013qna}. However, in the revised version of their response they conceded that ``Of course, our purely classical analysis does not include strong quantum gravity effects near the singularities because the technology does not yet exist to do those computations.''

\

This pattern of making strong incorrect statements did not remain unnoticed, and scientists voted with their feet. For example, in a large review of modern cosmology associated with the planning of the CMBPol Mission, which was written in 2009 by a team of 60 cosmologists, a critical discussion of alternatives to inflation is delegated to the Appendix \cite{Baumann:2008aq}.  A large recent review of theoretical cosmology, CMB and large scale structure, a part of the Snowmass 2013 \cite{Abazajian:2013vfg}, is completely dedicated to analysis of observational consequences of inflationary theory. Alternative models are mentioned only in 3 lines in this large article written by more than 90 collaborators. Theoretical interpretation of the Planck 2013 data release \cite{Ade:2013uln} is almost entirely concentrated on the inflationary theory, which an important exception of the paper on non-Gaussianity, which says: ``Ekpyrotic/cyclic scenarios were shown to be under pressure from the Planck data.'' The paper concludes that ``With these results, the paradigm of standard single-field slow-roll inflation has survived its most stringent tests to-date.''

\begin{figure}[t!h!]
\centering
\includegraphics[scale=.65]{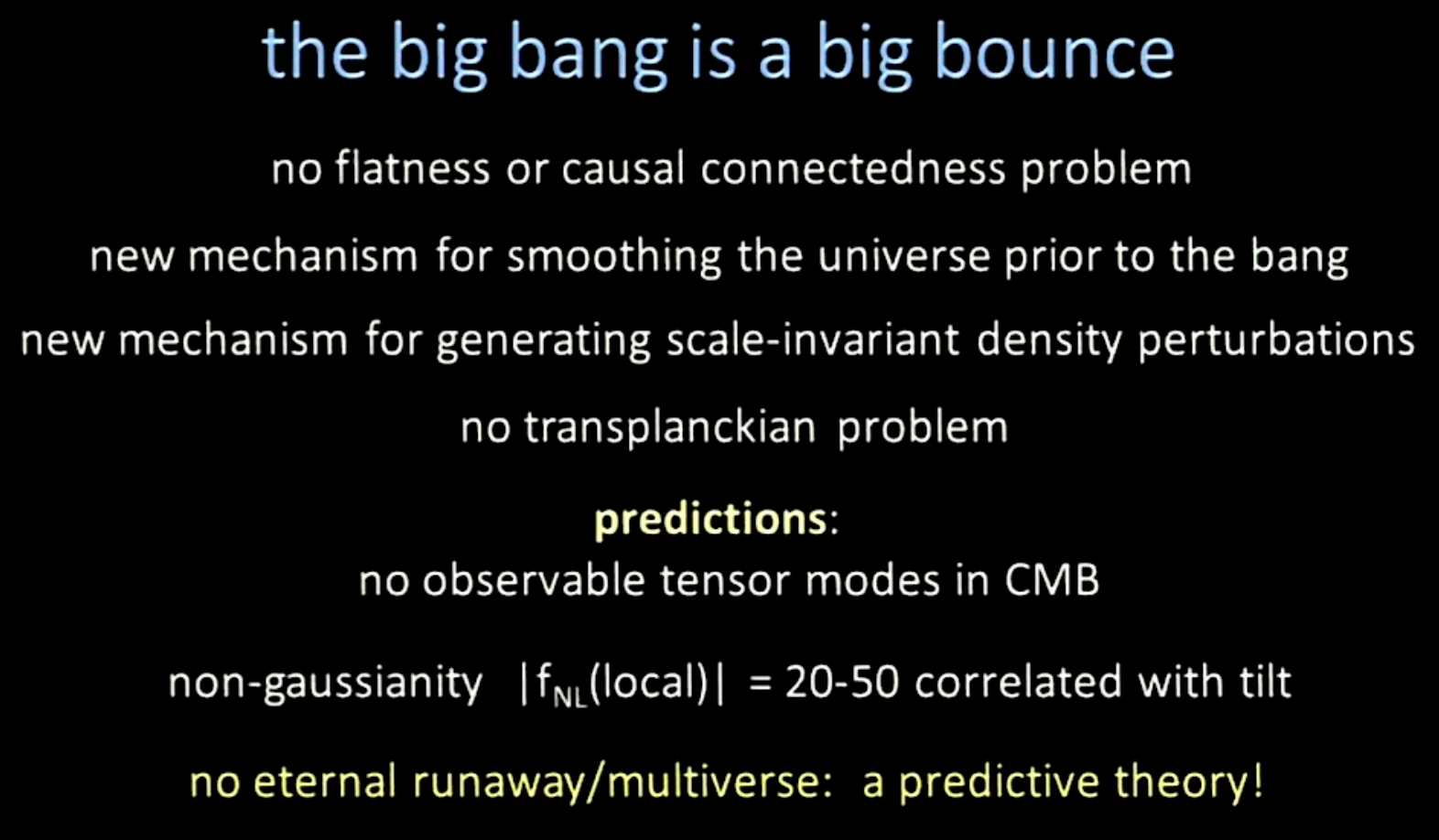} 
\caption{\footnotesize{From the talk given by Steinhardt at the Perimeter Institute, March 6, 2013, http://pirsa.org/13030079. }}
\label{stein}
\end{figure}

Indeed, according to the talk by Steinhardt \cite{RethinkingCosmology} two weeks before the Planck 2013 data release, if one does not allow accidental cancellations, the ekpyrotic/cyclic scenario predicts $f_{\rm NL}^{\rm local} = 20-50$, see Fig. \ref{stein}.
Meanwhile, according to the Planck data release,
 $f_{NL}^{\rm local} = 2.7 \pm 5.8$, which confirms  predictions of the simplest inflationary models with an accuracy $O(10^{{-4}})$ and rules out the models predicting $f_{\rm NL}^{\rm local} = 20 -  50$  \cite{Ade:2013uln}. 
 
The bet was made, and lost.  Three weeks after the Planck data release, Steinhardt and collaborators issued two new papers, where they made opposite claims. In one of them it was announced that Planck 2013 supports predictions of the cyclic scenario \cite{Lehners:2013cka}. Another paper \cite{Ijjas:2013vea} was entitled  ``Inflationary paradigm in trouble after Planck2013.''

In the absence of any convincing solution of the singularity problem in the cyclic scenario, it is hard to make any bets on what it actually predicts and why the authors changed their predictions so quickly, so I will not discuss it here. However, the paper on inflationary theory, contradicting the main conclusions of the Planck team, deserves a detailed analysis. The first response to  \cite{Ijjas:2013vea} was given in my talk at the KITP conference in April 2013, http://online.kitp.ucsb.edu/online/primocosmo-c13/linde/, and in my Les Houches lectures. Recently these issues were also addressed in the paper by Guth, Kaiser and Nomura \cite{Guth:2013sya}. Their conclusion was that ``cosmic inflation is on a stronger footing than ever before.''  I fully agree with it, and I would like to strengthen some of the points they made.

Let me briefly summarize the main arguments made in \cite{Ijjas:2013vea}: 

1) Planck data strongly disfavor convex potentials used in the simplest models of chaotic inflation. Other models, such as new inflation, or complicated versions of chaotic inflation, either suffer from the problem of initial conditions, or are ``unlikely.''

2) The curse of the multiverse: After Planck, eternal inflation becomes unavoidable, which is an unmitigated disaster.

3) LHC data suggest that the Higgs vacuum is metastable. Inflation would induce a catastrophic transition to the state with negative energy density.

\vskip 6pt
 
I will comment in reverse order.

\vskip 6pt

3) LHC data indeed suggest that in the simplest non-supersymmetric versions of the standard model the present vacuum state in metastable. However, the corresponding energy scale is many orders of magnitude smaller than the energy scale during inflation. Therefore the shape of the Higgs potential during inflation was completely different. The very existence of inflationary fluctuations is sufficient to restore the symmetry in the standard model, and the same can be achieved by the interaction of the inflaton field with the Higgs field, see e.g. \cite{Lebedev:2012sy}. 

The real threat appears only if the Higgs field itself plays the role of the inflaton field. In the Higgs inflation scenario, the initial value of the Higgs field is supposed to be very large. When inflation ends, it rolls down, but instead of rolling to our metastable vacuum state, it may fall to the true vacuum with $V<0$, in which case the universe collapses \cite{Degrassi:2012ry}. But this is not the general problem of inflation, this is just a problem of a very specific and rather exotic model. It may totally disappear once one considers a supersymmetric version of the Higgs inflation, which requires a specific supergravity implementation of the NMSSM model \cite{Ferrara:2010yw,Ferrara:2010in}. The existence of a metastable minimum was not established in this scenario. Most importantly, this problem does not appear at all in any of the models considered in the present lectures. So this is not a real problem for inflation.

\vskip 5pt

2) Some people do not like eternal inflation, multiverse and the anthropic principle. In particular, according to Steinhardt, the emergence of the concept of the inflationary multiverse ``is the failure of two favorite theoretical ideas - inflationary cosmology and string theory'' \cite{SteinhardtEdge}.  Many others disagree and believe that the idea of the inflationary multiverse is a cornerstone of the new scientific paradigm which flourished during the last decade after its unification with string theory into the concept of the string theory landscape. 
Among those who share this view is Steven Weinberg. In his paper ``Living in the multiverse'' \cite{Weinberg:2005fh} he compared the recent developments with the invention of special theory of relativity and said ``Now we may be at a new turning point, a radical change in what we accept as a legitimate foundation for a physical theory.''

My own opinion about eternal inflation and the theory of the multiverse is expressed in Section \ref{land}, and I will not repeat it here.  In \cite{Steinhardt:2011zza,Ijjas:2013vea}, the authors mostly concentrated on the problems with one of the two probability measures proposed back in 1993 \cite{Linde:1993xx}. However, they did not mention that this particular problem was resolved long ago in \cite{Linde:2007nm,Linde:2008xf}, and the second of the two measures proposed in \cite{Linde:1993xx} did not suffer from this problem. For a recent discussion of the probability measure in the multiverse, I refer the readers to the recent paper by Guth, Kaiser and Nomura  \cite{Guth:2013sya}. 

It is quite interesting that the proponents of the ekpyrotic/cyclic theory are talking now not only about the problems of inflationary theory, but about ``the failure of two favorite theoretical ideas - inflationary cosmology and string theory'' \cite{SteinhardtEdge}. It is a rather unexpected turn of events, since the original version of the ekpyrotic theory was supposed to be a part of string theory. But in all known versions of string theory with stable or metastable vacua one expects enormous number of dS vacua, which leads to the theory of the multiverse. Thus if the ekpyrotic/cyclic models were based on the modern versions of string theory, they would automatically become members of the `multiverse club' to which they do not want to belong.

In order to propose a true alternative to the theory of inflationary multiverse one should achieve several incredibly difficult goals: One should propose an alternative to inflation, and also an alternative to the most developed versions of string theory, explain why only one vacuum of string theory can actually exist and why all other $10^{500}$ vacua are forbidden. In addition, one should find an alternative solution of the cosmological constant problem and many other coincidence problems, which have been addressed so far only in the context of the theory of the multiverse. I am unaware of any proposal how one could simultaneously achieve all of these goals.

\vskip 5pt

1) Finally let us return to the problem of initial conditions. Is it true that the Planck data imply that the initial conditions required for inflation are improbable?

In the earlier discussion of this issue in  \cite{Steinhardt:2011zza},  the probability measure introduced in \cite{Gibbons:2006pa} was used to argue that initial conditions for inflation are improbable.  However, in \cite{Linde:2007fr} and also in section \ref{ini} of my lectures it was explained why this measure is flawed and cannot be applied for investigation of initial conditions for inflation. Closely related arguments can be also found in \cite{Schiffrin:2012zf} and in other papers. 

In  \cite{Ijjas:2013vea} the authors used my own approach proposed in \cite{Linde:1985ub} and described in section \ref{ini} of these lectures, where I argued that the simplest versions of inflationary cosmology, the ones where inflation can start at the Planck density, do not suffer from the problem of initial conditions. The authors of  \cite{Ijjas:2013vea} admitted that these arguments are almost universally accepted, but then they claimed that the models of chaotic inflation where inflation can start at the Planck density either contradict the Planck data, or are  ``unlikely.''

An answer to this statement can be found in section \ref{chinflsugra} of these lectures, where an example of a chaotic inflation model is presented which has a simple polynomial potential \rf{polyn} and makes predictions perfectly consistent with the Planck data. Moreover, this simple model was implemented in the context of supergravity. Inflation in this model begins at the Planck density, so it satisfies the required conditions formulated in section \ref{ini}. Is this model ``unlikely''? Beauty is in the eye of the beholder, but I wonder who would consider a model consisting of a mixture of Galileons, ghost condensate, higher order terms required for the cosmological bounce and additional higher order terms required for stabilization of the theory \cite{Koehn:2013upa} more ``likely'' than a simple model with a polynomial potential.

What about other models, such as the Starobinsky model, the model $\lambda\phi^{4}/4+\xi\phi^{2}R/2$, or the cosmological attractor models described in my lectures? Inflation in these models begins at $V\ll 1$. Does it mean that inflation in these models is impossible or unlikely? This question was addressed in section \ref{torus}, where three different mechanisms are discussed which can provide natural initial conditions for inflation in the models of this type. These mechanisms are not new, they have been discussed in the inflationary literature for years, so it is a bit surprising that none of these mechanisms was even mentioned in \cite{Steinhardt:2011zza,Ijjas:2013vea,Ijjas:2013sua}. Many other possibilities appear in the context of the string theory landscape, see e.g.  \cite{Guth:2013sya}. 

\vskip 5pt

To conclude,  at the moment it is hard to see any real alternative to inflationary cosmology, despite an active search for such alternatives. All of the proposed alternatives are based on various attempts to solve the singularity problem: One should either construct a bouncing nonsingular cosmological solution, or learn what happens to the universe when it goes through the singularity. This problem  bothered cosmologists for nearly a century, so it would be great to find its solution, quite independently of the possibility to find an alternative to inflation. None of the proposed alternatives can be consistently formulated until this problem is solved. 

In this respect, inflationary theory has a very important advantage: it works practically independently of the solution of the singularity problem. It can work equally well after the singularity, or after the bounce, or after the quantum creation of the universe. This fact is especially clear in the  eternal inflation scenario: Eternal inflation makes the processes which occurred near the big bang practically irrelevant for the subsequent evolution of the universe. 

\

{\bf Note added:}

After this paper was submitted to the arXiv, Ijjas, Steinhardt and Loeb (ISL) responded in \cite{Ijjas:2014nta} to the critical evaluation of their work in the paper  by Guth, Kaiser and Nomura  \cite{Guth:2013sya} and also in my lectures. The main part of their response consists of a philosophical debate of the multiverse, mostly concentrated on the ``youngness problem'' related to one of the probability measures proposed in 1993 in \cite{Linde:1993xx}. Just as in their previous papers, ISL did not mention that this problem was solved long ago in \cite{Linde:2007nm,Linde:2008xf}. Similarly, they did not mention that the second of the two volume-weighted measures proposed in \cite{Linde:1993xx} also does not suffer from the problem they discuss. These facts have been emphasized twice in my lectures to which ISL supposedly respond. And this problem does not appear at all in many other more sophisticated versions of the probability measure presently considered in the literature.

ISL also discuss the problem of initial conditions for inflation. In section  \ref{ini} of my lectures  it  was argued that for inflation to begin in the simplest versions of chaotic inflation scenario, it is sufficient to have a relatively uniform domain of a Planckian size having all different energy forms of the same order. This is the absolutely minimal requirement one can imagine.  In their previous paper \cite{Ijjas:2013vea}  ISL admitted that this conclusion is ``almost universally accepted,'' but in their latest paper \cite{Ijjas:2014nta} instead of using these considerations they concentrated on the probability measure introduced in \cite{Gibbons:2006pa}. The reason why this particular probability measure cannot tell us anything about initial conditions for inflation is explained in \cite{Linde:2007fr,Schiffrin:2012zf}, and in section  \ref{ini} of my lectures. Then ISL repeated their statement that inflation at small energy density is exponentially improbable. In section  \ref{torus} of these lectures I described three different solutions of the problem of initial conditions for the low-scale inflation. These solutions are not even mentioned by ISL in \cite{Ijjas:2014nta}. It is difficult to avoid the impression that ISL have formed their opinion about the present status of inflationary cosmology without paying enough attention to what actually happens there.

\end{document}